\documentclass{ws-ijmpe}

\usepackage[dvips]{color}

\newcommand{\be}{\begin{equation}}
\newcommand{\ee}{\end{equation}}
\newcommand{\bea}{\begin{eqnarray}}
\newcommand{\eea}{\end{eqnarray}}

\begin{document}

\markboth{Linnyk, Bratkovskaya, Cassing}{Charm in heavy-ion collisions}

\catchline{}{}{}{}{}

\title{OPEN AND HIDDEN CHARM IN PROTON-NUCLEUS AND HEAVY-ION COLLISIONS}

\author{O.~LINNYK}

\address{Frankfurt Institute for Advanced Studies, Ruth-Moufang-Str. 1\\
60438 Frankfurt am Main, Germany\\
Linnyk@fias.uni-frankfurt.de}

\author{E.~L.~BRATKOVSKAYA}

\address{Frankfurt Institute for Advanced Studies, Ruth-Moufang-Str. 1;\\
Institut f\"ur Theoretische Physik, Johann Wolfgang Goethe University, Max-von-Laue-Str. 1\\
60438 Frankfurt am Main, Germany\\
Elena.Bratkovskaya@th.physik.uni-frankfurt.de}

\author{W.~CASSING}

\address{Institut f\"ur Theoretische Physik, Universit\"at Giessen, Heinrich-Buff-Ring 16\\
35392 Giessen, Germany\\
Wolfgang.Cassing@theo.physik.uni-giessen.de}



\maketitle

\begin{history}
\received{(received date)}
\revised{(revised date)}
\end{history}

\begin{abstract}
We review the collectivity and the suppression pattern of charmed
mesons - produced in proton-nucleus and nucleus-nucleus collisions
at SPS ($\sim$ 158 A$\cdot$GeV) and RHIC energies ($\sim$ 21
A$\cdot$TeV) - in comparison to dynamical and thermal models. In
particular, we examine  the charmonium `melting'  and the `comover
dissociation' scenarios - implemented in a microscopic transport
approach - in comparison to the available data from the SPS and
RHIC. The analysis shows that the dynamics of $c, \bar{c}$ quarks
at RHIC are dominated by partonic or `pre-hadronic' interactions
in the strongly coupled plasma stage and can neither be modeled by
`hadronic' interactions nor described appropriately by color
screening alone. Both the `charmonium melting' and the hadronic `comover absorption
and recreation model' are found, however,  to be compatible with
the experimental observation at SPS energies; the experimental
ratio of $\Psi^\prime/J/\Psi$  versus centrality clearly favors
the `hadronic comover' scenario. We find that the collective flow
of charm in the purely hadronic Hadron-String Dynamics (HSD)
transport appears compatible with the data at SPS energies, but
substantially underestimates the data at top RHIC energies. Thus,
the large elliptic flow $v_2$ of $D$-mesons and the low
$R_{AA}(p_T)$ of $J/\Psi$ seen experimentally have to be
attributed to early interactions of non-hadronic degrees of
freedom. Simultaneously, we observe that non-hadronic interactions
are mandatory in order to describe the narrowing of the $J/\Psi$
rapidity distribution from $pp$ to central $Au+Au$ collisions at
the top RHIC energy of $\sqrt{s} = 200$~GeV.  We demonstrate
additionally that the strong quenching of high-$p_T$ $J/\Psi$'s in
central $Au+Au$ collisions indicates that a large fraction of
final $J/\Psi$ mesons is created by a coalescence mechanism close
to the phase boundary. Throughout this review we, furthermore,
provide predictions for charm observables from Au+Au collisions at
FAIR energies of 25 - 35 A$\cdot$GeV.
\end{abstract}

\newpage
\tableofcontents

\markboth{Linnyk, Bratkovskaya, Cassing}{Charm in heavy ion collisions}


\section{Introduction}

The formation of a quark-gluon plasma and its transition to
interacting hadronic matter has motivated a large community for
almost three decades~\cite{QM01}. According to current
understanding, the universe in the `Big Bang' scenario has evolved
from a quark-gluon plasma (QGP) to color neutral hadronic states
within the first second of its lifetime. In this context, the
phase transition from partonic degrees of freedom (quarks and
gluons) to interacting hadrons is a central topic of modern
high-energy physics. In order to understand the dynamics and
relevant scales of this transition laboratory experiments under
controlled conditions are performed with relativistic
nucleus-nucleus collisions. The study of nuclear matter under
extremely high baryon density and temperature -- where according
to lattice quantum chromodynamics (QCD)~\cite{Fodor} the hadronic
matter transforms to a strongly interacting quark-gluon plasma
(sQGP) -- is the aim of a variety of experiments at current and
future facilities: NA38, NA50 and NA60 at the
Super-Proton-Synchrotron (SPS)~\cite{NA38,NA50O,NA60}; PHENIX,
STAR, PHOBOS and BRAHMS at the Relativistic-Heavy-Ion-Collider
(RHIC)~\cite{STARS}; ALICE at the Large-Hadron-Collider
(LHC)~\cite{heavyAlice,LHC.general}; CBM and PANDA at the Facility
for Antiproton and Ion Research (FAIR)~\cite{FAIR.general}; NA61
at the SPS Heavy-Ion and Neutrino Experiment (SHINE)~\cite{Shine};
MPD at the Nuclotron-based Ion Collider Facility
(NICA)~\cite{Nica1}.

Relativistic nucleus-nucleus collisions have been studied so far
at beam energies from 0.1 to 2 $A\cdot$GeV at the SIS
(SchwerIonen-Synchrotron), from 2 to 11.6 $A\cdot$GeV at the AGS
(Alternating Gradient Synchrotron) and from 20 to 160 A$\cdot$GeV
at the SPS \cite{NA49_new,NA49_T}. While part of these programs
are closed now,  the heavy-ion research has been extended at RHIC
with Au+Au collisions at invariant energies $\sqrt{s}$ from $\sim
20$ to 200 GeV (equivalent energies in a fixed target experiment:
0.2 to 21.3 A$\cdot$TeV).  In the near future, further insight
into the physics of matter at even more extreme conditions will be
gained at the LHC, which will reach center-of-mass energies of the
TeV scale. Apart from LHC, the SPS successor SHINE will operate at
CERN in order to scan the 10A-158A$\cdot$GeV energy range with
light and intermediate mass nuclei~\cite{Shine}. At FAIR, which is
expected to start operation in 2015, collisions of gold nuclei
from  5 A$\cdot$GeV up to 35 A$\cdot$GeV will be studied, thus
exploring the high baryon density region of the nuclear matter
phase diagram. At NICA it is planned to start the experimental
program of  colliding Au and/or U ions as well as polarized light
nuclei at energies up to of 5 A$\cdot$GeV in 2013 (an upgrade to 9
A$\cdot$GeV is foreseen~\cite{Nica2}).

In fact, estimates based on the Bjorken formula~\cite{bjorken} for the energy density achieved in
central Au+Au collisions suggest that the critical energy density for the formation of a QGP is by
far exceeded during a few fm/c in the initial phase of the collision at RHIC energies ($\sqrt{s}$
up to 200~GeV), but sufficient energy densities ($\sim$ 0.7-1 GeV/fm$^3$, {\it cf.}
Ref.~\cite{Karsch}) might already be achieved at  AGS energies of
$\sim$ 10 A$\cdot$GeV~\cite{HORST,Cass00,Arsene} and thus also for the energy regime of FAIR. The
crucial question is, however, whether the partonic system really reaches thermal and chemical
equilibrium in relativistic nucleus-nucleus collisions. Nonequilibrium models are needed to
trace the entire collision history, so that one can study the nature of the transition and extract
the characteristics of the partonic phase from data.

Currently many properties of the new sQGP phase are still under
debate and practically no dynamical concepts are available to
describe the freeze-out of partons to color neutral hadrons that
are subject to experimental detection. Early concepts of the QGP
were guided by the idea of a weakly interacting system of partons
(quarks, antiquarks and gluons), because the entropy density and
energy density were found in lattice QCD to be close (within 20\%
accuracy) to the Stefan Boltzmann limit for a relativistic
noninteracting system~\cite{Karsch}. However, this notion had to
be given up in the last years, since experimental observations at
RHIC indicated that the new medium created in ultrarelativistic
Au+Au collisions interacted even more strongly than hadronic
matter. Moreover, in line with earlier theoretical studies in
Refs.~\cite{Thoma:2005uv,Thoma:2005aw,Andre,Shuryak} the medium showed phenomena of an
almost perfect liquid of partons~\cite{STARS,Miklos3}, {\it i.e.}
a strongly coupled system showing high collectivity. The most
intriguing questions have become: what degrees of freedom are
relevant in this new sQGP state and how is the transition from the
partonic liquid to the gas of interacting hadrons realized?

The $c, \bar{c}$ quark degrees of freedom are  of particular
interest in context with the phase transition to the sQGP. The
heavy flavor sector is important for unraveling the actual
dynamics from the experimental side since the high masses of the
charm and especially bottom quarks provide independent (and new)
energy scales. The hadronic bound states (with a $c$ or $\bar{c}$
quark) have a much larger mass than the ordinary hadrons, and it
is expected that $c\bar{c}$ pairs can only be formed in the very
early phase of the heavy-ion
collision~\cite{Satz,Satznew,Satzrev,KSatz}. As has been proposed
in Ref.~\cite{Rafelski.Mueller} the strangeness degrees of freedom
 might play an
important role in distinguishing hadronic and partonic dynamics.
The kaons (and antikaons), indeed, have been proven to provide a
suitable probe for the compression phase in heavy-ion reactions at
SIS energies. One expects that, due to the new scale introduced by
the charm flavor, $D$-mesons can be used in a similar way to probe
the high density matter at FAIR around the threshold for open
charm production~\cite{CBM.book}. Moreover, heavy quarkonia
($c\bar{c}$, $b\bar{b}$) might no longer be bound at high
temperature. Due to different binding energies, quarkonium states
should `melt' at different temperatures and might provide the
cleanest ``measurement'' for the energy densities
achieved~\cite{Satz}.

Apart from the total and relative abundances of charmonia and open
charm mesons also their collective properties are of interest.
Here especially the transverse momentum (or mass) spectra are
expected to provide valuable insight into the dynamics in either
the very early or late phase~\cite{Dumitru,NXU,Cass01,Rapp04}. We
recall that the collective properties of open charm mesons go
along with a suppression (relative to scaled $pp$ reactions) of
high transverse momentum $p_T$ particles as observed at RHIC by
the STAR and PHENIX Collaborations~\cite{PHENIX_v2DD} as well as
the suppression of charmonia ($J/\Psi$, $\Psi'$). Since the charm
quarks are produced early in the reaction, their rescattering --
reflected in high $p_T$ suppression -- and collectivity (in the
elliptic flow $v_2(p_T)$) signal more sensitively the dynamics of
the early phase. All the arguments above qualify charm-flavored
particles as practical and promising probes for an exploration of
QCD matter.

The present review is structured as follows: We shall start in
Section~\ref{elementary} with a description of elementary
reactions involving `charm' and introduce the transport models
that will be employed for a dynamical description of
proton-nucleus and nucleus-nucleus collisions, respectively. The
latter dynamical models allow for an overview of the local energy
densities achieved in Au+Au collisions from FAIR to RHIC energies
(Section~\ref{energy}).  We continue with `cold nuclear matter'
effects in proton-nucleus reactions in Section~\ref{pA} and
provide information about the `hadronic environment' of charmonia
encountered in relativistic heavy-ion reactions in
Section~\ref{abundancies}. The leading concepts/models for
charmonium dynamics are presented in Section 6 as well as their
actual implementation in the microscopic transport approach. The
relative abundances of charmonia and open charm mesons are
presented in Section 7 together with a discussion of the anomalous
suppression of charmonia ($J/\Psi$, $\Psi'$) in comparison to the
experimental data available. The issue of charmonium chemistry is
examined in Section~\ref{chemi} in comparison with the statistical
hadronization model. Sections~\ref{mT} and \ref{quenching} are
devoted to the energy loss of charmonia in the dense medium, i.e.
in particular to the transverse mass spectra and high $p_T$
quenching. The collective flow properties of open charm and
charmonia are studied for the elliptic flow $v_2$ in
Section~\ref{flow}. We conclude with a summary and discussion of
open questions in Sections~\ref{conclusions} and~\ref{open}.

\section{Transport models and elementary reactions}
\label{elementary}

The main difficulty in the study of nuclear matter under extreme
conditions as created in relativistic heavy-ion collisions is that
the information about the initial sQGP stage of matter can be
obtained only {\em indirectly} from the measurement of hadronic
observables. The sQGP signal might be strongly distorted by the
hadronization process and final state interactions of the hadrons.
In order to reliably subtract the ``trivial"  hadronic
contribution from the sQGP signal, one needs a microscopic
transport dynamical approach. Also, the possible equilibration of
quark-gluon matter is on central interest. Such issues of
equilibration phenomena are traditionally examined within
nonequilibrium relativistic transport theory
\cite{Cass99,HORST,Cassing:1990dr,Cassing:1991cy,Koreview,UrQMD1}.

Ideally, one would use a microscopic transport model containing
the proper degrees of freedom -- quarks and gluons in the initial
phase and hadrons in the final phase -- and parton dynamics which
are consistent with the Lattice QCD equation of state. The
development of such approaches is in progress
\cite{PHSD,Ko_AMPT,PHSD.transport}. However, a proper
understanding of the transport properties of the partonic phase is
still lacking (see Section~\ref{open} for details). One has to
keep in mind that most of the results in this review have been
obtained based on hadron-string or hydrodynamical models without
including an explicit phase transition to the QGP.
As we will see, there are a number of questions that can be
answered before the development of a nonperturbative parton-hadron
transport approach is completed. For example, the interactions of
$J/\Psi$'s with formed hadrons in the late stages of the collision
(when the energy density falls below a critical value of about 1
GeV/fm$^3$ corresponding roughly to the critical energy density
for a parton/hadron phase transition) gives a sizable contribution
to its anomalous suppression at all beam energies as demonstrated
in
Refs.~\cite{Olena.SPS,Olena.RHIC,Olena.SQM07,Cass01,brat03,brat04}.
Accordingly, this more obvious hadronic contribution has to be
incorporated when comparing possible models for QGP-induced
charmonium suppression to experimental data.

In any case, the link of the underlying physics to the heavy ion
experiments should be provided by dynamical transport models, such
as UrQMD~\cite{UrQMD1,UrQMD2}, HSD~\cite{HSD96,Cass99},
GiBUU~\cite{Effe99gam,Larionov}, RQMD~\cite{Sorge},
QGSM~\cite{QGSM,QGSM2}, or AMPT~\cite{Ko_AMPT}. we recall that
microscopic (pre-)hadronic transport models describe the formation
and distributions of many hadronic particles from SIS to SPS
energies reasonably well~\cite{Weber}. Furthermore, the nuclear
equation of state has been extracted by comparing transport
calculations to hadron flow data up to AGS
energies~\cite{Andronic03,Andronic01,Soff99,Csernai99,Sahu1,Sahu2}.
In particular, at SIS energies, microscopic transport models
reproduce the data on the excitation function of the proton
elliptic flow $v_2$ when incorporating a soft, momentum-dependent
equation of
state~\cite{Andronic:2000cx,Andronic:2004cp,Andronic99,Larionov}.
In addition to nucleus-nucleus collisions from SIS to SPS
energies~\cite{Weber,Bratkovskaya:2004kv}, the HSD transport
approach is found to work reasonably well also at RHIC energies
for the `soft' hadron abundances, so that the `hadronic
environment' for open charm mesons and charmonia should be
sufficiently realistic~\cite{brat03}. One also finds generally a
good agreement among the predictions of different transport models
for hadron multiplicities. Modest differences between the HSD and
UrQMD transport results for pion and kaon multiplicities can be
attributed to different implementations of string formation and
fragmentation, which are not sufficiently controlled by
experimental data.

\begin{figure}
\centerline{\psfig{file=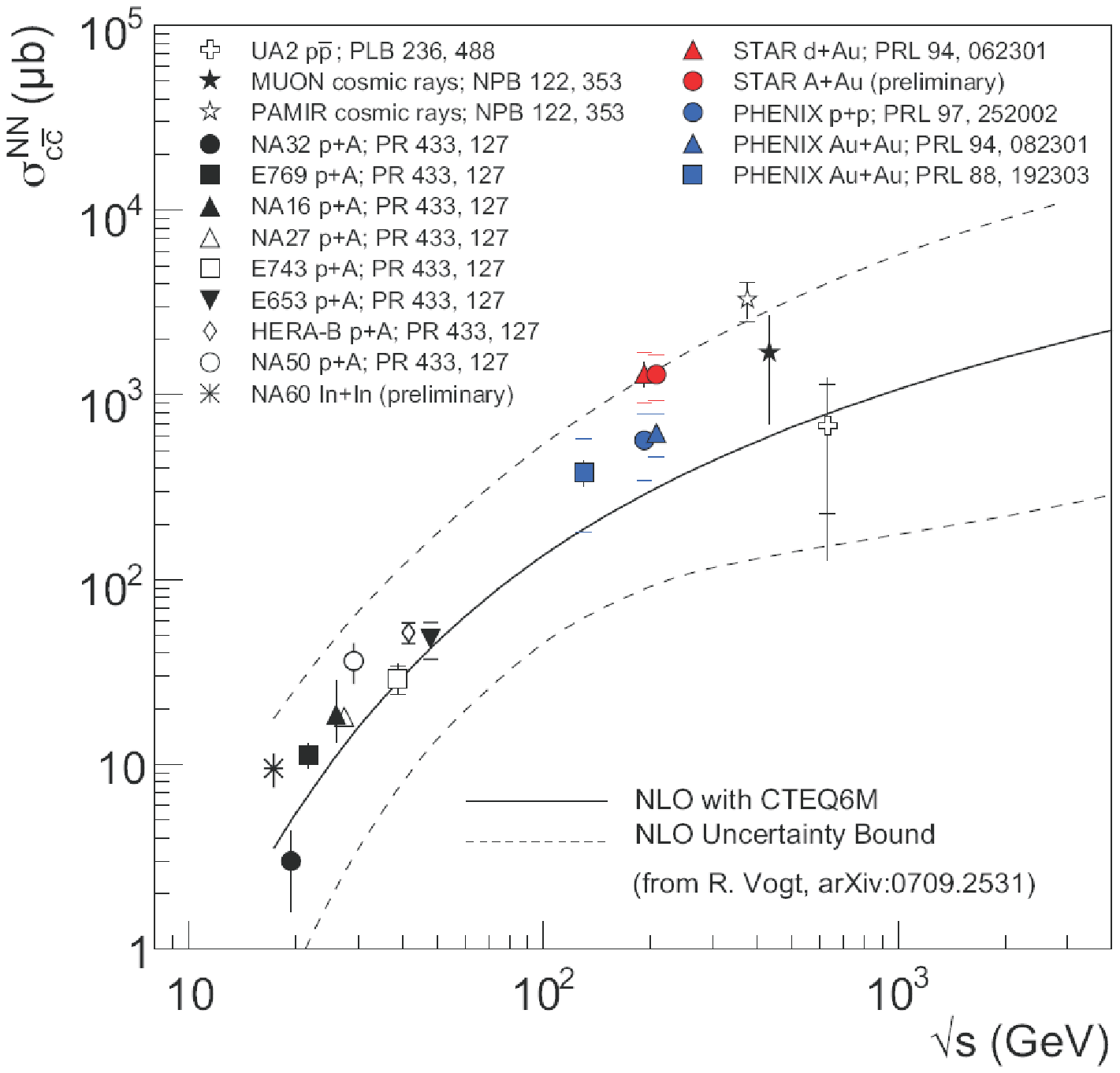,width=0.75\textwidth}}
\caption{Comparison of total cross section measurements for
$c\bar{c}$ pairs as a function of the invariant energy $\sqrt{s}$.
The STAR and PHENIX results are given as cross section per binary
collisions; vertical lines reflect the statistical errors,
horizontal bars indicate the systematic uncertainties (where
available). The figure is taken from
Ref.~\protect\cite{Frawley:2008kk}.}
\label{fig:charm_sigma_vs_energy}
\end{figure}

The precision of such models depends crucially on the elementary
input, i.e. the knowledge of the elementary reaction cross
sections. In order to tackle the production and dynamics of
charmed hadrons in heavy-ion collisions, one needs data from $NN$
and $\pi N$ reactions at different energies to establish the
underlying cross sections and distributions. One can apply
perturbative QCD (pQCD) to calculate the total cross section for
$c\bar c$ pair production. The results of next-to-leading order
pQCD~\cite{rvjoszo} calculations are shown in
Fig.~\ref{fig:charm_sigma_vs_energy} in comparison to $pp$ data as
well as to $pA$ ($AA$) measurements divided by the number of
binary collisions $N_{coll}$ (charm pairs, due to their high mass,
are expected to be produced only in initial nucleon-nucleon
collisions and their yield should scale with $N_{coll}$). The
discrepancy of a factor of $\sim 2$ between the PHENIX and STAR
cross sections is not yet understood. Moreover, the large
theoretical uncertainty implies that there is little predictive
power in the pQCD total charm cross section. Therefore, the use of
phenomenological parametrizations of the world data for the total
charm cross sections from elementary reactions is legitimate so
far. This approach is followed up in transport models.

\begin{figure}
\psfig{file=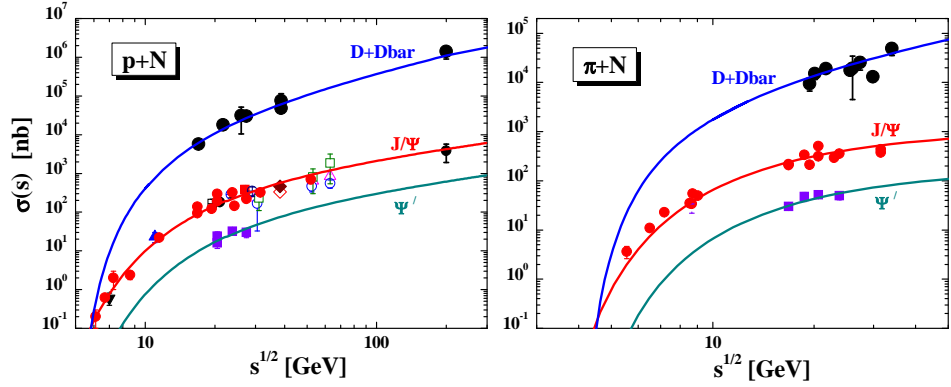,width=\textwidth} \caption{The cross section for $D+\bar D$, $J/\Psi$ and
$\Psi^\prime$ meson production in  $pN$ (left part) and $\pi N$ reactions (right part). The solid
lines show the parametrizations used in HSD, whereas the symbols stand for the experimental data
from Refs.
\protect\cite{NA16,NA27,E743,E653,E789,NA32,E769,WA92,E791,PHENIX_pp}. Note that the $J/\Psi$ cross
sections include the decay from $\chi_c$ mesons. The figure is taken from
Ref.~\protect\cite{Olena.SPS}.} \label{xs_pp_pip}
\end{figure}

In the rest of this Section we present the implementation of charm
production from elementary hadron collisions in the
Hadron-String-Dynamics (HSD)~\cite{Cass99} transport
model\footnote{The open source code is available from
Ref.~\cite{open}}. The total cross sections for the elementary
production channels including the charmed hadrons $D, \bar{D},
D^*$, $\bar{D}^*$, $D_s$, $\bar{D}_s$, $D_s^*$, $\bar{D}_s^*$,
$J/\Psi, \Psi(2S), \chi_{2c}$ from $NN$ and $\pi N$ collisions
were fitted in Refs.
\cite{Cass99,Olena.SPS,Cass01,brat03,Cass00,Olena.RHIC} to
PYTHIA~\cite{PYTHIA} calculations above $\sqrt{s}$ = 10 GeV and
extrapolated to the individual thresholds, while the absolute
strength of the cross sections was adjusted to the experimental
data (cf.  Ref.~\cite{Cass01}). The actual results are displayed
in Fig.~\ref{xs_pp_pip} for $p+N$ and $\pi + N$ reactions. The
total charmonium cross sections ($i = \chi_c, J/\Psi,
\Psi^\prime$) from $NN$ collisions as a function of the invariant
energy $\sqrt{s}$ are approximated by the expression
\begin{eqnarray}
\sigma_i^{NN}(s) = f_i \ a \ \left(1 - \frac{m_i}{\sqrt{s}}\right)^\alpha \
\left(\frac{\sqrt{s}}{m_i}\right)^\beta \theta(\sqrt{s}-\sqrt{s_{0i}}),
 \label{fitj}
\end{eqnarray}
where $m_i$ denotes the mass of charmonium $i$ while $\sqrt{s_{0i}}=m_i+2 m_N$ is the threshold in
vacuum. The parameters in Eq. (\ref{fitj}) have been fixed to describe the $J/\Psi$ and $\Psi^\prime$
data up to the RHIC energy $\sqrt{s}=200$~GeV ({\it cf.} Ref.~\cite{Olena.SPS}). We use $a=0.16$ mb,
$\alpha$ = 10, $\beta =0.775$.

\begin{figure}
\centerline{\psfig{file=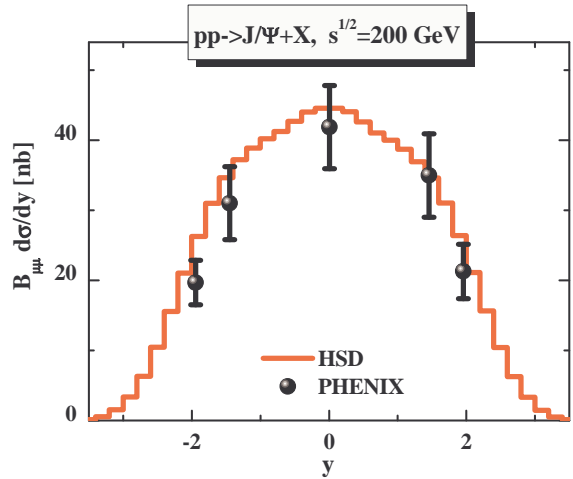,width=0.6\textwidth}} \caption{Cross section for the
differential  $J/\Psi$ production in rapidity (times the branching ratio to di-muons $B_{\mu \mu}$)
in $p p$ collisions at $\sqrt{s}=200$~GeV. The HSD (input) parametrization (solid line) is compared
to the PHENIX data (symbols) from Ref.~\protect\cite{PHENIXppY}. The figure is taken from
Ref.~\protect\cite{Olena.RHIC.2}.} \label{JPsiInpp}
\end{figure}

The parameters $f_i$ are fixed as $f_{\chi_c}=0.636, \ f_{J/\Psi}=0.581,\ f_{\Psi^\prime}=0.21$ in
order to reproduce the experimental ratio $$\frac{B(\chi_{c1}\to J/\Psi)\sigma_{\chi_{c1}}
 +B(\chi_{c2}\to J/\Psi)\sigma_{\chi_{c2}}}
 {\sigma^{exp}_{J/\Psi}}=0.344\pm 0.031$$
measured in $pp$ and $\pi N$ reactions~\cite{E705_93,WA11_82} as well as the averaged $pp$ and $pA$
ratio $(B_{\mu\mu}(\Psi^\prime)\sigma_{\Psi^\prime})
 / (B_{\mu\mu}(J/\Psi)\sigma_{J/\Psi})\simeq 0.0165$ ({\it cf.} the compilation
of experimental data in Ref.~\cite{NA50_03}). The experimentally measured $J/\Psi$ cross section
includes the direct $J/\Psi$ component $(\sigma_{J/\Psi})$ as well as the decays of higher
charmonium states $\chi_{c}$ and $\Psi^\prime$, {\it i.e.}
\begin{eqnarray}
\sigma^{exp}_{J/\Psi}=\sigma_{J/\Psi}+B(\chi_{c}\to J/\Psi)\sigma_{\chi_{c}} +B(\Psi^\prime\to
J/\Psi)\sigma_{\Psi^\prime}. \ \label{xsexp}\end{eqnarray} Note, we do not distinguish the
$\chi_{c1}(1P)$ and $\chi_{c2}(1P)$ states. Instead, we use only the $\chi_{c1}(1P)$ state (which
we denote as $\chi_c$), however, with an increased  branching ratio for the decay to $J/\Psi$ in
order to include the contribution of $\chi_{c2}(1P)$, {\it i.e.}  $B(\chi_{c}\to J/\Psi) = 0.54$.
Furthermore, we adopt $B(\Psi^\prime\to J/\Psi)=0.557$ from Ref.~\cite{PDG}.

We recall that (as in Refs. \cite{Cass01,brat03,Geiss99,Cass97,CassKo}) the charm degrees of
freedom in the HSD approach are treated perturbatively and that initial hard processes (such as
$c\bar{c}$ or Drell-Yan production from $NN$ collisions) are `pre-calculated' to achieve a scaling
of the inclusive cross section with the number of projectile and target nucleons as $A_P \times
A_T$ when integrating over impact parameter. For fixed impact parameter $b$, the $c\bar{c}$ yield
then scales with the number of binary hard collisions $N_{coll}$ ({\it cf.} Fig. 8 in
Ref.~\cite{Cass01}). To implement this scaling, we separate the production of the hard and soft
processes: The space-time production vertices of the $c\bar{c}$ pairs are 'precalculated' in each
transport run by neglecting the soft processes, i.e. the production of light quarks and associated
mesons, and then reinserted in the dynamical calculation at the proper space-time point during the
actual calculation that includes all soft processes. As shown in Ref. \cite{Cass01} this
prescription is very well in line with Glauber calculations for the production of hard probes at
fixed impact parameter, too. We mention that this 'precalculation' of $c \bar{c}$ production might
be modified at RHIC energies due to changes of the gluon structure functions during the heavy-ion
reaction or related shadowing phenomena \cite{Schmitt:2000yc}. The amount of shadowing at RHIC
energies will be discussed in more detail in Section~\ref{pA}.

In addition to primary hard $NN$ collisions, the open charm mesons or charmonia may also be
generated by secondary meson-baryon ($mB$) reactions. Here we include all secondary collisions of
mesons with baryons by assuming that the open charm cross section (from Section 2 of
Ref.~\cite{Cass01}) only depends on the invariant energy $\sqrt{s}$ and not on the explicit meson
or baryon state. Furthermore, we take into account all interactions of `formed' mesons -- after a
formation time of $\tau_F \approx$ 0.8 fm/c (in their rest frame)~\cite{Geiss} -- with baryons or
diquarks. For the total charmonium cross sections from meson-baryon (or $\pi N$) reactions we use
the parametrization (in line with Ref. \cite{Vogt99}):
\begin{eqnarray}
\sigma_i^{\pi N} (s) = f_i \ b \ \left(1 - \frac{m_i}{\sqrt{s}}\right)^\gamma
\label{fitpin}\end{eqnarray} with $\gamma=7.3$ and $b=1.24$~mb, which describes the existing
experimental data at low $\sqrt{s}$ reasonably well, as seen in Fig.~\ref{xs_pp_pip}.

\begin{figure}
\centerline{\psfig{file=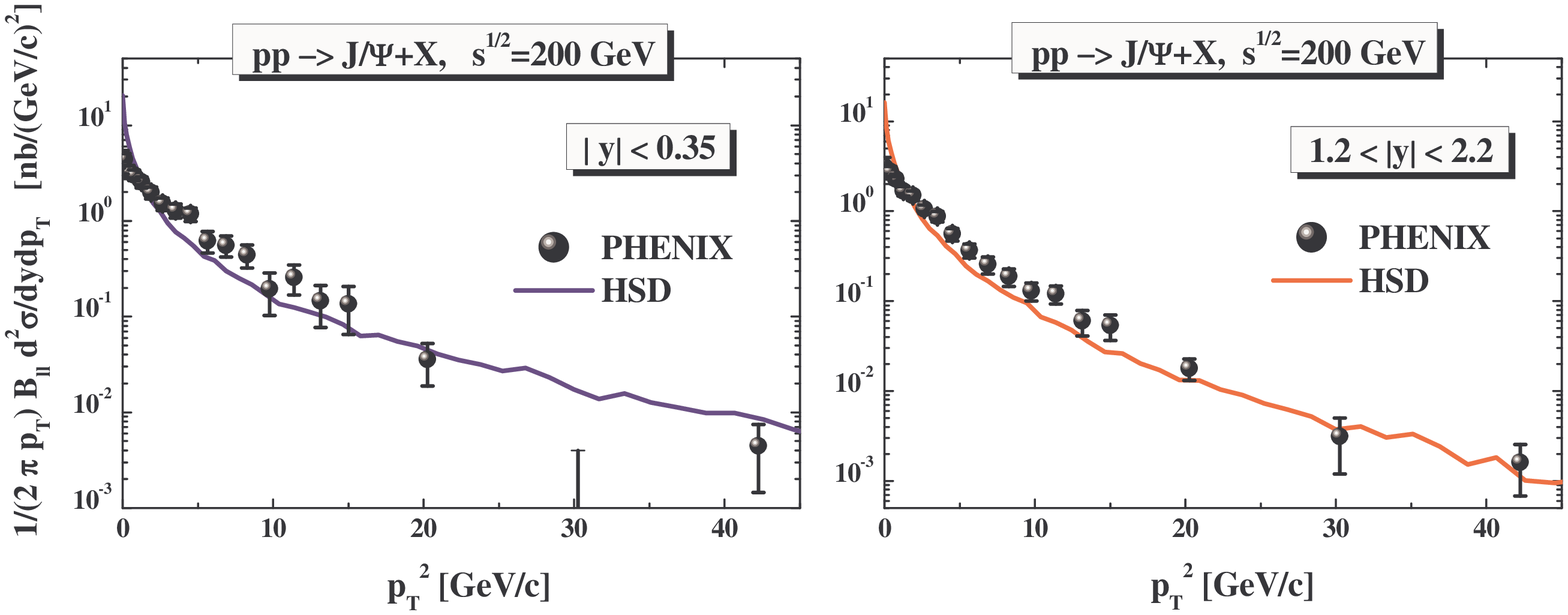,width=\textwidth}} \caption{Differential cross section of
$J/\Psi$ production in $pp$ collisions at $\sqrt{s}=200$~GeV at mid-rapidity ($|y|<0.35$, l.h.s.)
and at forward rapidity ($1.2<|y|<2.2$, r.h.s.) {\it vs} the transverse momentum squared $p_T^2$ as
implemented in HSD (solid line) compared to the PHENIX data from Ref.~\protect\cite{PHENIXppY}
(dots). The figure is taken from Ref.~\protect\cite{Olena.RHIC.2}.} \label{ppPT2mid}
\end{figure}

Apart from the total cross sections for charmonia  we also need the differential distributions of
the produced mesons in the transverse momentum $p_T$ and the rapidity $y$ (or Feynman $x_F$) from
each individual collision. We recall that $x_F = p_z/p_z^{max} \approx 2 p_z/\sqrt{s}$ with $p_z$
denoting the longitudinal momentum. For the differential distribution in $x_F$ from $NN$ and $\pi
N$ collisions we use the ansatz from the E672/E706 Collaboration~\cite{E672} and for the $p_T$
distribution: an exponential
\begin{equation}
\frac{dN}{dx_F dp_T} \sim (1 - |x_F|)^c \ \exp(-b_{p_T} p_T)
\end{equation}
with $b_{p_T}=2.08$ GeV$^{-1}$ at SPS energies or a power law parametrization from
Ref.~\cite{brat05} which has been fixed by the STAR data~\cite{STAR04} at high energies relevant to
RHIC, {\it i.e.}
\begin{equation}
\frac{dN}{dx_F dp_T} \sim (1 - |x_F|)^c \ \left(1 + \frac{p_T}{b_{p_T}}\right)^{c_{p_T}}
\end{equation}
with $b_{p_T}=3.5$ GeV/$c$ and $c_{p_T}=-8.3$. The exponent $c$ is given by $c= a/(1+b/\sqrt{s})$
and the parameters $a, b$ are chosen as $a_{NN}=16$, $b_{NN}=24.9$ GeV for $NN$ collisions and
$a_{\pi N}=4.11$, $b_{\pi N}=10.2$ GeV for $\pi N$ collisions~\cite{brat03,Cass01}. Note that the
parametrizations of the differential cross sections are taken as in Refs.~\cite{Cass01,brat03},
apart from a readjustment of the parameter $a_{NN}$ in order to reproduce the recently measured
rapidity distribution of $J/\Psi$'s in $p+p$ reactions at $\sqrt{s}=200$~GeV by
PHENIX~\cite{PHENIXppY} in Ref.~\cite{Olena.RHIC}.

The resulting rapidity distribution for $J/\Psi$ production in $pp$ collisions at $\sqrt{s}=200$
GeV is shown in Fig.~\ref{JPsiInpp}. We also present the $p p \to J/\Psi+X$ differential cross
section in $p_T^2$ at mid-rapidity ($|y|<0.35$) and at forward rapidity (averaged in the
interval $1.2<|y|<2.2$) in
Fig.~\ref{ppPT2mid}.  Both kinematical distributions (in $y$ and $p_T$) are in line with
the data from  Ref.~\cite{PHENIXppY} within error bars.

\section{Energy-density evolution in relativistic heavy-ion collisions}
\label{energy}

The HSD approach \cite{Cass99} provides the space-time geometry of
nucleus-nucleus reactions and a rather reliable estimate for the
local energy densities achieved, since the production of secondary
particles with light and a single strange quark/antiquark is
described well from SIS to RHIC
energies~\cite{Weber,Bratkovskaya:2004kv} (see also Section 5). In
the transport approach the local energy density is calculated from
the energy-momentum tensor $T_{\mu \nu}(x)$ for all space-time
points $x$ in the local rest frame: $\varepsilon(x) =
T_{00}^{loc}(x)$, where $T_{00}^{loc}(x)$ is calculated from
$T_{\mu \nu}(x)$ by a Lorentz boost to the local rest frame. In
order to exclude contributions to $T_{\mu \nu}$ from
noninteracting nucleons in the initial phase all nucleons without
prior interactions are discarded in the rapidity intervals
$[y_{tar}-0.4, y_{tar}+0.4]$ and $[y_{pro}-0.4, y_{pro}+0.4]$
where $y_{tar}$ and $y_{pro}$ denote projectile and target
rapidity, respectively. Note that the initial rapidity
distributions of projectile and target nucleons are smeared out
due to Fermi motion by about $\Delta y \approx \pm 0.4$. Some
comments on the choice of the grid in space-time are in order
here:  In the actual calculation (for Au+Au collisions) the
initial grid has a dimension of 1 fm $\times$ 1 fm $\times$
1/$\gamma_{cm}$ fm, where $\gamma_{cm}$ denotes the Lorentz
$\gamma$-factor in the nucleon-nucleon center-of-mass system.
After the time of maximum overlap $t_m$ of the nuclei the
grid-size in beam direction $\Delta z_0 = 1/\gamma_{cm}$ [fm] is
increased linearly in time as $\Delta z = \Delta z_0 + a (t-t_m)$,
where the parameter $a$ is chosen in a way to keep the particle
number in the local cells of volume $\Delta V(t) = \Delta x \Delta
y \Delta z(t)$ roughly constant during the longitudinal expansion
of the system. In this way local fluctuations of the energy
density $\varepsilon(x)$ due to fluctuations in the particle
number are kept low. Furthermore, the time-step is taken as
$\Delta t = 0.2 \Delta z(t)$ and increases in time in analogy to
$\Delta z(t)$.  This choice provides a high resolution in space
and time for the initial phase and keeps track of the relevant
dynamics throughout the entire collision history.

\subsection{SPS energies}

\begin{figure}
\centerline{\psfig{file=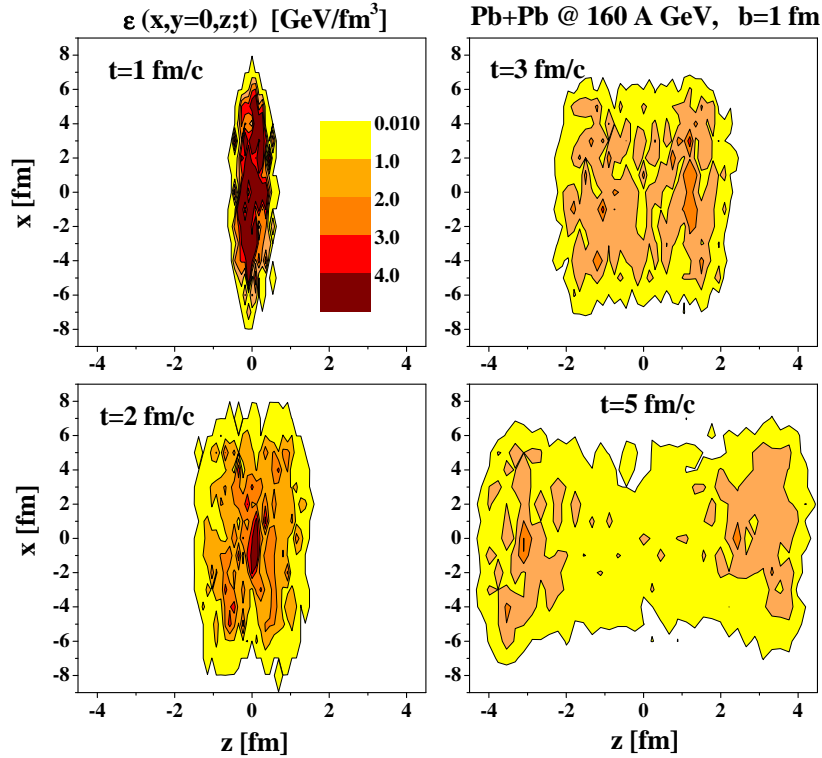,width=0.85\textwidth}} \caption{The energy density
$\varepsilon(x,y=0,z;t)$ from HSD for a Pb+Pb collision at 160 A$\cdot$GeV and impact parameter
$b=1$ fm in terms of contour lines (0.01, 1, 2, 3, 4 GeV/fm$^3$) for times of 1, 2, 3 and 5 fm/c
(from contact). Note that noninteracting nucleons have been discarded in the actual calculation of
the energy-momentum tensor. The figure is taken from Ref.~\protect\cite{Olena.SPS}.} \label{F1}
\end{figure}

\begin{figure}
\centerline{\psfig{file=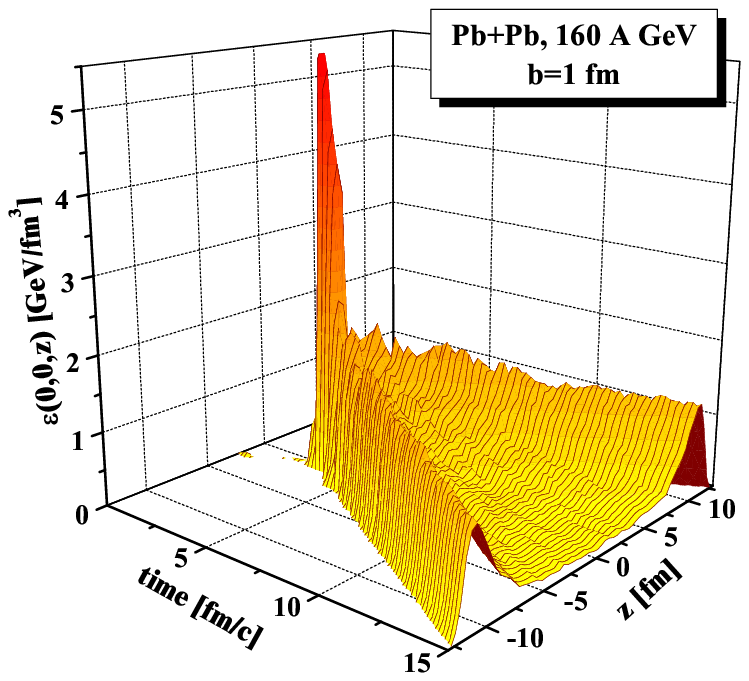,width=0.95\textwidth}} \caption{The energy density
$\varepsilon(x=0,y=0,z;t)$ from HSD for a Pb+Pb collision at 160 A$\cdot$GeV and impact parameter
$b=1$ fm on a linear scale. Note that noninteracting nucleons have been discarded in the actual
calculation of the energy-momentum tensor such that $\varepsilon(x)\ne 0$ only after contact of the
two Pb nuclei which is $\sim$ 2 fm/c. The figure is taken from Ref.~\protect\cite{Olena.SPS}.}
\label{energy.SPS}
\end{figure}

As a first example we display in Fig. \ref{F1} the energy density $\varepsilon(x,y=0,z;t)$ for a
Pb+Pb collision at 160 A$\cdot$GeV and impact parameter $b=1$ fm in terms of contour lines for
times of 1, 2, 3 and 5 fm/c (from contact). It is clearly seen that energy densities above 4
GeV/fm$^3$ are reached in the early overlap phase of the reaction and that $\varepsilon(x)$ drops
within a few fm/c below 1 GeV/fm$^3$ in the center of the grid. On the other hand the energy
density in the region of the leading particles - moving almost with the velocity of light - stays
above 1 GeV/fm$^3$ due to Lorentz time dilatation since the time $t$ here is measured in the
nucleon-nucleon center-of-mass system. Note that in the local rest frame of the leading particles
the eigentime $\tau$ is roughly given by $\tau \approx t/\gamma_{cm}$ with $\gamma_{cm} \approx
9.3$ (at 160 A$\cdot$GeV).

Another view of the space time evolution of the energy density is given in Fig.~\ref{energy.SPS}
where we display $\varepsilon(x=0,y=0,z;t)$ for the same system as in Fig. \ref{F1} on a linear
scale. The contact time of the two Pb nuclei here is 2 fm/c and the overlap phase of the Lorentz
contracted nuclei is identified by a sharp peak in space-time which is essentially given by the
diameter of the nuclei divided by $\gamma_{cm}$. As noted before, the energy density in the center
of the reaction volume ($z \approx$ 0) drops fast below 1 GeV/fm$^3$ whereas the ridges close to
the light-cone basically stem from the leading ends of the strings formed in the early
nucleon-nucleon collisions. In these space-time regions all reaction rates are reduced by the
factor $\sim 1/\gamma_{cm}$ such that the transport calculations have to be carried out to large
times of several hundred fm/c in order to catch the dynamics and decays in these regions. In the
central regime, however, all interaction rates vanish after about 15 fm/c. Since the $c,\bar{c}$
pairs are produced dominantly at midrapidity with a small spread in rapidity ($\sigma_y \approx
0.8$ at 160 A$\cdot$GeV) it is the central region that is of primary interest for this study.

\subsection{RHIC energies}

\begin{figure}
\centerline{\psfig{file=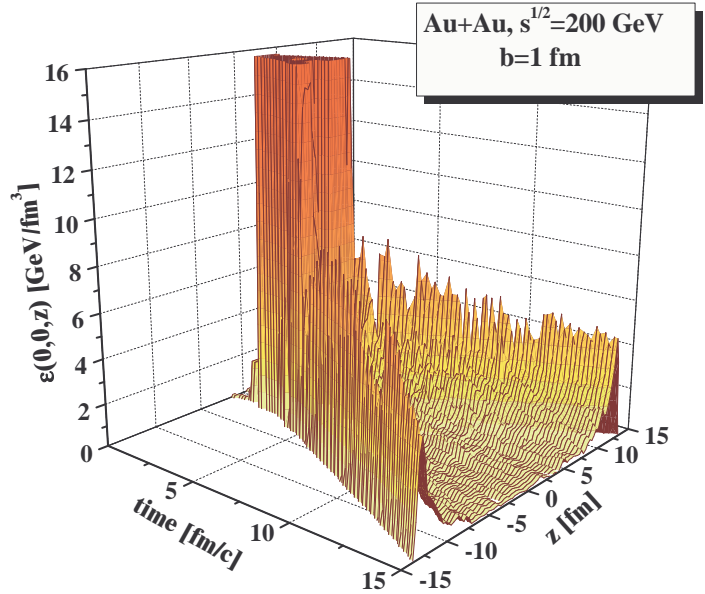,width=0.75\textwidth}} \caption{The energy density
$\varepsilon(x=0,y=0,z;t)$ from HSD for a central Au+Au collision at $\sqrt{s}$ = 200 GeV. The time
$t$ is given in the nucleon-nucleon center-of-mass system. The figure is taken from
Ref.~\protect\cite{Olena.RHIC.2}.} \label{3D}
\end{figure}

The energy density $\varepsilon({\bf r};t)$ becomes very high in a central Au+Au collision at
$\sqrt{s}$ = 200 GeV as shown in Fig. \ref{3D} (in analogy to  Fig.~\ref{energy.SPS} for the top SPS
energies). Fig.~\ref{3D} shows the space-time evolution of the energy density
$\varepsilon(x=0,y=0,z;t)$ for a Au+Au collision at 21300~AGeV or $\sqrt{s} $ = 200 GeV. It is
clearly seen that energy densities above 16 GeV/fm$^3$ are reached in the early overlap phase of
the reaction and that $\varepsilon(x)$ drops after about 6~fm/c (starting from contact) below 1
GeV/fm$^3$ in the center of the grid. On the other hand the energy density in the region of the
leading particles - moving almost with the velocity of light - stays above 1 GeV/fm$^3$ due to
Lorentz time dilatation since the time $t$ in the transport calculation is measured in the
nucleon-nucleon center-of-mass system. As seen from Fig.~\ref{3D}, the energy density in the local
rest frame is a rapidly changing function of time in nucleus-nucleus collisions. For orientation
let us quote the relevant time scales (in the cms reference frame):

-- The $c\bar{c}$ formation time $\tau_c \approx 1/M_\perp$ is
about 0.05 fm/c for a transverse mass of 4 GeV; the transient time
for a central Au+Au collision at $\sqrt{s}$ = 200 GeV is $t_r
\approx 2 R_A/\gamma_{cm} \approx 0.13$ fm/c.  According to
standard assumptions, the $c\bar{c}$ pairs are produced in the
initial hard $NN$ collisions dominantly by gluon fusion in the
time period $t_r$.  In fact, the formation time $\tau_c$ is
significantly smaller than $t_r$, which implies that the $c$ or
$\bar{c}$ quarks may interact with the impinging nucleons of the
projectile or target for times $t \leq t_r$.

-- Using the Bjorken estimate for the energy density and employing the time-scale $t_r = 0.13$~fm/c,
the energy density -- after the nuclei have punched through each other -- amounts to about
$5/0.13>30$~GeV/fm$^3$ (as quoted also in the HSD calculations in
Refs.~\cite{Olena.RHIC,Olena.RHIC.2}). Even when adding the $c \bar{c}$ formation time, this gives an
energy density $\sim 5/0.18 \approx 28$ GeV/fm$^3$.  So the numbers in Fig.~\ref{3D} agree with
transparent and simple estimates and illustrate the high initial densities after $c\bar{c}$
production from primary interactions.

\begin{figure}
\centerline{\psfig{file=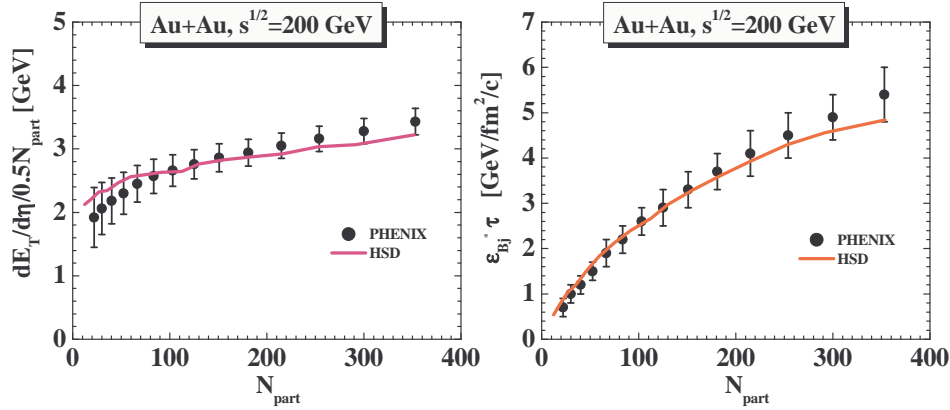,width=\textwidth}} \caption{Left part: The transverse energy $E_T$
per pseudorapidity interval $d \eta$ divided by the number of participant pairs $ (0.5 N_{part})$
from HSD (solid line) in comparison to the PHENIX data (dots)~\protect\cite{Et}. Right part: The
Bjorken energy density $\varepsilon_{Bj}\cdot \tau$ from HSD (solid line) for Au+Au collisions at
$\sqrt{s}$ = 200 GeV in comparison to the PHENIX data (dots)~\protect\cite{Et}. The figure is taken
from Ref.~\protect\cite{Olena.RHIC.2}.} \label{ET}
\end{figure}

The energy densities quoted above are considerably different from the estimate
\begin{equation} \label{bjorken} \tau \cdot \epsilon_{Bj} =
\frac{<E_T> {dN \over d\eta}}{\pi R_T^2},
\end{equation}
where $<E_T>$ is the average transverse energy per particle, $dN/d
\eta$ the number of particles per unit of pseudorapidity, and
$\tau$ a formation time parameter often used as $\tau = 1$ fm/c.
Furthermore, $\pi R_T^2$ denotes the overlap area for the
corresponding centrality. Is is important to point out that the
estimate (\ref{bjorken}) is only well defined for the product
$\tau \cdot \epsilon_{Bj}$! The question naturally arises, if the
transport calculations follow the corresponding experimental
constraints.

To this aim we show $dE_T/d\eta$ (divided by half the number of
participants $N_{part}$) from HSD in Fig. \ref{ET} (l.h.s.) in
comparison to the measurements by PHENIX~\cite{Et}.  Accordingly,
the Bjorken energy density $\epsilon _{Bj}$ -- multiplied by the
time-scale $\tau$ (\ref{bjorken})-- from HSD is shown additionally
in the r.h.s. of Fig. \ref{ET} in comparison to the PHENIX data as
a function of $N_{part}$. The similarity between the calculated
quantities and the experimental data demonstrates that the
space-time evolution of the energy-momentum tensor $T_{\mu \nu}$
in HSD is sufficiently well under control also at RHIC energies.

\subsection{FAIR energies}

\begin{figure}
\centerline{\psfig{file=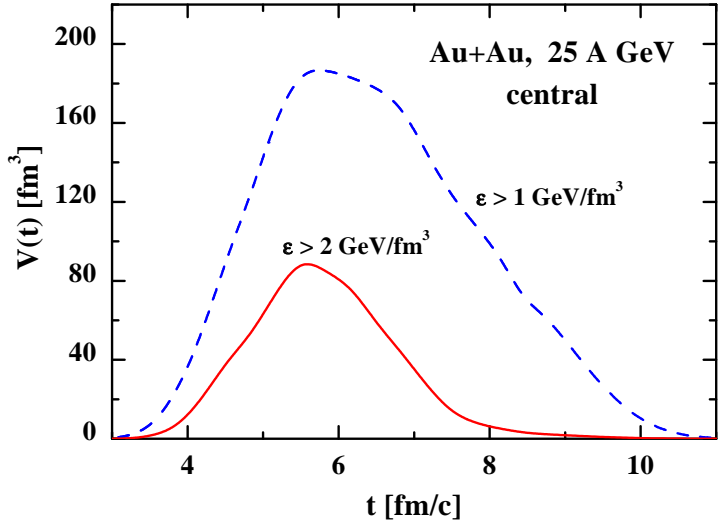,width=0.75\textwidth}} \caption{Time evolution for the volume with
energy density $\varepsilon \geq$ 1 GeV/fm$^3$ (dashed line) and $\geq$ 2 GeV/fm$^3$ (solid line)
in the HSD approach for a central $Au+Au$ reaction at 25 A$\cdot$GeV. The figure is taken from
Ref.~\protect\cite{Cass01}.} \label{FairEcrit}
\end{figure}

The question emerges if central collisions of e.g. Au+Au at the
future FAIR facility might be also suited to explore a possible
phase transition to the sQGP. For a quantitative orientation we
display in Fig.~\ref{FairEcrit} the volume (in the nucleus-nucleus
center-of-mass) with  energy densities above 1 GeV/fm$^3$ and 2
GeV/fm$^3$ as a function of time for a central Au+Au collision at
25 A$\cdot$GeV, where only interacting and produced hadrons have
been counted as explained in the beginning of this Section. It is
important to note that in HSD the high energy density is
essentially build up from `strings', i.e. `unformed' hadrons. The
absolute numbers in Fig.~\ref{FairEcrit} have to be compared to
the volume of a Au-nucleus in the moving frame which -- for a
Lorentz $\gamma$-factor of 3.78 -- gives $\approx $ 330 fm$^3$. In
this case, the overlap phase of the nuclei lasts for about 3.7
fm/c during which energy densities above 2 GeV/fm$^3$ are seen in
Fig.~\ref{FairEcrit} in a sizeable volume. Thus also at 25
A$\cdot$GeV the phase boundary to a QGP might be probed in a
sizeable volume for time scales of a few fm/c. Contrary to central
collisions at the SPS and RHIC, these volumes are characterized by
a high net quark density; for such configurations we presently
have only `hints' from lattice QCD calculations rather than solid information.

\section{Proton-nucleus reactions: Cold nuclear matter effects}
\label{pA}

Before coming to charm production and propagation in heavy-ion
reactions it is mandatory to explore the charm dynamics in
proton-nucleus reactions. Such reactions are a first step beyond
the elementary $pp$ reactions and provide an additional reference
with respect to the heavy-ion case. Since in $p+A$ reactions only
subnormal nuclear densities are achieved and the target nucleus
remains approximately in its geometrical shape for the first few
fm/c, these reactions are much easier to treat and allow to
separate `cold nuclear matter' (CNM) effects from those induced by
the new partonic medium encountered in relativistic
nucleus-nucleus collisions.  As we will show in the following, one
can observe e.g. gluon shadowing at RHIC energies by it's
influence on charmonium production in $d+Au$ collisions.
Additionally, the amount of `normal' nuclear charmonium
dissociation by the target nucleons can be probed. Thus $p+A$
reactions provide a necessary base-line for the heavy-ion case and
can independently be controlled by experimental data.

It is found experimentally that the yield of $J/\Psi$ in $p+A$ and $A+A$ reactions is modified
compared to that in $p+p$ collisions - scaled with the number of
initial binary scatterings $N_{coll}$~\cite{NA60,PHENIXNov06}.
Indeed, the produced $c\bar c$ can be dissociated or absorbed on
either the residual nucleons of the projectile or target or on
light co-moving particles (usually on mesons or, at high energy,
on partons) produced in the very early phase. Alternatively, the
initial production of $c\bar c$ pairs by gluon fusion might be
suppressed due to shadowing (at RHIC energies).  In particular,
charmonium absorption on {\em baryons} is the leading suppression
mechanism in $d+A$ ($p+A$) scattering at SPS energies and is an
important base-line for the investigation of charmonium
absorption.

\subsection{SPS energies}

\begin{figure}
\centerline{\psfig{file=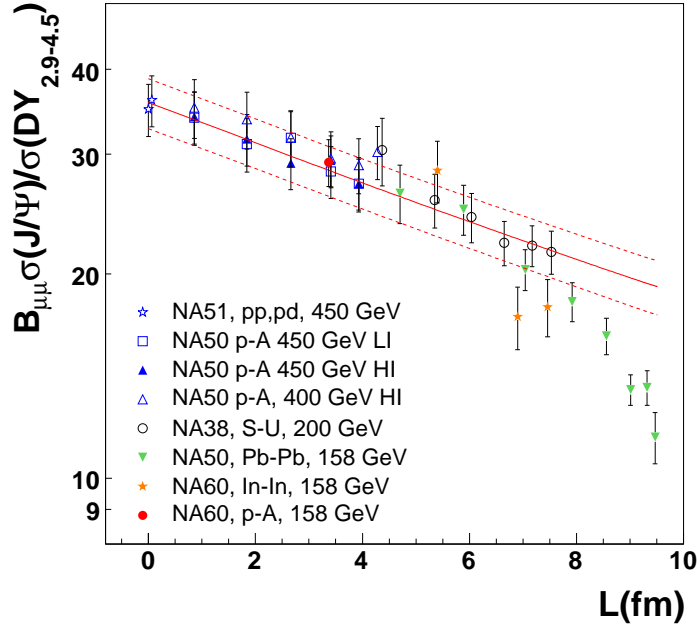,width=0.75\textwidth}} \caption{The ratio $\sigma_{\rm
J/\psi}/\sigma_{\rm DY}$ (times the branching ratio $B_{\mu \mu}$)  measured in \mbox{p-A} and
nucleus-nucleus collisions at the SPS, compiled and re-scaled, when necessary, to 158 GeV incident
energy by the NA60 Collaboration. The quantity $L$ denotes the effective path length of the
$J/\Psi$ as extracted within the Glauber model. The lines indicate the results of a Glauber fit to
the \mbox{p-A} data and the size of the error as calculated by the NA60 Collaboration. The figure
is taken from Ref.~\protect\cite{NA50O}.} \label{SPS.Glauber}
\end{figure}

The NA50 and NA60 Collaborations present their results on $J/\Psi$ suppression as the ratio of the
dimuon decay of the $J/\Psi$ relative to the Drell-Yan background from 2.9 - 4.5 GeV invariant mass
as a function of the transverse energy $E_T$, or alternative, as a function of the number of
participants $N_{{\rm part}}$, {\it i.e.} %
\begin{equation} \label{rat} B_{\mu\mu}\sigma(J/\Psi) /
\sigma(DY)|_{2.9-4.5},
\end{equation}
where $B_{\mu\mu}$ is the branching ratio for $J/\Psi\to
\mu^+\mu^-$. In order to compare our calculated results to
experimental data (see below) we need an extra input, i.e. the
normalization factor $B_{\mu\mu}\sigma_{NN}(J/\Psi) /
\sigma_{NN}(DY)$, which defines the $J/\Psi$ over Drell-Yan ratio
for elementary nucleon-nucleon collisions. We will adopt
$B_{\mu\mu}\sigma_{NN}(J/\Psi) / \sigma_{NN}(DY) = 36$ in line
with the NA60 compilation~\cite{NA60} (at the SPS energy of 158
GeV).

In Fig.~\ref{SPS.Glauber} we show the combined $p+A$, $S+U$, $Pb+Pb$ and $In+In$ data (from NA50
and NA60) for the ratio $B_{\mu\mu}\sigma(J/\Psi) / \sigma(DY)$ at 158 A$\cdot$GeV as a function of
centrality - reflected in the effective path length $L$ -  together with the Glauber-model fit by
the NA50/60 collaboration. The `default' interpretation of the experimental results in
Fig.~\ref{SPS.Glauber} is that for a $J/\Psi$ path length $L$ below about 7 fm dominantly `normal'
$J/\Psi$ dissociation with target nucleons is seen while for $L >$ 7 fm an `anomalous' suppression
sets in. Since this `anomalous' suppression only is observed in central $In+In$ and $Pb+Pb$
collisions it is attributed to a `hot matter' effect in contrast to the `normal' absorption (`cold
nuclear matter' effect). However, in order to distinguish more clearly such `hot' and `cold
nuclear' matter effects it is mandatory to employ non-equilibrium transport.

In order to study the effect of charmonium rescattering on projectile/target nucleons, we adopt in
HSD the following dissociation cross sections of charmonia with baryons independent of the energy
(in line with the most recent NA50 and NA60 compilations~\cite{NA60,NA50pA}):
\begin{eqnarray}
&& \sigma_{c\bar{c}B} = 4.18 \ {\rm mb}; \label{sigmacB} \\
&&\sigma_{J/\Psi B} = 4.18 \ {\rm mb}; \ \sigma_{\chi_c B} = 4.18 \ {\rm mb}; \ \sigma_{\Psi^\prime
B} = 7.6 \ {\rm mb}. \nonumber\end{eqnarray}
The applicability of the Glauber picture to the baryon-induced suppression (at $L<7$~fm) as illustrated in Fig.~\ref{SPS.Glauber} suggests that the produced $c\bar c$ pair can be absorbed on baryons already in its pre-resonant state.
In (\ref{sigmacB}) the cross section
$\sigma_{c\bar{c}B}$ stands for a (color dipole)  pre-resonance ($c\bar{c})$ - baryon cross
section, since the $c\bar{c}$ pair produced initially cannot be identified with a particular
charmonium due to the uncertainty relation in energy and time. For the life-time of the
pre-resonance $c\bar{c}$ pair (in it's rest frame) a value of $\tau_{c\bar{c}}$ = 0.3 fm/c is
assumed following Ref.~\cite{Kharz}. This time scale corresponds to the mass difference of the
$\Psi^\prime$ and $J/\Psi$ according to the uncertainty relation.
Note that -- in contrast to the absorption on primordial baryons $B$ (nucleons of the
incoming nuclei) -- interactions with secondary particles created in the nucleus-nucleus
collision (mesons or secondary baryons) are only allowed after the local energy-density
has dropped  below $1$~GeV/fm$^3$ in order to assure that the interaction is hadronic.

\begin{figure}
\centerline{\psfig{file=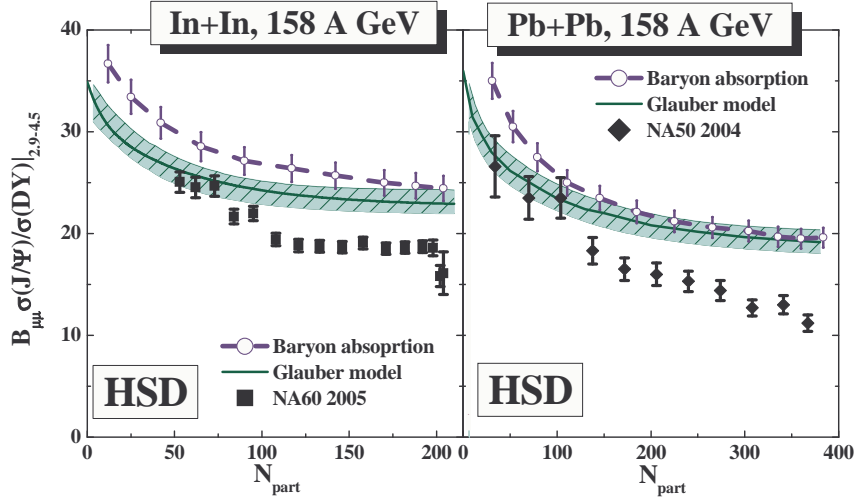,width=0.9\textwidth}} \caption{The ratio
$B_{\mu\mu}\sigma(J/\Psi) / \sigma(DY)$  as a function of the number of participants in In+In
(l.h.s.) and Pb+Pb reactions (r.h.s.) at 158 A$\cdot$GeV. The full symbols denote the data from the
NA50 and NA60 Collaborations (from Refs. \protect\cite{NA50O,NA60,NA50PsiPrime}), while the dashed
(blue) lines represent the HSD calculations including only dissociation channels with nucleons. The
(dashed blue-green) bands in the upper parts of the figure give the estimate for the normal nuclear
$J/\Psi$ absorption as calculated by the NA60 Collaboration. The vertical lines on the graphs
reflect the theoretical uncertainty due to limited statistics of the calculations. The figure is
taken from Ref.~\protect\cite{Olena.SPS}.} \label{SPS.Baryon}
\end{figure}

In Fig.~\ref{SPS.Baryon} we present the results of the HSD
calculations for the observable (\ref{rat}) for Pb+Pb and In+In
collisions  in the nuclear suppression scenario, {\it i.e.} with
only baryonic absorption and no additional, meson- or
parton-induced suppression. Instead of the (model-dependent) path
length $L$ we display this ratio as a function of the number of
participants $N_{part}$, which can directly be taken from the
transport calculations. The dashed (blue) lines stand for the HSD
results while the (green-blue) bands give the estimate for the
normal nuclear $J/\Psi$ absorption as calculated by the NA60
Collaboration in the Glauber model~\cite{NA50O}. The normal
nuclear suppression from HSD is seen to be slightly lower than the
(model dependent) estimate from NA60, however, agrees quite well
with their  calculations for more central reactions. The various
experimental data points have been taken from
Refs.~\cite{NA50O,NA60,NA50PsiPrime}.

\begin{figure}
\centerline{\psfig{file=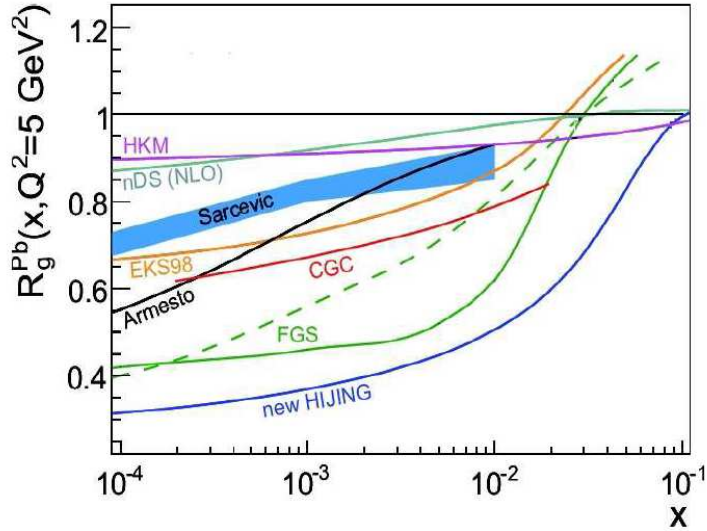,width=0.75\textwidth}} \caption{Ratio of gluon densities in
lead nucleus and in proton as a function of Bjorken $x$ at a fixed hard scale
$Q^2=5$~GeV$^2$ from various parton distribution fits. The figure is taken from
Ref.~\protect\cite{d'Enterria:2008ge}.} \label{gPDFs}
\end{figure}

Apart from the statistical uncertainties in the calculations --
reflected by the vertical lines on the theoretical graphs in Fig.
\ref{SPS.Baryon} -- some dependence on the model parameters enters
the actual numbers in Fig. \ref{SPS.Baryon}. The charmonium
nuclear absorption cross section is considered to be `fixed' by
the NA50/NA60 compilations and we have taken the same cross
section for the `pre-resonance' cross section for the $J/\Psi$ and
$\chi_c$. Accordingly, the life-time of the pre-resonance state
($\tau_{c\bar{c}}$ = 0.3 fm/c) has no impact on the absorption
with baryons as far as the $J/\Psi$ and $\chi_c$ mesons are
concerned. Only for $\Psi^\prime$ collisions with baryons this
plays a role, since the $\Psi^\prime$ + baryon cross section is
larger (7.6~mb). Consequently, the $J/\Psi$ suppression (including
the feed down from $\chi_c$) does not depend on $\tau_{c\bar{c}}$.
Within these systematic uncertainties we will now be able to
separate `cold nuclear' and `hot' matter effects in relativistic
nucleus-nucleus collisions at SPS energies.

\subsection{RHIC energies}

The  cross sections $\sigma_{J/\Psi N}, \sigma_{c\bar c N}$ at
RHIC energies are currently debated in the literature. On one
side, all the data on $J/\Psi$ production in $p+A$ reactions at
energies $\sqrt{s} \le 40$~GeV are found to be consistent with an
energy-independent cross section of the order of $4-7$ {\rm
mb}~\cite{NA60,NA50pA,Borges,csPsiN1,csPsiN2}; on the other hand,
at the much higher energy of $\sqrt{s}=200$~GeV some  part of the
suppression might be attributed to other (initial-state) `cold
nuclear matter' effects such as gluon
shadowing~\cite{Bravina,CapellaGluon,Vogt}, radiative gluon energy
loss in the initial state or multiple gluon
rescattering~\cite{Armesto:2003fi,Capella:2005cn,Ferreiro:2008qj}.
We recall that `shadowing' is a depletion of the low-momentum
parton distribution in a nucleon embedded in a nucleus compared to
the population in a free nucleon; this also leads to a lowering in
the (scaled) charmonium production cross section in $p+A$ relative
to $pp$ reactions. The reasons for depletion, though, are
numerous, and models of shadowing vary accordingly. There is,
therefore, a considerable (about a factor of 3) uncertainty in the
amount of shadowing predicted at
RHIC~\cite{Bravina,CapellaGluon,Vogt,Kopeliovich,Capellanew}.
Indeed, there is currently a lack of precise differential data,
which would allow to constrain the gluon distribution in nuclei at
low Bjorken $x$; this region is probed by charmonium production by
gluon fusion processes. Fig.~\ref{gPDFs} shows an overview of the
various model concepts which demonstrates that at low Bjorken $x$
(of the order of $10^{-2}$-$10^{-4}$) the nuclear gluon
distribution function has large uncertainties.

\begin{figure}
\centerline{\psfig{file=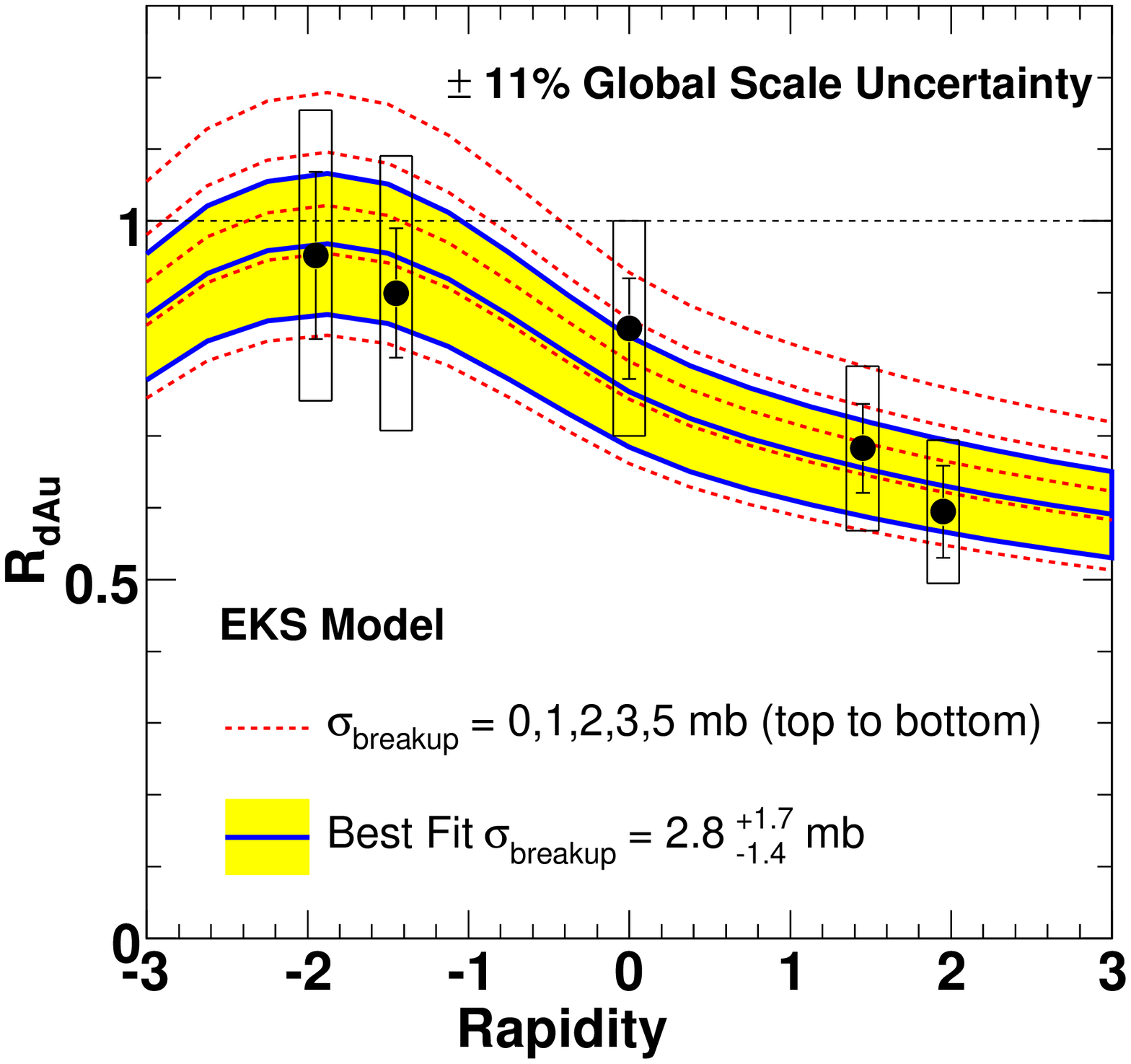,width=0.32\textwidth} \psfig{file=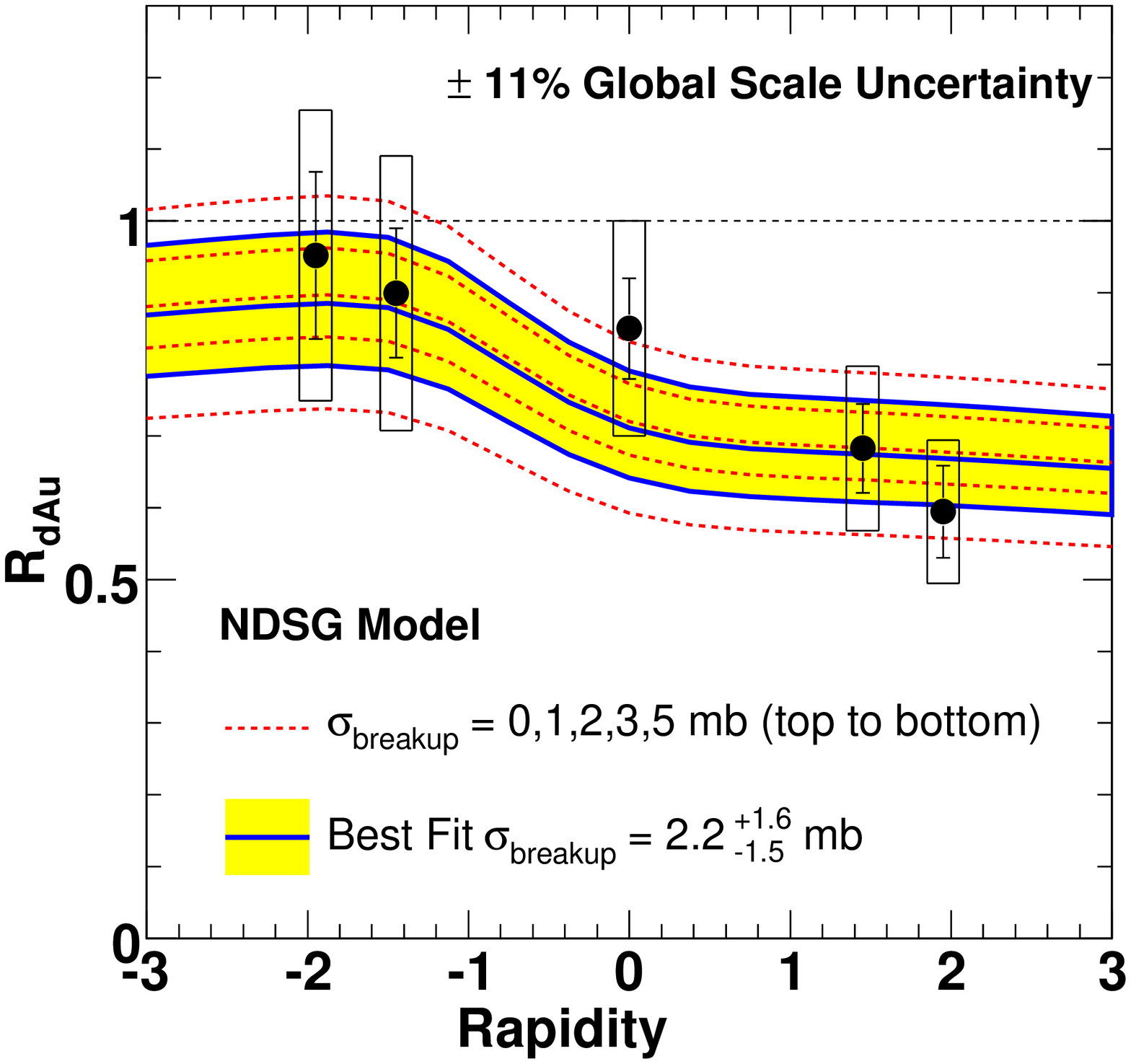,width=0.32\textwidth} \psfig{file=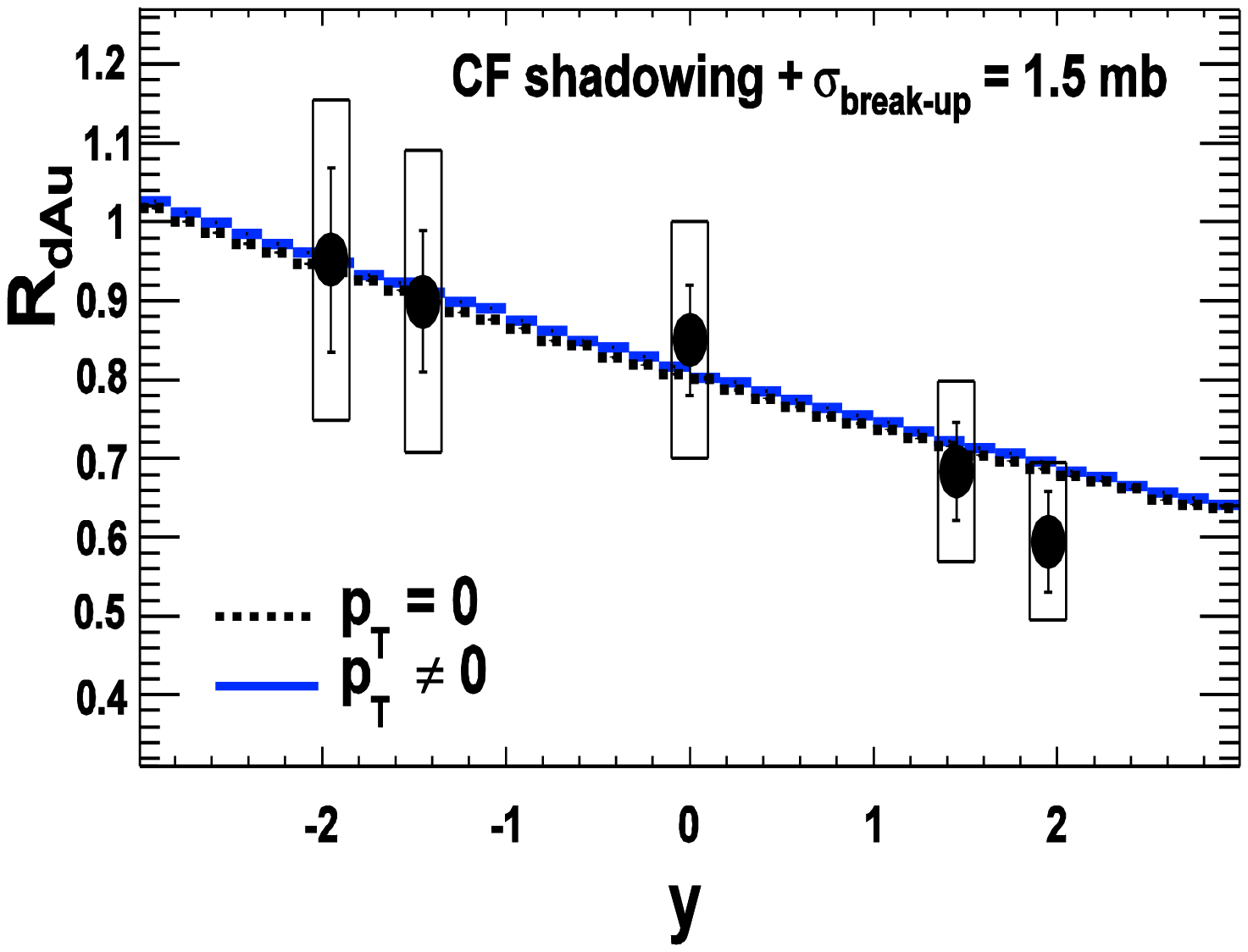,width=0.35\textwidth}}
\caption{ The $J/\psi$ nuclear modification factor in d~+~Au collisions at $\sqrt{s}=200$~GeV as a function of rapidity~\protect\cite{PHENIX08} compared with calculations including shadowing on top of the Glauber model nuclear absorption with an adjusted $J/\psi N$ breakup cross section. Left panel: the data are compared with calculations~\protect\cite{Vogt} using the EKS98 parameterization~\protect\cite{EKS1,EKS2} of shadowing. Middle panel: same as the left panel, but the calculations~\protect\cite{Vogt} employ the nDSg shadowing parameterization~\protect\cite{NDSG}. Right panel: the calculations~\protect\cite{Rakotozafindrabe:2008ms} employ the CF picture of shadowing~\protect\cite{Vogt,Capella:2005cn,Ferreiro:2008qj,Armesto:2003fi}. The figures are taken from Refs.~\protect\cite{Frawley:2008kk,Rakotozafindrabe:2008ms}.} \label{vogt}
\end{figure}

In the analysis of the $d+Au$ data at $\sqrt{s}=200$~GeV, in which
the maximum estimate for the effect of  shadowing was
made~\cite{Vogt,dA}, the additional absorption on baryons allowed
by the data was found to lead to $\sigma_{J/\Psi N}=1-3$~mb or
higher, if some contribution of anti-shadowing is present. The
authors of Ref.~\cite{Vogt} advocate at least $\sigma_{J/\Psi
N}=3$~mb in order to preserve the agreement with the data of the
Fermilab experiment E866. The PHENIX Collaboration \cite{PHENIX08}
finds a breakup cross section of $2.8^{+1.7}_{-1.4}$ mb (using EKS
shadowing) which still overlaps with the CERN-SPS value of 4.18 mb
(though with large error bars).

Fig.~\ref{vogt} demonstrates the variation in the rapidity
dependence and the amount of initial state interaction between different
implementations of shadowing for the nuclear modification factor defined as
\begin{equation} R_{dA} \equiv \frac{d N^{d Au} _{J/\Psi} / d y }{
\langle N_{coll} \rangle \cdot d N^{p p} _{J/\Psi} / d y }.
\label{FigRdA} \end{equation} In Eq. (\ref{FigRdA})  $d N^{d Au}
_{J/\Psi} / d y$ is the $J/\Psi$ invariant yield in $d+A$
collisions, $d N^{pp} _{J/\Psi} / d y$ is the $J/\Psi$ invariant
yield in $p+p$ collisions; $\langle N_{coll} \rangle$ is the
average number of binary collisions for the same rapidity bin
($\langle N_{coll} \rangle=7.6\pm 0.3$ according to the PHENIX
estimate~\cite{PHENIX08}). There is an additional large
theoretical uncertainty in the results shown in Fig. \ref{vogt}
 since in the works above only an approximate model for
baryonic absorption was applied and not a microscopic transport
approach that e.g. also includes secondary production channels of
charm pairs as described in Section~\ref{elementary}.

\begin{figure}
\centerline{\psfig{file=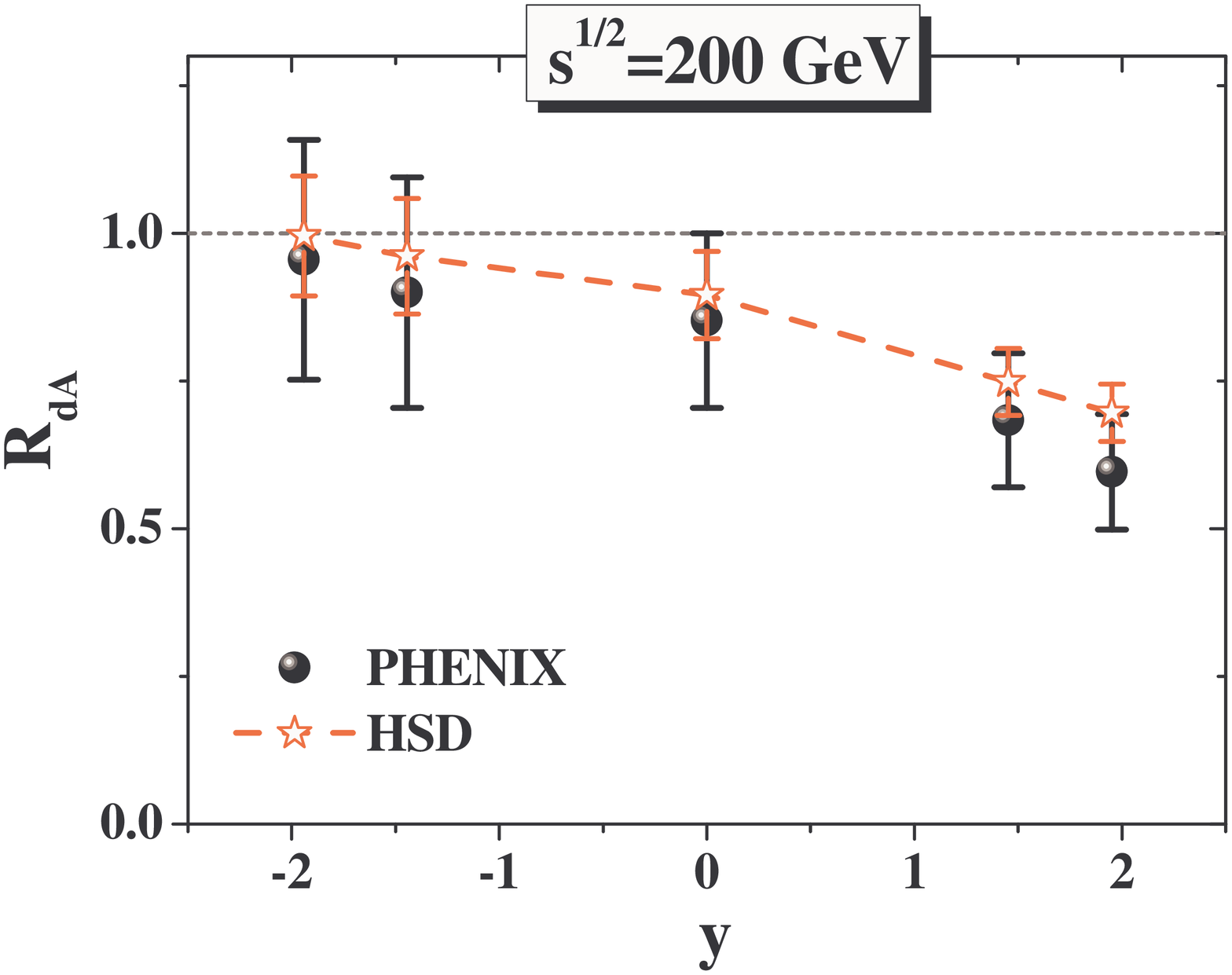,width=0.75\textwidth}} \caption{$J/\Psi$ production cross section
in $d+Au$ collisions relative to that in $p+p$ collisions (see text for the definition of $R_{dA}$)
in HSD (red stars) -- neglecting gluon shadowing -- as compared to the PHENIX
data~\protect\cite{PHENIX08} (full dots). The figure is taken from Ref.
~\protect\cite{Olena.RHIC.2}.} \label{dAu}
\end{figure}

We continue our investigation of `cold nuclear matter' effects at
RHIC energies for $d+Au$ reactions employing the same cross
sections  for baryonic absorption (\ref{sigmacB}) as at SPS
energies (cf. Fig.~\ref{SPS.Baryon}). In Fig.~\ref{dAu} we compare
the HSD result (neglecting shadowing) for the $J/\Psi$ production
in $d+Au$ collisions at $\sqrt{s}=200$~GeV to the PHENIX
data~\cite{PHENIX08}. It is seen from Fig.~\ref{dAu} that the
calculations follow approximately the decrease in $R_{dA}$ with
rapidity, however, with a tendency to overshoot at forward
rapidity. Within error bars we find the values of $\sigma_{c\bar c
B}$ from (\ref{sigmacB}) to be compatible with the inclusive RHIC
measurement as well as with the lower energy data~\cite{Borges}.
This finding is also in line with the analysis of the PHENIX
Collaboration in Ref.~\cite{PHENIX08}.

\begin{figure} \centerline{\psfig{file=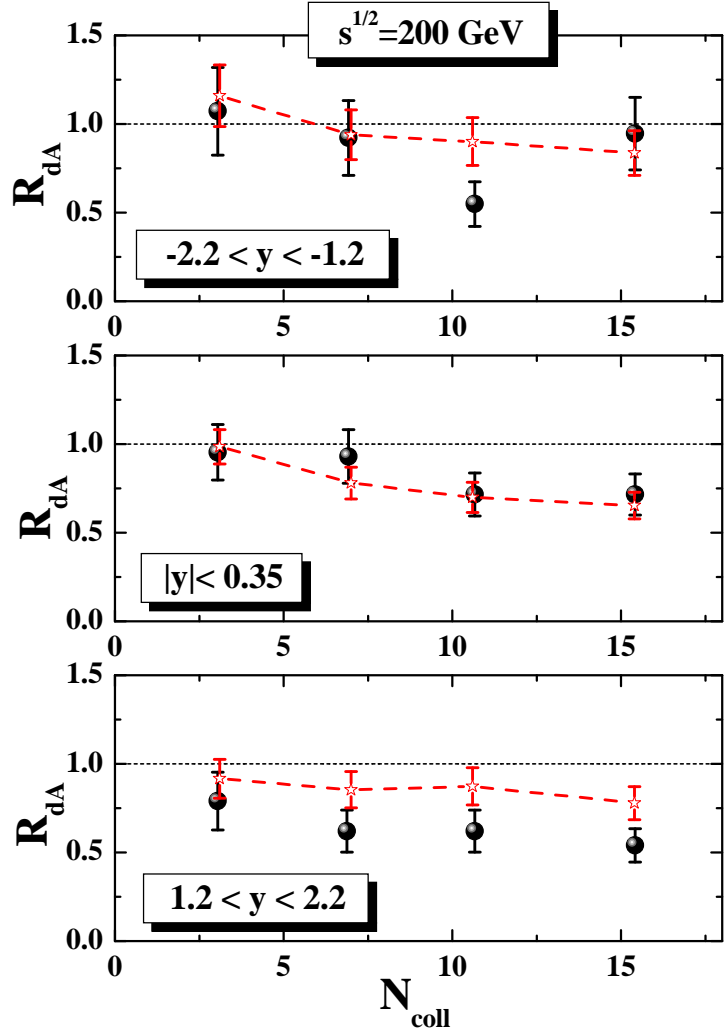,width=0.75\textwidth}} \caption{The ratio
$R_{dA}$ (\ref{FigRdA}) for backward, central and forward rapidity bins as a function of the number
of binary collisions $N_{coll}$ for $d+Au$ at $\sqrt{s}$ = 200 GeV. The experimental data have been
taken from Ref.~\protect\cite{PHENIX08}. The HSD results (stars connected by red dashed lines) show
calculations without including low-$x$ gluon shadowing and slightly overestimate $R_{dA}$ in the
forward interval 1.2 $< y <$ 2.2. The theoretical error bars are due to the finite statistics of
the calculation. The figure is taken from Ref.~\protect\cite{Olena.RHIC.2}.} \label{dAu2}
\end{figure}

In order to shed some further light on the role of shadowing, we compare our calculations for
$R_{dA}$ in different rapidity bins as a function of the centrality of the $d+Au$ collision, which
in Fig.~\ref{dAu2} is represented by the number of binary collisions $N_{coll}$. The latter number
is directly taken from the number of binary hard $NN$ collisions in the transport calculation while
the comparison with experiment is based on a Glauber model analysis of the data similar to that
performed in Ref.~\cite{Gran}. The actual results displayed in Fig.~\ref{dAu2} (stars connected by
dashed lines) and the PHENIX data from Ref.~\cite{PHENIX08} are roughly compatible for the rapidity
intervals -2.2 $< y <$ -1.2 and $|y| <$ 0.35, but demonstrate that the suppression at forward
rapidity (1.2 $< y < $2.2) is underestimated in the color-dipole dissociation model with a constant
cross section of 4.18 mb. This clearly points to the presence of shadowing effects at least at
forward rapidities which is not so pronounced in the inclusive data set in Fig.~\ref{dAu}. A more
serious question is a quantification of the shadowing, which is extremely challenging because of the limited statistics of both the experimental data and the calculations. Here we do not attempt to attribute a fixed number for the shadowing effect but merely point out that independent high statistics data will be necessary to fix this unsatisfactory situation from the experimental side.

Nevertheless, some note of caution is appropriate for the further
analysis of charmonium suppression in $Au+Au$ collisions: There
are `cold nuclear matter effects' such as `gluon shadowing' beyond
those incorporated in the HSD transport calculations, and
especially quantitative statements about any `agreement with data'
might have to be reconsidered. In case of $Au+Au$ reactions the
shadowing from projectile/target will show up symmetrically around
$y=0$ and in part contribute to the stronger $J/\Psi$ suppression
at forward/backward rapidities. Nevertheless, following Granier de
Cassagnac~\cite{Gran}, an anomalous suppression of $J/\Psi$ beyond
`cold nuclear matter' effects is clearly present in the $Au+Au$
data (to be investigated below).

\subsection{FAIR energies}

At FAIR energies shadowing is not expected to show any sizeable
effect (as at SPS energies). Accordingly, the charmonium
suppression will be driven by dissociation reactions with baryons,
mesons etc. as will be discussed in  Section 6. As we will see in
Subsection~\ref{Fair.S}, HSD predicts that the dissociation on
nucleons will actually be the leading mechanism for $J/\Psi$
suppression at FAIR energies.

\section{Hadron abundances from heavy-ion collisions}
\label{abundancies}

\begin{figure}
\centerline{\psfig{file=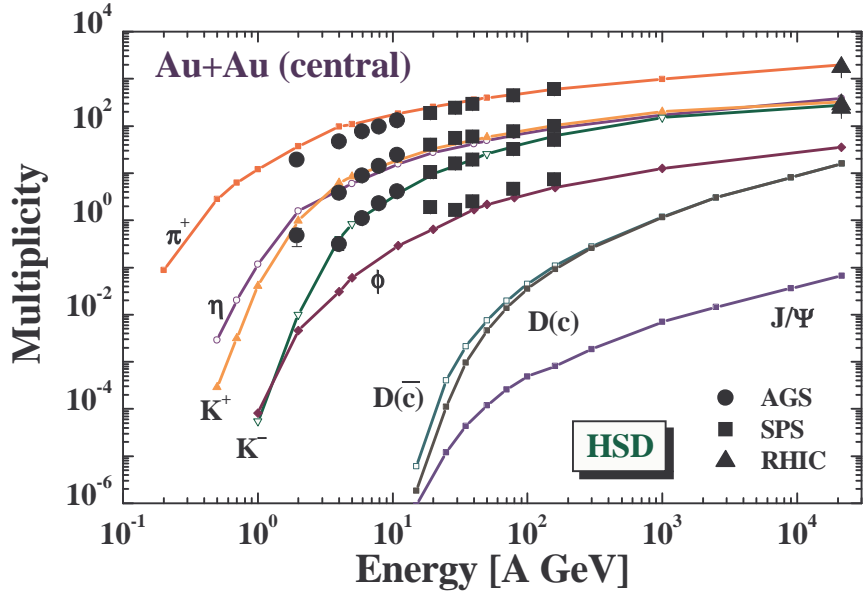,width=0.9\textwidth}} \caption{Overview on the experimental
meson abundances and HSD predictions for the multiplicities of $\pi^+, \eta, K^+, K^-$, $\phi$,
$D,\bar{D}$ and $J/\Psi$-mesons for central collisions of Au+Au as a function of bombarding energy
from SIS to RHIC energies. The figure is taken from Ref.~\protect\cite{Cass01} while the
experimental data have been added recently.} \label{Fig5s}
\end{figure}

We here recall the information gained from previous experimental
(and theoretical) studies on hadron production in heavy-ion
reactions. A general overview on the experimental meson and
strange baryon abundances from central nucleus-nucleus collisions
(Au+Au or Pb+Pb) is given in Fig.~\ref{Fig5s}, which shows the
 meson abundances from central Au+Au reactions as
predicted by HSD transport calculations in 2000~\cite{Cass01} from
SIS to RHIC energies. The experimental data at AGS (dots), SPS
(squares) and RHIC energies (triangles) have been added recently.
All meson multiplicities show a monotonic increase with bombarding
energy which is only very steep at `subthreshold' energies, i.e.
at bombarding energies per nucleon below the threshold in free
space for $NN$ collisions. Note that the HSD transport results are
also in a fair agreement with results from the UrQMD transport
approach and data (from AGS to RHIC energies) for the hadron
rapidity distributions~\cite{Weber}. Only in case of transverse
mass spectra both transport approaches underestimate the
experimental slopes from lower SPS to RHIC energies
\cite{Bratkovskaya:2004kv}.

Some comments with respect to the charm production in heavy-ion reactions are in order:  The mass of the $J/\Psi$ - for the states of interest here - gives the lowest scale of 3.097 GeV.
Accordingly, the formation of a $J/\Psi$ from an initial $c\bar{c}$ pair is the only allowed process (in vacuum) close to the charm threshold, because the $D+\bar{D}$ channel, {\it i.e.} $N\!+N\!\to\! D\!+\!\bar{D}\!+N\!+\!N$, has an effective invariant mass of 3.739~GeV. The associated production of a
$D(\bar{c})$ meson with a charmed hyperon $\Lambda_c (\Sigma_c$), {\it i.e.} $N\!+\! N\!\to\!\Lambda_c\!+\!D\!+\!N$, is more favorable due to effective invariant mass of 3.216 GeV (3.386
GeV), which is lower than for the production of a $D\bar{D}$ pair. This explains, why in Fig. \ref{Fig5s} the $D(\bar{c})$ cross section is larger than the $D(c)$ cross section close to threshold energies, and the $J/\Psi$ formation dominates in the far subthreshold domain. At higher bombarding energies the meson
abundances group according to their quark content, i.e. the multiplicities are reduced (relative to $\pi^+$) by about a factor of 5 for a strange quark, a factor of $\approx 2 \cdot 5^2$ = 50 for $s\bar{s} \equiv \phi$, a factor of $\approx 100$ for $D$ ($\bar{D}$), and $\approx 2\cdot 10^{4}$ for $c\bar{c} \equiv
J/\Psi$.

According to the arguments given above the $\bar{D}$-mesons with a $\bar{c}$ are produced more
frequently at low energies (due to an associated baryon $\Lambda_c, \Sigma_c$). At roughly 15
A$\cdot$GeV the cross sections for open charm and charmonia are expected to be of similar magnitude,
while at higher energies the ratio of open charm to charmonium bound states increases rapidly with
energy. Since the excitation function for open charm drops very fast with decreasing bombarding
energy, experiments around 25 A$\cdot$GeV, {\it e.g.} at the future FAIR
facility~\cite{FAIR.general}, will be a challenging task, because the multiplicity of the other
mesons is higher by orders of magnitude.

\begin{figure}
\centerline{\psfig{file=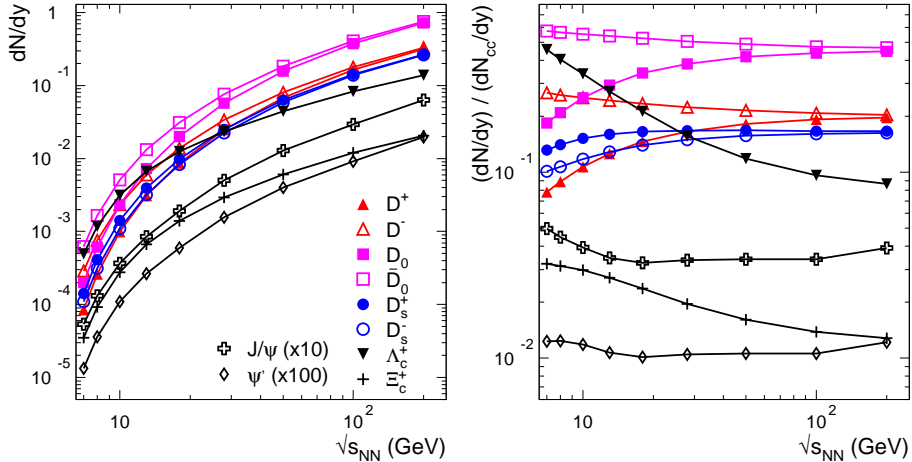,width=\textwidth}}
\caption{Energy dependence of charmed hadron production in
heavy-ion collisions at midrapidity in the statistical
hadronization model of Ref.~\protect\cite{PBM}. Left panel:
absolute yields, right panel: yields relative to the number of
$c\bar{c}$ pairs. Note (in both panels) the scale factors of 10
and 100 for $J/\Psi$ and $\Psi'$ mesons, respectively. The figure
is taken from Ref.~\protect\cite{PBM2}.} \label{pbmLambda}
\end{figure}

In the statistical hadronization model  of Ref.~\cite{PBM} the
production of $c \bar{c}$ pairs is assumed to proceed by hard
initial nucleon-nucleon scattering - as in HSD - but the
redistribution of charm quarks and antiquarks in the hadronization
process is assumed to follow statistical laws determined by a
chemical freeze-out temperature $T_{cfr}$ and the baryon chemical
potential $\mu_B$. The latter parameters are taken from the
experimental  systematics of the chemical freeze-out line in the
($T, \mu_B$) diagram~\cite{PBM}. The resulting energy dependence
of charmed hadron production in heavy-ion collisions (at
midrapidity) is displayed in Fig.~\ref{pbmLambda}, where the
absolute yields are shown on the l.h.s. whereas the yields
relative to the number of $c\bar{c}$ pairs produced are given in
the right panel. Note (in both panels) the scale factors of 10 and
100 for $J/\Psi$ and $\Psi'$ mesons, respectively. The relative
ratios of charmed baryons to charmed mesons differ significantly
from those in the HSD approach where full chemical equilibrium is
not achieved in the charm sector (cf. Section 8). On the other
hand it is unknown whether the statistical model is applicable to
the charm sector in the FAIR energy range (particularly for
non-central collisions). Future experiments at FAIR are expected
to clarify this issue.

As pointed out in Ref.~\cite{Grandchamp:2003uw}, dropping $D,
\bar{D}$ masses with baryon density (and/or temperature) might
lead to an increase of $J/\Psi$ absorption  and to a net lowering
of the $\Psi^{\prime}/J/\Psi$ ratio for central collisions. Thus
the $\Psi^{\prime}/J/\Psi$ ratio could also qualify as a probe of
$D$-meson in-medium effects. On the other hand, medium
modification of D-mesons is more complex than a simple picture of
dropping masses suggests. Microscopic G-matrix studies indicate
that the drop of the D-meson masses with baryon density is only
very moderate and as a leading effect one should expect a spectral
broadening~\cite{laura07,laura08}.

Note, however,  that the elementary cross sections for open charm
and charmonia in $pp$ and $\pi N$ reactions have to be measured in
the relevant kinematical regimes before reliable conclusions can
be drawn about charm dynamics in the nucleus-nucleus case.
Experimental data in the 20 - 30~A$\cdot$GeV with light and heavy
systems will have to clarify, furthermore, if the quasi-particle
picture of open charm mesons at high baryon density is applicable
at all or if the dynamics is already governed by partonic degrees
of freedom rather than hadronic ones (see below).

The production of `ordinary' hadrons (with `u,d,s' flavor) as well as open charm and charmonium  at
SPS and RHIC energies has been calculated so far within the AMPT\cite{Kojpsi}, HSD
\cite{Cass01,brat03,Cassing:2003nz} and UrQMD~\cite{Spieles} transport approaches using
parametrizations for the elementary production channels as described in  Section~\ref{elementary}.
Backward channels `charm + anticharm meson $\rightarrow$ charmonia + meson' are treated in HSD via
detailed balance in a schematic interaction model with a single  matrix element $|M|^2$ that is fixed by the $J/\Psi$ suppression data from the NA50 collaboration~\cite{Ramello} at SPS energies (cf. Ref.~\cite{brat03} and Section 6.1). The independent transport approaches provide in general very similar results for energy densities, baryon densities, meson densities etc. such that the bulk dynamics of relativistic nucleus-nucleus collisions is known to a sufficient extent (cf. Ref.~\cite{Arsene} for a more detailed comparison).

\section{Anomalous suppression of $J/\Psi$: The basic models}
\label{suppression}

In the past, the charmonia $J/\Psi$, $\chi_c$, $\Psi^\prime$ have
been discussed in context of the phase transition to the QGP,
since $c\bar{c}$ states might no longer be formed due to color
screening~\cite{Satznew}. However, more recent calculations within
lattice QCD (lQCD) have shown that at least the $J/\Psi$ survives
up to $\sim$ 1.5 $T_c$ ($T_c \approx$ 0.17 - 0.19 GeV) such that
the lowest $c\bar{c}$ states remain bound up to energy densities
of about 5 GeV/fm$^3$ (see Refs.~\cite{KarschJP,HatsudaJP}). It is
presently not clear, if also the $D$ or $D^*$ mesons survive at
temperatures above $T_c$, but strong correlations between a light
quark (antiquark) and a charm antiquark (quark) are likely to
persist. One may speculate that similar correlations survive also
in the light quark sector above $T_c$, such that `pre-hadronic
comovers' - most likely with different spectral functions - might
show up also at energy densities above 1 GeV/fm$^3$, which is
taken as a characteristic scale for the critical energy density.

On the other hand, it is well known that the baryonic (normal)
absorption alone cannot explain the suppression of charmonia in
heavy-ion collisions with increasing centrality~\cite{rev1} ({\it
cf.} Fig.~\ref{SPS.Baryon}). Different mechanism for the
additional (anomalous) suppression or formation of charmonia have
been suggested in the past, i.e. charmonia might be `melting'
according to the scenario advocated in Ref.~\cite{Satznew} (their
formation be suppressed due to plasma screening~\cite{KoO}),
absorbed on co-moving mesons in hot hadronic
matter~\cite{Capellanew,Capella} or they could be absorbed early
by neighboring strings~\cite{Geiss99}. Moreover, charmonia might
also be generated in a statistical fashion at the phase boundary
between the QGP and an interacting hadron gas such that their
abundance could be in statistical (chemical) equilibrium with the
light and strange hadrons as suggested in
Refs.~\cite{MG**2,PBM97}. The latter picture is expected to lead
not to a suppression but to an enhancement of $J/\Psi$ mesons at
the full RHIC energy if compared to the scaled $J/\Psi$
multiplicity from $pp$ collisions~\cite{Rafelski}.

\subsection{`Comover' suppression (and recombination)}
\label{comover}

First of all let us stress that the interactions with `comoving' mesons lead not only to the
dissociation of charmonia, but also to their recreation via the inverse recombination process
$D+\bar D \to c \bar c + m$, where $m= \{ \pi, \rho, \omega, K, ... \}$.
 As already pointed out before, the $J/\Psi,
\chi_c, \Psi^\prime$ formation cross sections by open charm mesons
or the inverse `comover' dissociation cross sections are not well
known and the significance of these channels is discussed
controversially in the literature~\cite{KoO,BMS,Rafelski,Bernd,Redlich,I2,I3}. In HSD --
following the concept of Refs.~\cite{brat03,brat04} --  a simple
2-body transition model is employed with a single parameter
$|M_0|^2$ that allows to implement the backward reactions uniquely
by exploiting detailed balance for each individual channel. We
briefly review this concept in the following.

\begin{figure}
\centerline{\psfig{file=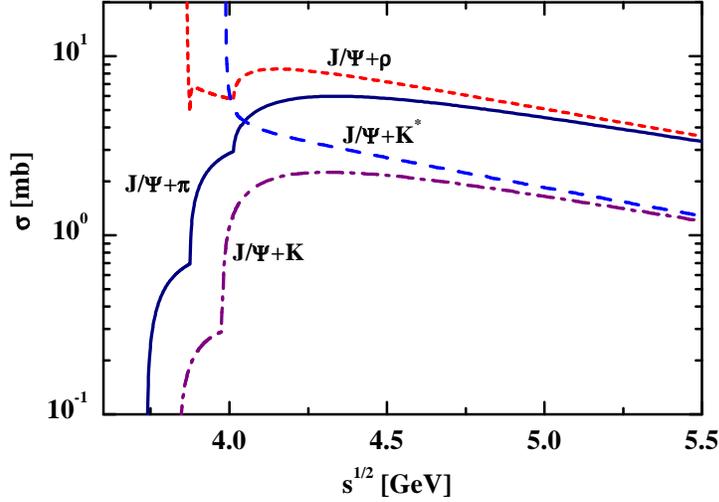,width=0.75\textwidth}} \caption{The $J/\Psi$ dissociation cross
sections with $\pi$, $\rho, K$ and $K^*$ mesons as specified in the text (Eqs. (\protect\ref{model})-(\protect\ref{channels})).} \label{bild6}
\end{figure}

Since the charmonium-meson dissociation and backward reactions typically occur with low relative
momenta (`comovers'), it is legitimate to write the cross section for the process $1+2\to 3+4$ as
\begin{equation}
\label{model}
 \sigma_{1+2\to 3+4}(s) = 2^4 \frac{E_1 E_2 E_3 E_4}{s}
|\tilde M_i|^2 \left(\frac{m_3+m_4}{\sqrt{s}}\right)^6 \frac{p_f}{p_i},
\end{equation}
 where $E_k$ denotes the energy of hadron $k$
$(k=1,2,3,4)$, respectively. The initial and final momenta for fixed invariant energy  $\sqrt{s}$
are given by
\begin{eqnarray}
p_i^2 = \frac{(s-(m_1+m_2)^2)(s-(m_1-m_2)^2)}{4s}, \nonumber\\
p_f^2 = \frac{(s-(m_3+m_4)^2)(s-(m_3-m_4)^2)}{4s}, \label{moment}
\end{eqnarray}
where $m_k$ denotes the mass of hadron $k$. In (\ref{model}) $|\tilde M_i|^2$ ($i=\chi_c, J/\Psi,
\Psi^\prime$) stands for the effective matrix element squared, which for the different 2-body
channels is taken of the form
\begin{eqnarray} \label{channels}
&&\hspace*{-3mm}|\tilde M_i|^2 =|M_i|^2  \ \ {\rm for} \
    \ (\pi,\rho)+(c\bar c)_i \to D+\bar{D} \label{mod}\\
&&\hspace*{-3mm}|\tilde M_i|^2 = 3 |M_i|^2  \ \ {\rm for} \
    \ (\pi,\rho)+(c\bar c)_i  \to D^*+\bar{D}, \ D+\bar{D}^*, \ D^* + \bar{D}^* \nonumber\\
&&\hspace*{-3mm}|\tilde M_i|^2 = \frac{1}{3} |M_i|^2 \ \ {\rm for}\
    \ (K,K^*)+(c\bar c)_i  \to D_s + \bar{D}, \ \bar{D}_s + D \nonumber \\
&&\hspace*{-3mm}|\tilde M_i|^2 =  |M_i|^2  \ \ {\rm for} \
    \ (K,K^*)+(c\bar c)_i  \to D_s + \bar{D}^*, \ \bar{D}_s + D^*,\ D^*_s + \bar{D}, \nonumber \\
&&\phantom{|\tilde M_i|^2 =  |M_i|^2  \ \ {\rm for} \ \ (K,K^*)+(c\bar c)_i  \to D_s + \bar{D}^*, \
}     \bar{D}^*_s + D, \ \bar{D}^*_s + D^* \nonumber
\end{eqnarray}
The relative factors of 3 in (\ref{mod}) are guided by the sum
rule studies in Ref.~\cite{korean} which suggest that the cross
section is increased whenever a vector meson $D^*$ or $\bar{D}^*$
appears in the final channel while another factor of 1/3 is
introduced for each $s$ or $\bar{s}$ quark involved. The factor
$\left( {(m_3+m_4)}/{\sqrt{s}} \right)^6 $ in (\ref{model})
accounts for the suppression of binary channels with increasing
$\sqrt{s}$ and has been fitted to the experimental data for the
reactions $\pi + N \rightarrow \rho+N, \omega+N, \phi+N, K^+
+\Lambda$ in Ref.~\cite{CaKo}.

\begin{figure}
\centerline{\psfig{figure=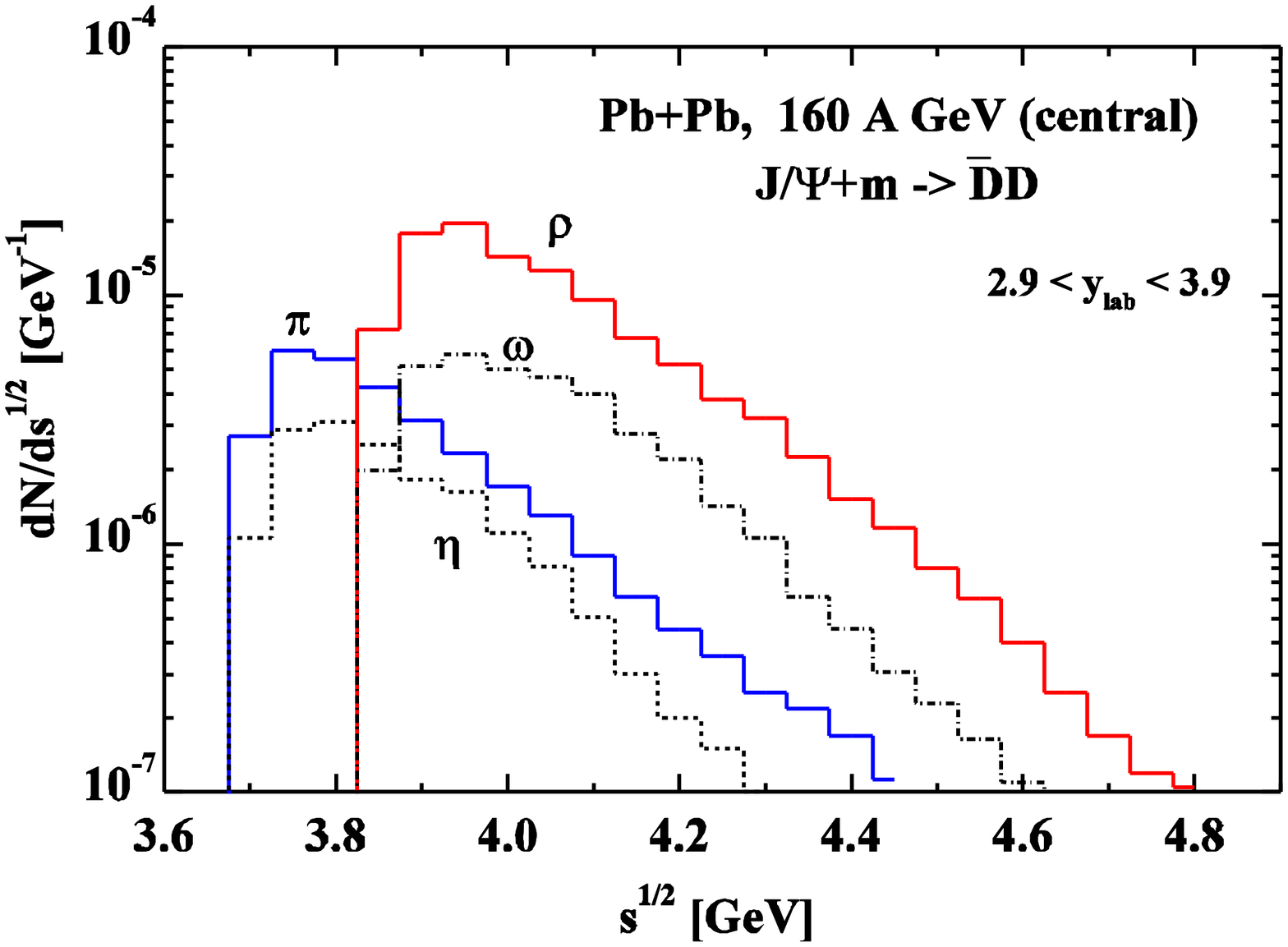,width=0.75\textwidth}} \caption{The distribution in the
invariant collision energy $\sqrt{s}$ for $J/\Psi$ absorption on $\pi$-, $\eta$-, $\rho$- and
$\omega$-mesons in central ($b$ = 2 fm) Pb + Pb collisions at 160 A$\cdot$GeV within the HSD
transport approach. The figure is taken from Ref.~\protect\cite{Cass99}.} \label{Ch7_Fig13}
\end{figure}

\begin{figure}
\centerline{\psfig{file=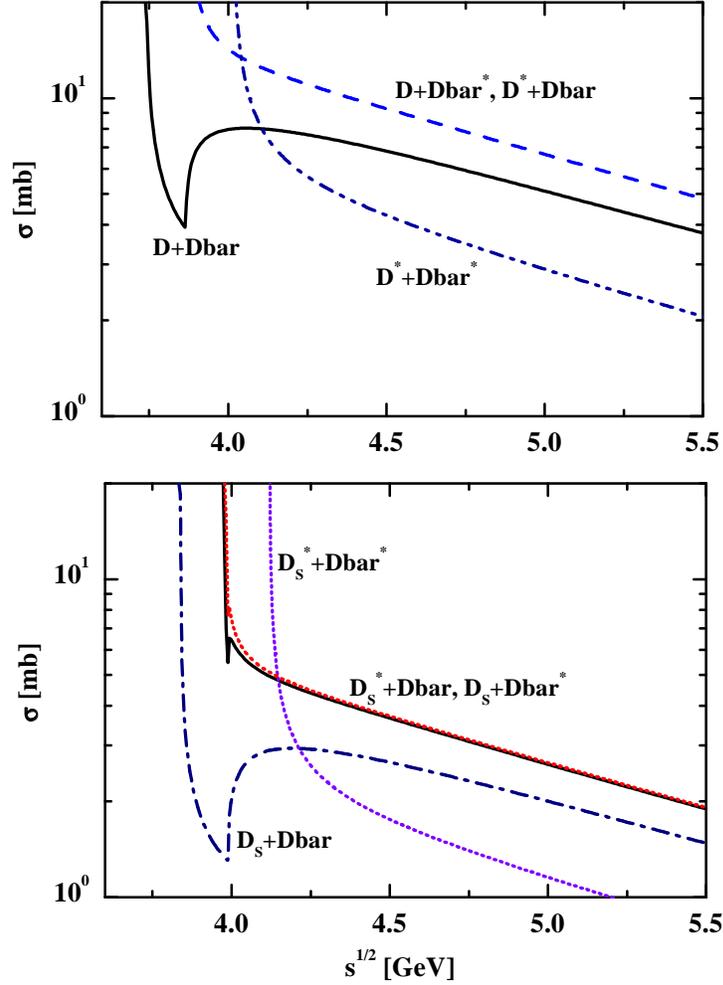,width=0.75\textwidth}} \caption{The cross sections for the
channels $D+\bar{D}, D+\bar{D}^*, D^*+\bar{D}, D^*+\bar{D}^* \rightarrow J/\Psi$ + meson (upper
part) and the channels involving $s$ or $\bar{s}$ quarks $D_s+\bar{D}$, $D_s+\bar{D}^*$,
$D^*_s+\bar{D}$, $D^*_s+\bar{D}^+ \rightarrow J/\Psi +(K,K^*)$ (lower part) as a function of the
invariant energy $\sqrt{s}$ according to the model described in the text (Eq. (\protect\ref{balance})).} \label{bild7}
\end{figure}

We use  the same matrix elements for the dissociation of all charmonium states $i$ ($i=\chi_c,
J/\Psi, \Psi^\prime$) with mesons:
\begin{eqnarray}
 |M_{J/\Psi}|^2 = |M_{\chi_c}|^2 = |M_{\Psi^\prime}|^2 = |M_0|^2.
\label{MatrElem}
\end{eqnarray}
The best fit for $|M_0|^2$ (in comparison to the latest NA50 and
NA60 analysis~\cite{NA60,NA50pA}) is obtained for
$|M_0|^2=0.18$~fm$^2$/GeV$^2$ ~\cite{Olena.SPS}; this value will
be employed in the HSD calculations for all bombarding energies
and systems.

The resulting $J/\Psi$ dissociation cross sections with $\pi$,
$\rho$, $K$ and $K^*$ mesons are displayed in Fig.~\ref{bild6} .
The final state includes all binary channels compatible with charm
quark and charge conservation. Note that for the comover
absorption scenario essentially the regime 3.8 GeV $\leq \sqrt{s}
\leq$ 4.8 GeV is of relevance (cf. Fig.~\ref{Ch7_Fig13}) where the
dissociation cross sections are on the level of a few mb. We note
that the explicit channel $J/\Psi + \pi \rightarrow D+ \bar{D}$,
which has often been calculated in the
literature~\cite{I2,I3,Konew,Ko}, is below 0.7 mb in our model. A
somewhat more essential result is that the $J/\Psi$ dissociation
cross section with $\rho$-mesons is in the order of 5-7 mb as in
the calculations of Haglin~\cite{Haglin} used before in
Ref.~\cite{Cass01}, since this channel was found to dominate the
$J/\Psi$ dissociation at SPS energies~\cite{Cass99}. Indeed, the
calculations show that in the realistic fireball evolution the
$J/\Psi$'s are dominantly absorbed in collisions with
$\rho$-mesons as seen in Fig.~\ref{Ch7_Fig13}, where the
distribution in the invariant collision energy $\sqrt{s}$ is
plotted for $J/\Psi$ absorption on $\pi, \eta, \rho$ and $\omega$
mesons in central ($b$ = 2 fm) Pb+Pb collisions at 160
A$\cdot$GeV. It should be pointed out that the `comover'
dissociation channels for charmonia are described in HSD with the
proper individual thresholds for each channel in contrast to the
more schematic `comover' absorption model in
Refs.~\cite{Capellanew,Capella}.

\begin{figure}
\centerline{\psfig{file=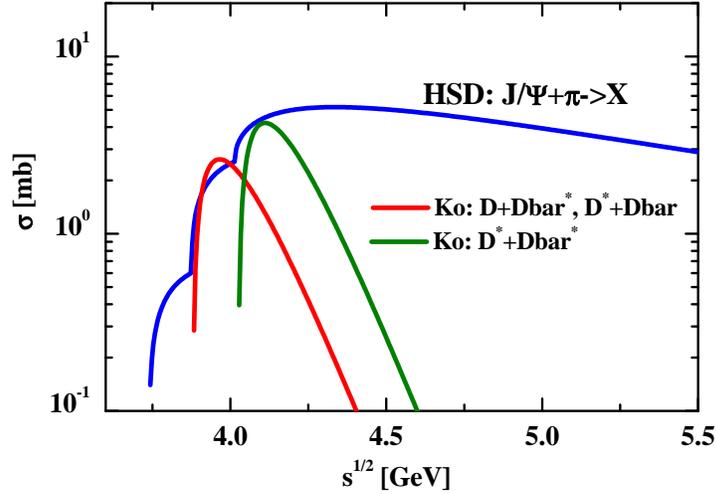,width=0.75\textwidth}}
\caption{The $J/\Psi$ dissociation cross section specified in
(\protect\ref{model})-(\protect\ref{channels}), upper blue line,
compared to the exclusive cross sections with $D\bar D^*$,
$D^*\bar D^*$ final states from Ref.~\protect\cite{KoO} (red and green lines). }
\label{xsJP_ko}
\end{figure}

The advantage of the model (\ref{model})  is that detailed balance
for the binary reactions can be employed strictly for each
individual channel, {\it i.e.}
\begin{eqnarray}
\!\!\sigma_{3+4 \rightarrow 1+2}(s) =
 \sigma_{1+2 \rightarrow 3+4}(s)
\frac{(2S_1+1)(2S_2+1)}{(2S_3+1)(2S_4+1)} \ \frac{p_i^2}{p_f^2}, \ \label{balance}
\end{eqnarray}
 and the role of the backward reactions ($(c\bar c)_i$+meson formation by $D+\bar{D}$ flavor exchange)
 can be explored without introducing any additional parameter once $|M_0|^2$ is fixed.
 In Eq.~(\ref{balance}) the quantities $S_j$
denote the spins of the particles, while ${p_i^2}$ and ${p_f^2}$
denote the cms momentum squared in the initial and final channels,
respectively. The uncertainty in the cross sections
(\ref{balance}) is of the same order of magnitude as that in
Lagrangian approaches using {\it e.g.} $SU(4)_{flavor}$
symmetry~\cite{Konew,Ko}, since the form factors at the vertices
are essentially unknown~\cite{korean}. The cross sections for
these backward channels -- summed up again over all possible
binary final states -- are displayed in Fig.~\ref{bild7}
separately for the `non-strangeness' (upper part) and
`strangeness' channels (lower part).

The regeneration of charmonia by recombination of $D$ ($D^*$)
mesons in the hadronic phase was first studied by C.M.~Ko and
collaborators in Ref.~\cite{KoO}. The conclusion at that time was
that this process was unlikely  at RHIC
energies~\cite{KoO,Redlich,Redlich2}. On the other hand, it was
shown within HSD~\cite{brat03} that the contribution of the
$D+\bar D$ annihilation to the produced $J/\Psi$ at RHIC is
considerable. Moreover, the equilibrium in the reaction $J/\Psi+m
\leftrightarrow D\bar D$ is reached (i.e. the charmonium
recreation is comparable with the dissociation by `comoving'
mesons). The reason for such differences is that the pioneering
study~\cite{KoO} within the hadron gas model was confined to
$J/\Psi$ reactions with pions only and into two particular $D\bar
D$ channels ($D+\bar D^*$ and $D^*+\bar D^*$). As one can see in
Fig.~\ref{xsJP_ko}, the cross sections used in
Refs.~\cite{KoO,Redlich2} --
 as obtained in the quark exchange model~\cite{I2} --
 in the two dissociation channels $J/\Psi+\pi\to D+\bar D^*$
 and $J/\Psi+\pi\to D^*+\bar D^*$ agree with the parametrization (\ref{model}).
However, in HSD the interactions with all mesons into all possible
combinations of $D\bar D$ states have been taken into account (cf.
Fig.~\ref{bild6}). Note that the $\rho$-meson density at RHIC is
large such that the channel with the most abundant $\rho$-meson
resonance is {\em dominant}. Furthermore, in Ref.~\cite{brat03}
the feed down from $\chi _c$ and $\Psi'$ decays has been
considered. The results of Ref.~\cite{brat03} are in accordance with
independent studies in Refs.~\cite{DD1,DD2,DD3,DD4} that stress
the importance for $D\bar D$ annihilation in the late (purely
hadronic) stages of the collisions.

\begin{figure}
\psfig{figure=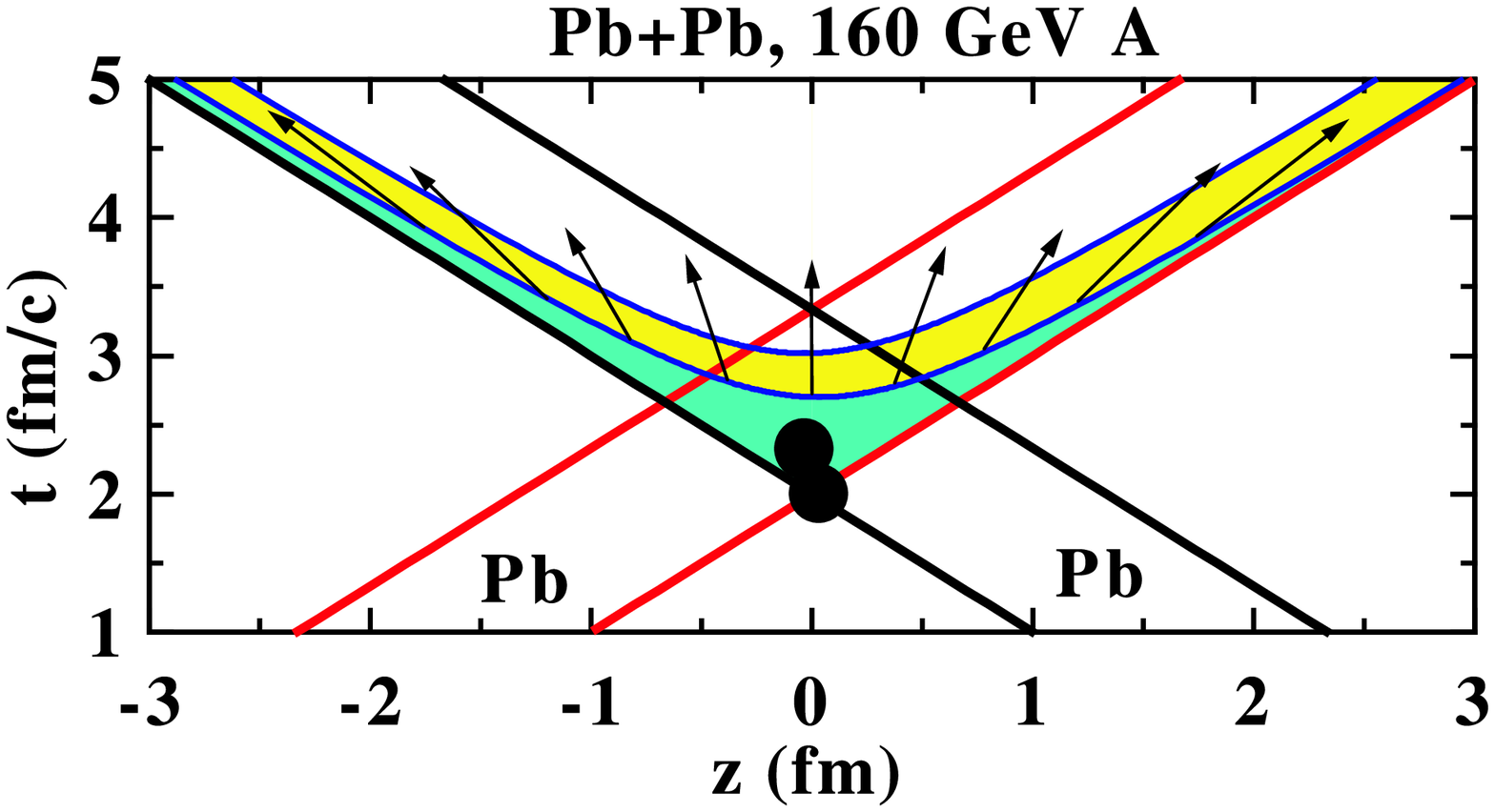,width=0.9\textwidth} \caption{Schematic representation of a Pb~+~Pb collision
at 160~A$\cdot$GeV in space-time.  The full dots represent early hard collision events (for
Drell-Yan and $c\bar{c}$-pairs) while mesons ($\pi, \eta, \rho, \omega$, etc. -- arrows) only
appear after a respective formation time $t_F$. The overlap area (inner rectangle) specifies the
space-time region of hard production events. The figure is taken from Ref.~\protect\cite{Cass97}.}
\label{LightCone}
\end{figure}

Note that in the default HSD approach (i.e. in the hadronic
comover dissociation and recombination scenario) only formed
comoving mesons participate in dissociation or $D\bar D$
recombination reactions (cf. Subsection~\ref{prehadron}).
Being hard probes, $c\bar c$ pairs are created in the early stage
of the collision, while the comoving mesons are formed at a later
stage. This situation is illustrated in Fig.~\ref{LightCone} for a
central Pb+Pb collision at 160~A$\cdot$GeV for freely streaming
baryons (thick black and red lines). The initial string formation
space-time points are indicated by the full dots; the mesons
(indicated by arrows) hadronize after a time delay $t_F \approx$
0.8 fm/c  as shown by the first hyperbola. A $c \bar{c}$-pair
produced in the initial hard nucleon-nucleon collision cannot be
absorbed by mesons in the (green) shaded areas in space and time;
however, a sizeable fraction of $c\bar{c}$ pairs (which should be
produced within the inner rectangles) can also be produced in a
dense mesonic environment. The upper hyperbolas in
Fig.~\ref{LightCone} represent the boundaries for the appearance
of mesons from the second interaction points (full dots) which
appear somewhat later in time; they stand for a representative
further nucleon-nucleon collision during the reaction.

\begin{figure}
\centerline{\psfig{file=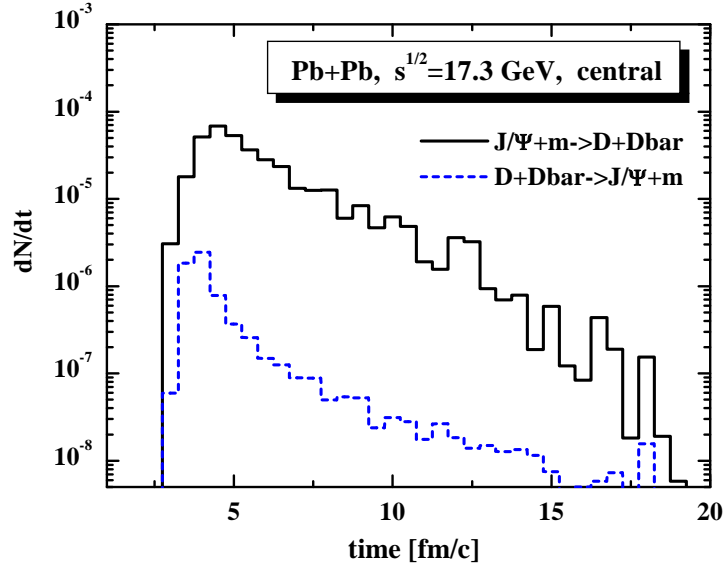,width=0.75\textwidth}}
\caption{The  calculated rate of $J/\Psi$ dissociation reactions
with mesons (solid histogram) for central $Pb+Pb$ collisions at
$\sqrt{s}$ = 17.3 GeV in comparison to the rate of backward
reactions of open charm pairs to $J/\Psi$ + meson (dashed
histogram). The figure is taken from Ref.~\protect\cite{brat03}.}
\label{bild10n}
\end{figure}
\begin{figure}
\centerline{\psfig{file=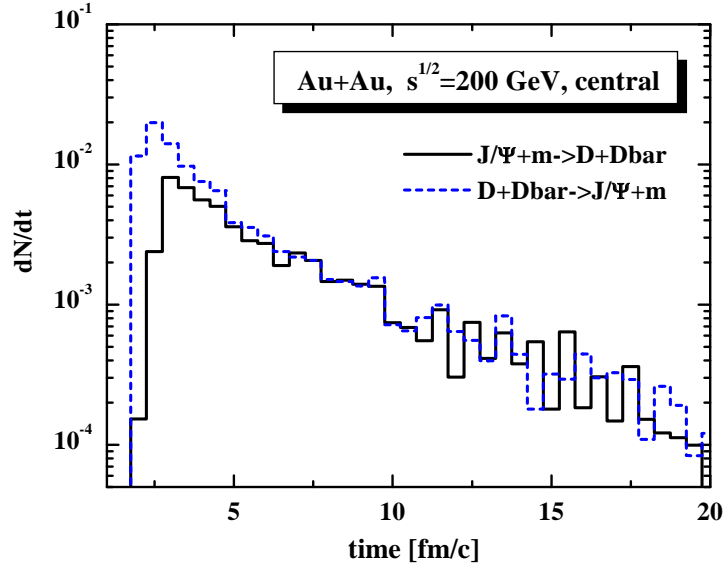,width=0.75\textwidth}}
\caption{The  calculated rate of $J/\Psi$ dissociation reactions
with mesons (solid histogram) for central $Au+Au$ collisions at
$\sqrt{s}$ = 200 GeV in comparison to the rate of backward
reactions of open charm pairs to $J/\Psi$ + meson (dashed
histogram). The figure is taken from Ref.~\protect\cite{brat03}.}
\label{bild11n}
\end{figure}

Thus, the interaction with comoving mesons can occur only in the late stages
of the reaction, i. e. after $t_F$ in their rest frame. This is taken into account
in HSD by treating  `formed' particles and `pre-hadrons' differently:
\begin{itemize}
\item A `formed' meson (baryon) is a quark-antiquark (quark-diquark) correlation -- produced at time
$t_0$ -- with hadronic quantum numbers that satisfies the two
constrains: \label{formed}
\bea \mbox{the time since production } & t-t_0 > t_F=\gamma \tau_F
\nonumber \\ \mbox{{\em and}} & \nonumber \\ \mbox{the local
energy density } & \varepsilon<1\mbox{~GeV/fm}^3, \nonumber \eea
where $\tau_F \approx$ 0.8 fm/c is the formation time, and
$\gamma$ the Lorentz $\gamma$-factor w.r.t. to the calculational
frame.
\item A `pre-hadron' is a correlation with hadronic quantum numbers
for
\bea & t-t_0 < t_F =\gamma \tau_F \nonumber & \\ & \mbox{{\em or}}
\nonumber  &\\ &\varepsilon > 1\mbox{GeV/fm}^3 . \nonumber & \eea
\item leading quarks (diquarks) of the strings do not carry hadronic quantum numbers; they interact with
reduced cross sections, which are set to 1/3 (2/3) of the hadronic
ones in line with constituent quark number scaling.
\end{itemize}

The result for the total $J/\Psi$ comover absorption rate (solid histogram) in central $Pb~+~Pb$
collisions at 160 A~GeV is shown in Fig.~\ref{bild10n} in comparison to the $J/\Psi$ reformation
rate (dashed histogram) that includes all backward channels. Since the rates differ by about 2
orders of magnitude, the backward rate for $J/\Psi$ formation can  be neglected at SPS
energies even for central $Pb+Pb$ reactions. For central $Au+Au$ collisions at $\sqrt{s}$ = 200
GeV, however, the multiplicity of open charm pairs should be $\sim$ 16, i.e. by about 2 orders of
magnitude larger, such that a much higher $J/\Psi$ reformation rate ($\sim N_{c\bar{c}}^2$) is
expected at RHIC energies (cf. Ref.~\cite{Rappnew}). In Fig.~\ref{bild11n} we display the total
$J/\Psi$ comover absorption rate (solid histogram)  in comparison to the $J/\Psi$ reformation rate
(dashed histogram) as a function of time in the center-of-mass frame. Contrary to Fig.~\ref{bild10n}
now the two rates become comparable for $t \geq$ 4-5 fm/c and suggest that in central
collisions at the
full RHIC energy of $\sqrt{s}$ = 200 GeV the $J/\Psi$ comover dissociation is no longer important
since the charmonia dissociated in this channel are approximately recreated in the backward
channels.

\subsection{`Threshold melting' (and recombination)}
\label{threshold}

The `threshold melting' scenario is based on the idea of a sequential dissociation of charmonia
with increasing temperature~\cite{Satz,Satznew,Satzrev,KSatz}, {\it i.e.} of charmonium melting in
the QGP due to color screening as soon as the fireball temperature reaches the dissociation
temperatures ($\approx 2 T_c$ for $J/\Psi$, $\approx 1.1-1.2 T_c$ for excited states, where $T_c$
stands for the critical temperature of the deconfinement phase transition). In the geometrical
Glauber model of Blaizot et al.~\cite{Blaizot} as well as in the percolation model of
Satz~\cite{Satzrev,DigalSatz}, it is assumed that the QGP suppression  sets in rather abruptly as
soon as the energy density exceeds a threshold value $\varepsilon_c$, which is a free parameter.
These models are motivated by the idea that the charmonium dissociation rate is drastically larger
in a quark-gluon-plasma (QGP)  than in a hadronic medium~\cite{Satzrev} such that further
(hadronic) comover absorption channels might be neglected.

We modify the standard sequential dissociation model of Refs.~\cite{Blaizot,Satzrev,DigalSatz} in
two aspects: (i) the energy density is calculated locally and microscopically instead of using
schematic estimates ({\it cf.} Section~\ref{energy}); (ii) the model incorporates a charmonium
regeneration mechanism (by $D\bar D$ annihilation processes as described in Section 6.1). The
`threshold scenario' for charmonium dissociation is implemented in HSD in a straight forward way:
whenever the local energy density $\varepsilon(x)$ is above a threshold value $\varepsilon_j$,
where the index $j$ stands for $J/\Psi, \chi_c, \Psi^\prime$, the charmonium is fully dissociated
to $c + \bar{c}$. The default threshold energy densities adopted are
\begin{equation} \mbox{ $\varepsilon_{J/\Psi} = 16$ GeV/fm$^3$ ,
$\varepsilon_{\chi_c} = 2$ GeV/fm$^3$, and
$\varepsilon_{\Psi^\prime} =2 $ GeV/fm$^3$.} \label{melt}
\end{equation}
The dissociation of charmonia has been widely studied using
lattice QCD (lQCD) in Refs.~\cite{Aarts,Petreczky,lQCD1,lQCD2,lQCD3}
in order to determine the
dissociation temperature (or energy density) via the maximum
entropy method. On the other hand one may use potential models -
reproducing the charmonium excitation spectrum in vacuum - to
calculate Mott transition temperatures in a hot medium. Both
approaches have their limitations and the quantitative agreement
between the different groups is still unsatisfactory:
\begin{itemize}
\item (A) Potential models usually employ the static heavy quark-antiquark
pair free energy $F(T)$ - calculated on the lattice - to obtain
the charmonium spectral functions. This leads to the (low)
dissociation temperatures~\cite{Mocsy,Digal:2001iu}
$$T_{melt}(J/\Psi) \le 1.2 \, T_c, \ T_{melt}(\chi_{c}) \le T_c, \
T_{melt}(\Psi') \le T_c .$$ An alternative way is to use the
internal energy $U= F + TS$ as a quark-antiquark potential, which
due to large contributions from the entropy $S$ provides
dissociation temperatures closer to the estimate
(\ref{melt})~\cite{Rapp:2008qc,Shuryak:2004tx,Mannarelli:2005pz,Mocsy:2005qw,Alberico:2005xw,Cabrera:2006wh}.
It is presently unclear (and very much debated) how to extract a
proper quark-antiquark potential from lQCD calculations.
\item (B) The maximum entropy method is used to relate the
Euclidean thermal correlators of charmonia - calculated on the
lattice - to the corresponding spectral functions. This yields
higher dissociation temperatures~\cite{Aarts} $$T_{melt}(J/\Psi) =
1.7\! -\! 2 \, T_c, \ T_{melt}(\chi_{c}) = 1.1\! - \! 1.2 \, T_c$$
or~\cite{Petreczky} $$T_{melt}(J/\Psi) \ge 1.5 \, T_c, \
T_{melt}(\chi_{c}) \approx 1.1 \, T_c.$$
\end{itemize}
Since  the low values for the melting temperatures from the
potential models are already in conflict with the $J/\Psi$ data at
SPS, the values (\ref{melt}) are employed in the following (if not
stated otherwise).

\subsection{Charm interactions with pre-hadrons in the early phase}
\label{prehadron}

In the default HSD approach all newly produced hadrons (by string
fragmentation)  have a formation time of $\tau_F \approx$  0.8
fm/c $\approx 1/\Lambda_{QCD}$ in their rest frame and do not
interact during the `partonic' propagation. Furthermore,
hadronization is inhibited, if the energy density -- in the local
rest frame -- is above 1 GeV/fm$^3$, which roughly corresponds to
the energy density for QGP formation in equilibrium. This default
approach underestimates the elliptic flow of hadrons at RHIC
energies \cite{brat03}, the suppression of hadrons with high
transverse momentum $p_T$ \cite{HPT1} as well as the suppression
of the far-side jets \cite{HPT2} in central $Au+Au$ collisions at
$\sqrt{s}$ = 200 GeV. This failure has been attributed to the lack
of explicit partonic interactions in the early collision phase,
which corresponds to the phase of high energy density with the
majority of hadrons being still `under formation'. In order to
simulate partonic interaction effects the HSD approach has been
extended by explicit interactions of pre-hadrons with the
(perturbative) charm degrees of freedom \cite{Olena.RHIC.2}.

Accordingly, an additional scenario has been implemented in the HSD simulations which is closely related to the `comover suppression' scenario outlined in Section~\ref{comover}, which clearly separates `formed hadrons' (existing only at energy densities below the energy density $\varepsilon_{cut}=\varepsilon_c \approx 1$~GeV/fm$^3$) from possible pre-hadronic states at higher energy densities, i.e, above  the parton/hadron
phase transition. Indeed, it is currently not clear whether $D$- or $D^*$-mesons survive at energy densities above $\varepsilon_c$, but hadronic correlators with the quantum numbers of the hadronic
states are likely to persist above the phase transition~\cite{Rapp05}.  Thus `comovers' -- with
modified spectral functions -- could show up also at energy densities above $\varepsilon_c$. We recall that in HSD a {\it pre-hadron} is defined as a state with the quantum numbers of a hadron, if the local energy density is above $\approx 1$~GeV/fm$^3$ or if the state is still under `formation', i.e. if the time between production and hadronization is smaller than the formation time $\tau_F$ (in its rest frame). For a more detailed
description of the pre-hadron concept we refer the reader to Section~\ref{formed} and to Refs.~\cite{Falter1,HPT1,Falter2,HPT2}.

In line with the investigations in Refs.~\cite{HPT1,HPT2}, the
$J/\Psi$ production and absorption in $Au+Au$ collisions at
$\sqrt{s}=200$~AGeV has been studied assuming the absorption of
charmonia on pre-hadrons as well as their regeneration by
pre-hadrons~\cite{Olena.RHIC.2}. This adds additional interactions
of the particles with charm quarks (antiquarks) in the very early
phase of the nucleus-nucleus collisions as compared to the default
HSD approach. Since these pre-hadronic (color-dipole) states
represent some new degrees-of-freedom, the interactions of charmed
states with these objects have to be specified separately.

For notation, we define a pre-hadronic state consisting of a
quark-antiquark pair as {\it pre-meson} ${\tilde m}$ and a state
consisting of a diquark-quark pair as {\it pre-baryon} ${\tilde
B}$. The dissociation cross section of a $c{\bar c}$ color dipole
state with a pre-baryon is taken to be of the same order as with a
formed baryon,
\begin{equation} \label{sss1}
\sigma_{c{\bar c}{\tilde B}}^{diss} = 5.8 \ {\rm mb},
\end{equation}
whereas the cross section with a pre-meson follows from the additive quark model as
\cite{Falter1,Falter2}
\begin{equation} \label{sss2}
\sigma_{c{\bar c}{\tilde m}}^{diss} = \frac{2}{3} \sigma_{c{\bar c}{\tilde B}}^{diss}.
\end{equation}
Elastic cross sections are taken as
\begin{equation} \label{sss3}
\sigma_{c{\bar c}{\tilde B}}^{el} = 1.9 \ {\rm mb}, \hspace{2cm}
 \sigma_{c{\bar c}{\tilde m}}^{el}
= \frac{2}{3} \sigma_{c{\bar c}{\tilde B}}^{el}.
\end{equation}
Furthermore, elastic interactions of a charm quark (antiquark) are modeled by the scattering of an
unformed $D$ or $D^*$ meson on pre-hadrons with only light quarks as
\begin{equation} \label{sss4}
\sigma_{D{\tilde B}}^{el} = 3.9 \ {\rm mb}, \hspace{2cm}
 \sigma_{D {\tilde m}}^{el}
= \frac{2}{3} \sigma_{D{\tilde B}}^{el}.
\end{equation}
In this way one can incorporate in HSD some dynamics of
quark-antiquark pairs with a medium that has not yet formed the
ordinary hadrons. However, it has to be stressed that further
explicit partonic degrees of freedom, i.e. gluons and their mutual
interactions as well as gluon interactions with quarks and
antiquarks, have not been taken into account explicitly so far.

\subsection{Thermal and statistical models}

The statistical and thermal models are based on the assumption of
statistical equilibrium, where the physical system can be
characterized by a few Lagrange parameters that specify the
average energy as well as the average particle number, flavor
content, $etc$. Since the fireball created in relativistic
nucleus-nucleus collisions is rapidly expanding, equilibrium rate
equations for charmonium dissociation and regeneration are folded
over a thermally evolving background in time~\cite{NewRapp}. In a
simplified form, the rate equation for the time evolution of a
charmonium state $\psi$, $N_\psi(\tau)$ can be written as
\begin{equation}
\frac{dN_{\psi}}{d\tau} = -\Gamma_{\psi}\left[N_{\psi}-N_{\psi}^{eq}\right] \ . \label{eq:rate}
\end{equation}
The first key quantity in (\ref{eq:rate})  is the inelastic
charmonium reaction rate, $\Gamma_{\psi}$, which, by detailed
balance, governs both gain and loss terms~\cite{NewRapp}. This is
in quite analogy to the covariant transport models. In the QGP,
the leading-order (LO) process is the well-known
gluo-dissociation, $g+\psi\to c+\bar c$. However, as has been
first emphasized in Ref.~\cite{Rappnew}, the gluo-dissociation
process becomes inefficient for small $J/\psi$ binding energies as
expected due to color screening in the QGP (and even without
screening for $\psi'$ and $\chi_c$ states). Therefore, the
quasi-free dissociation process, $p+\psi\to c+\bar c +p$
($p=q,\bar q, g$) has been introduced~\cite{Rappnew}, which
naively is of next-to-leading order in $\alpha_s$, but provides a
much larger phase space, and, consequently, the dominant
dissociation rate for small charmonium binding (for
gluo-dissociation, the phase space vanishes in the limit of
vanishing binding energy). The other key quantity is the
charmonium equilibrium limit, $N_{\psi}^{eq}(\tau)$, which depends
on the charm content and temperature of the system. The typical
procedure is to assume charm production to be a hard process and
thus to be restricted to primordial $N$-$N$ collisions. The
statistical model is then used to distribute the fixed number of
$c\bar c$ pairs over the available charmed states in the system.
This introduces both temperature and volume dependencies into
$N_{\psi}^{eq}(T(\tau))$, as well as a sensitivity to medium
modifications of the charm states (e.g., reduced $D$-meson masses
lead to a reduction in the charmonium equilibrium
numbers)~\cite{Grandchamp:2003uw,Rapp:2003vj}. Alternative
absorption mechanisms might also play a role, such as gluon
scattering on color dipole states as suggested in
Refs.~\cite{Kojpsi,Rappnew,Blaschke1,Blaschke2} or charmonium
dissociation in the strong color fields of overlapping
strings~\cite{Geiss99}.

The statistical hadronization model~\cite{PBM} (originally
proposed in Ref.~\cite{BMS}) follows a very different idea. The
$c\bar{c}$ are produced by initially hard $NN$ scattering, but
charmonium bound states are generated in a statistical fashion at
the phase boundary from the QGP to the hadronic phase~\cite{BMS2}.
This is fundamentally different from a regeneration of charmonia
in the hadronic phase by $D+\bar{D}$ collisions and implies that
all charmonia are emerging from a QGP phase. Since this assumption
is questionable for peripheral nucleus-nucleus collisions, the
authors have later on introduced `corona effects' that mimic
non-QGP charmonium formation.  We will come back to actual results
and comparisons in Section 8.

Slightly later Thews et al. developed a coalescence model for
charmonium production in the QGP~\cite{Rafelski} which states that
the abundance of charmonia increases roughly with the $c\bar{c}$
density squared and accordingly becomes increasingly important at
higher bombarding energy (especially at LHC energies). While
actual data at RHIC do not strongly support this scenario, it will
be of substantial interest if the future data at LHC will provide
evidence for charmonium enhancement instead of
suppression~\cite{PBM,NewRapp}.

\section{Anomalous suppression of $J/\Psi$: Comparison to data}
\label{suppression.data}

\subsection{$J/\Psi$ and $\Psi^\prime$ suppression at SPS energies}
\label{dataSPS}

\begin{figure}
\centerline{\psfig{file=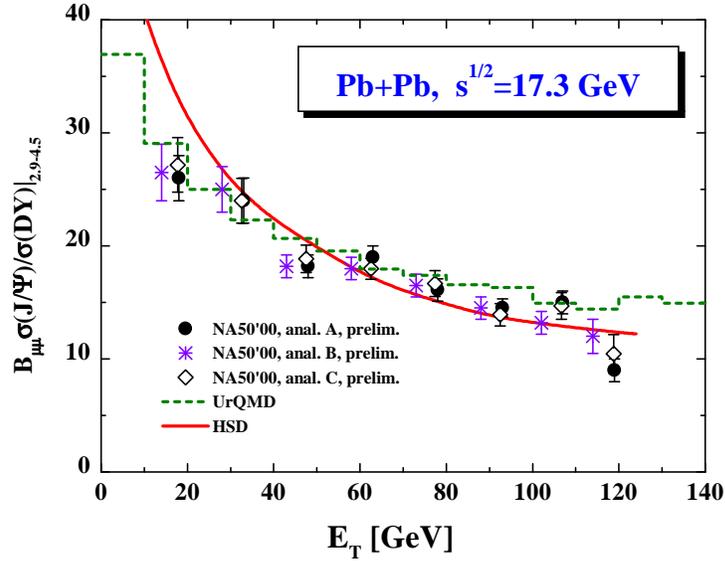,width=0.75\textwidth}} \caption{The $J/\Psi$ suppression as a
function of the transverse energy $E_T$ in Pb+Pb collisions at 160 A$\cdot$GeV. The solid line
shows the HSD result within the comover absorption scenario~\protect\cite{brat03,brat04}. The different
symbols stand for the NA50 data~\protect\cite{Ramello} from the year 2000 (ana\-lysis A,B,C) while
the dashed histogram is the UrQMD result~\protect\cite{Spieles}. The figure is taken from Ref.
~\protect\cite{Cassing:2003nz}.} \label{Fig2Pb00}
\end{figure}

As pointed out before, `cold nuclear matter' absorption effects
alone cannot reproduce the strong suppression of $J/\Psi$ observed
by the NA50 collaboration in central $Pb+Pb$ collisions ({\it cf.}
Fig~\ref{SPS.Baryon}). The extra suppression of charmonia in the
high density phase of nucleus-nucleus collisions then may be
attributed to the different `hot matter' scenarios outlined in
Section~\ref{suppression}.

As a reminder, we recall the early $J/\Psi$ suppression results
for Pb+Pb at 160 A$\cdot$GeV (in the comover suppression scenario)
both from the UrQMD and HSD transport calculations, which are in
line with the data of the NA50 Collaboration as demonstrated in
Fig.~\ref{Fig2Pb00}, where the calculated $J/\Psi$ suppression is
shown as a function of the transverse energy $E_T$. The solid line
stands for the HSD result (within the comover absorption
scenario)~\cite{brat03,brat04}, while the various data points
reflect the NA50 data from the year 2000 (analysis A,B,C) that
agree reasonably well with the HSD and UrQMD
calculations~\cite{Spieles} (dashed histogram in
Fig.~\ref{Fig2Pb00}).

The anomalous suppression observed in S+U and Pb+Pb collisions by
the NA38~\cite{NA38} and NA50 Collaborations~\cite{NA50O,Ramello}
has been experimentally confirmed by NA60~\cite{NA60}  in In+In
collisions at 158 A$\cdot$GeV. A couple of models have predicted
$J/\Psi$ suppression in In+In collisions as a function of
centrality at 158 A$\cdot$GeV based on the parameters fixed for
Pb+Pb reactions at the same bombarding energy. However, the
predictions within the `Glauber based' comover model and the
`threshold melting scenario' from
Refs.~\cite{NA60theo2,NA60theo1,DigalSatz} have failed to describe
the In+In data with sufficient accuracy (cf. Section 7.2).

\begin{figure}
\centerline{\psfig{file=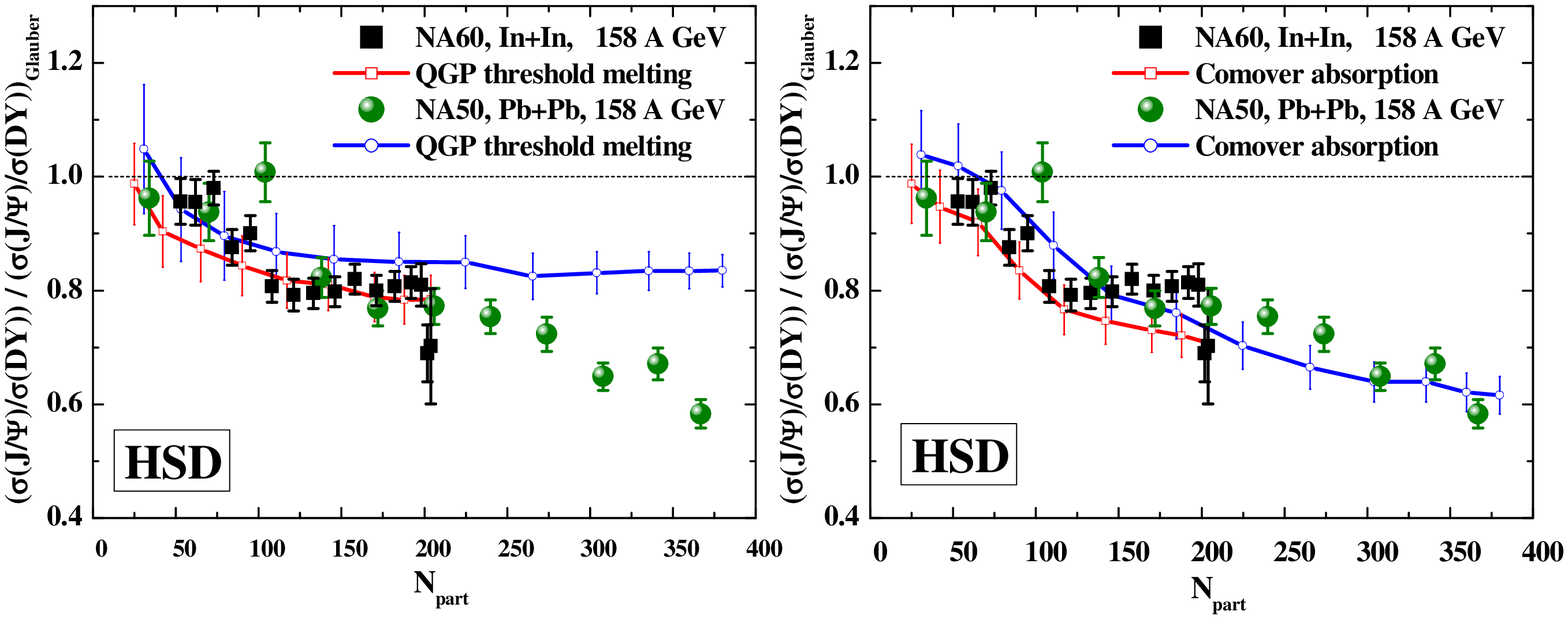,width=\textwidth}} \caption{The ratio
$B_{\mu\mu}\sigma(J/\Psi)/\sigma(DY)$ as a function of the number of participants $N_{part}$ in
In+In (red line with open squares) and Pb+Pb reactions (blue line with open circles) at 158
A$\cdot$GeV relative to the normal nuclear absorption given by the straight black line. The full
dots and squares denote the respective data from the NA50 and NA60 Collaborations. The model
calculations reflect the comover absorption model (right part) and the `QGP threshold scenario'
(left part) with $\varepsilon_{J/\Psi} = 16$ GeV/fm$^3$, $\varepsilon_{\chi_c} = 2$ GeV/fm$^3$,
$\varepsilon_{\Psi^\prime}$ = 2 GeV/fm$^3$ while discarding comover absorption. The figure is taken
from Ref.~\protect\cite{Olena.SPS}.} \label{ccQGP}
\end{figure}

\begin{figure}
\centerline{\psfig{file=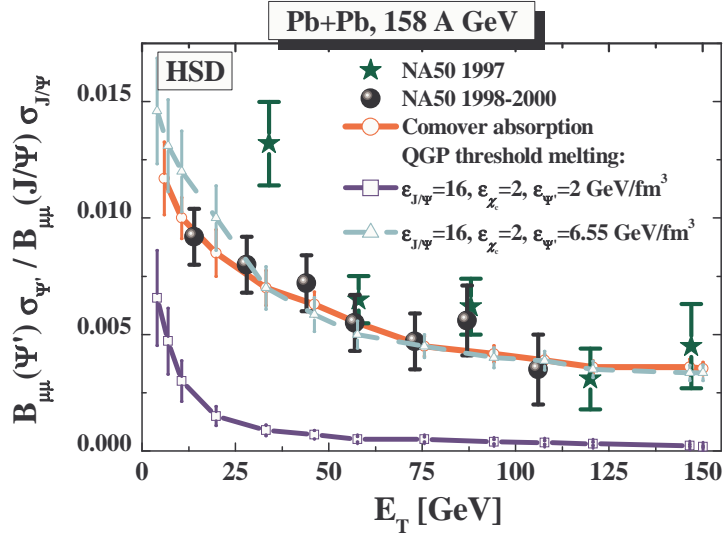,width=0.75\textwidth}} \caption{ The  $\Psi^\prime$ to
$J/\Psi$ ratio as a function of the transverse energy $E_T$ for Pb+Pb at 160 A$\cdot$GeV. The full dots
and stars denote the respective data from the NA50 Collaboration \protect\cite{NA50PsiPrime}. The
HSD result~\protect\cite{Olena.SPS} for the comover absorption model is shown as the red line,
whereas the blue line indicates the `QGP threshold scenario' with $\varepsilon_{J/\Psi} = 16$
GeV/fm$^3$, $\varepsilon_{\chi_c} = 2$ GeV/fm$^3$ = $\varepsilon_{\Psi^\prime}$ and the
light blue line with
$\varepsilon_{J/\Psi} = 16$ GeV/fm$^3$, $\varepsilon_{\chi_c} = 2$ GeV/fm$^3$ and
$\varepsilon_{\Psi^\prime} = 6.55$ GeV/fm$^3$  (discarding comover absorption). The figure is taken from Ref.~\protect\cite{Olena.SPS}.} \label{R_PPJP}
\end{figure}

The charmonium production and suppression in In+In and Pb+Pb reactions at SPS energies has been
 reinvestigated  in the HSD transport approach in 2006 within the `hadronic comover model' as well
as the `QGP threshold scenario'~\cite{Olena.SPS}. As found in Ref.~\cite{Olena.SPS}, the comover
absorption model -- with a single parameter $|M_0|^2$ for the matrix element squared for
charmonium-meson dissociation -- performs best with respect to all data sets for $J/\Psi$
suppression as well as for the $\Psi^\prime$ to $J/\Psi$ ratio for Pb+Pb ({\it cf.}
Figs.~\ref{ccQGP},\ref{R_PPJP}). We recall that the $\Psi^\prime$ suppression is presented
experimentally by the ratio
\begin{equation} \label{pis}
\frac{B_{\mu\mu}(\Psi^\prime\to \mu\mu)\sigma(\Psi^\prime)/\sigma(DY) } {B_{\mu\mu}(J/\Psi\to
\mu\mu)\sigma(J/\Psi) / \sigma(DY)}.
\end{equation}
In the HSD calculations this ratio is taken as 0.0165 for nucleon-nucleon collisions, which is
again based on the average over $pp, pd, pA$ reactions~\cite{NA50_03}. The centrality dependence of
the ratio is then a prediction of the model for different systems and bombarding energies.

The `QGP threshold scenario' roughly reproduces the $J/\Psi$ suppression for both systems at 158
A$\cdot$GeV (Fig.~\ref{ccQGP}, l.h.s.) apart from the very central Pb+Pb collisions. The comover
absorption model follows slightly better the fall of the $J/\Psi$ survival probability with
increasing centrality (Fig.~\ref{ccQGP}, r.h.s.), whereas the `threshold scenario'
leads to an approximate plateau in both reactions for high centrality.

On the other hand, the `threshold melting scenario' clearly fails
for the $\Psi^\prime$ to $J/\Psi$ ratio, since too many
$\Psi^\prime$ already melt away for a critical energy density of 2
GeV/fm$^3$ at 158 A$\cdot$GeV (cf. Fig.~\ref{R_PPJP}). Only when
assuming the $\Psi^\prime$ to dissolve  above $\sim 6.5$
GeV/fm$^3$ a reasonable description of all data is achieved in the
`QGP threshold scenario'; this threshold, however, is not in
accordance with lattice QCD calculations such that the `threshold
scenario' meets severe problems.

\begin{figure}
\centerline{\psfig{file=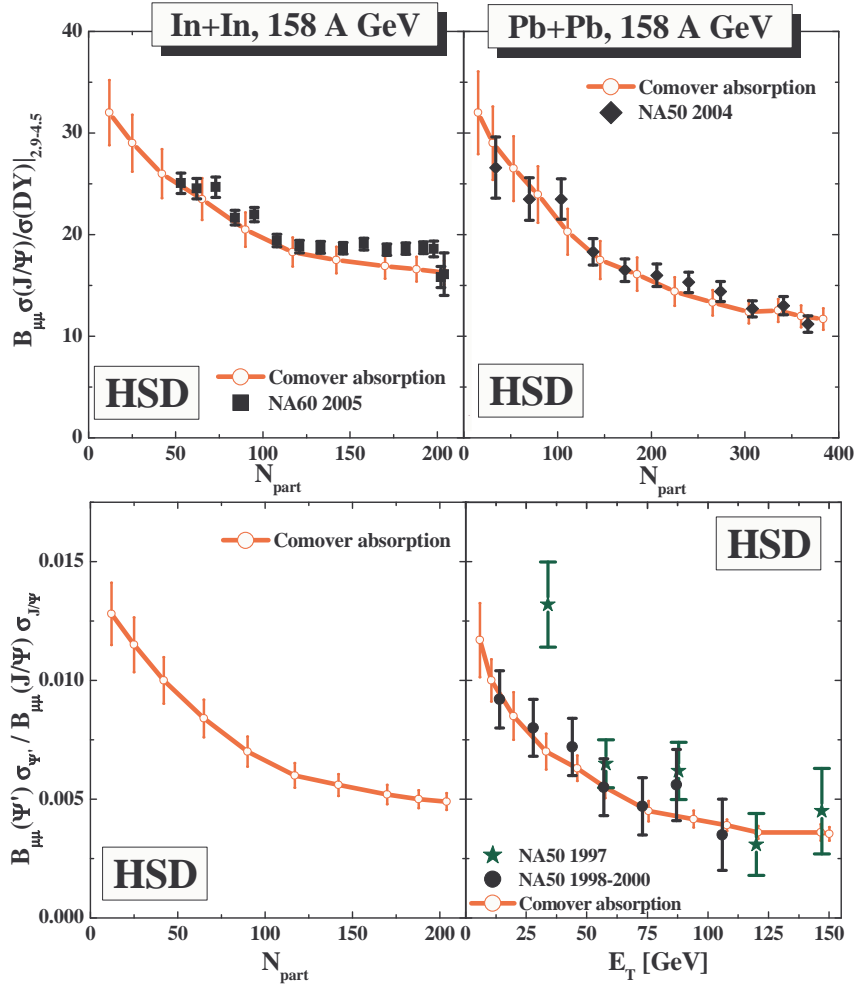,width=0.9\textwidth}} \caption{The ratio
$B_{\mu\mu}\sigma(J/\Psi)/\sigma(DY)$  as a function of the number of participants in In+In
(l.h.s.) and Pb+Pb reactions (r.h.s.) at 158 A$\cdot$GeV. The full symbols denote the data from the
NA50 and NA60 Collaborations (from Refs. \protect\cite{NA50O,NA60,NA50PsiPrime}). The solid (red)
lines show the HSD results for the comover absorption model with a matrix element squared $|M_0|^2$
= 0.18 fm$^2$/GeV$^2$. The lower parts of the figure show the HSD results in the same limit for the
$\Psi^\prime$ to $J/\Psi$ ratio as a function of $N_{part}$ (for In+In) or the transverse energy
$E_T$ (for Pb+Pb). The vertical lines on the graphs reflect the theoretical uncertainty due to
limited statistics of the calculations. The figure is taken from Ref.~\protect\cite{Olena.SPS}.}
\label{figure1}
\end{figure}

Indeed, the extra suppression of charmonia by comovers is seen in
Fig.~\ref{figure1} (solid red lines) to match the $J/\Psi$
suppression in In+In and Pb+Pb as well as the $\Psi^\prime$ to
$J/\Psi$ ratio (for Pb+Pb) rather well. The more recent data
(1998-2000) for the $\Psi^\prime$ to $J/\Psi$ ratio agree with the
HSD prediction~\cite{brat04} within error bars. This had been a
problem in the past when comparing to the 1997 data (dark green
stars). One may conclude that the comover absorption model so far
cannot be ruled out on the basis of the available data sets from
the SPS within error bars. The $\Psi^\prime$ to $J/\Psi$ ratio for
In+In versus centrality is not yet available from the experimental
side but the theoretical predictions (provided in
Fig.~\ref{figure1}) might be verified/falsified in future.

Some comments on the comover absorption model appear in place: As
shown in Fig. 7.2 of Ref.~\cite{Cass99}, the comover densities in
central Pb+Pb collisions at 158~A$\cdot$GeV become quite large and
almost reach 2~fm$^{-3}$ in the maximum, which appears high for
`free' mesons with an eigenvolume of about 1~fm$^3$. However, as
mentioned before, the quasi-particle mesons considered here
dynamically should not be identified with `free' meson states that
show a long polarization tail in the vacuum. As known from lattice
QCD, the correlators for pions and $\rho$-mesons survive well
above the critical temperature $T_c$, such that `dressed' mesons,
i.e spectral densities with the quantum numbers of the
pseudo-scalar and vector (isovector) modes, also show up at high
energy density (similar to the $J/\Psi$ discussed
above~\cite{KarschJP,HatsudaJP,Karsch2}). Such `dressed' mesons are
expected to have a more compact size in space.

\begin{figure}
\centerline{\psfig{file=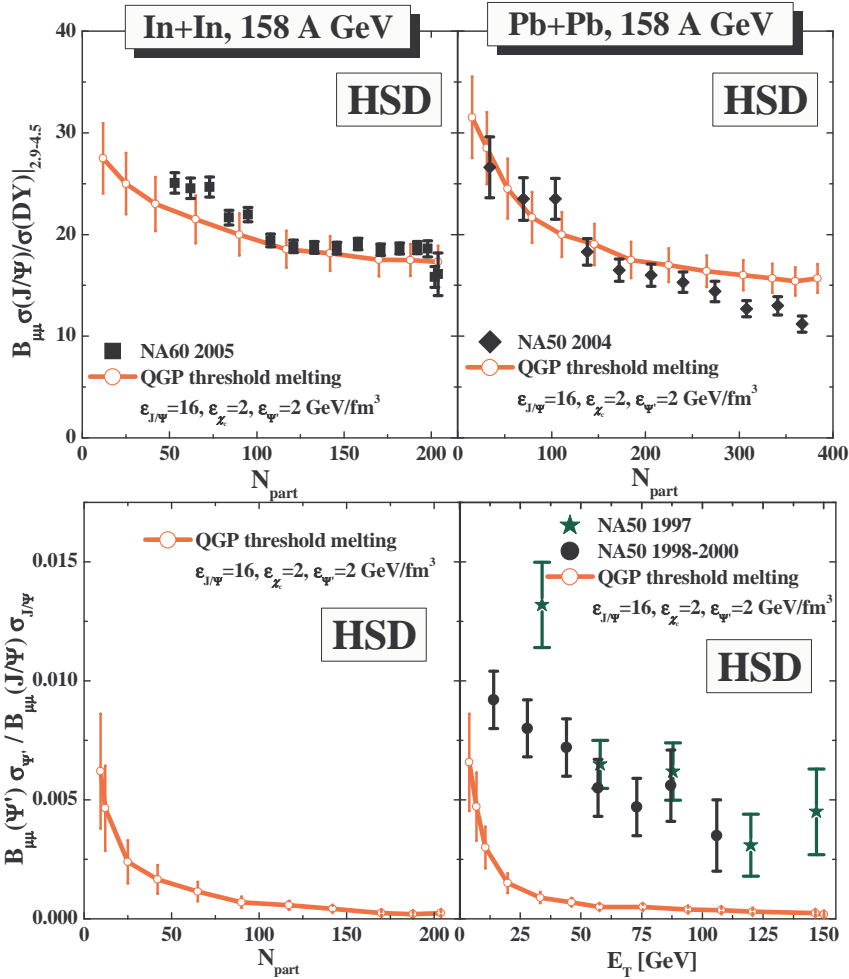,width=0.9\textwidth}} \caption{Same as Fig.~\protect\ref{figure1}
but for the `QGP threshold scenario' with $\varepsilon_{J/\Psi} = 16$ GeV/fm$^3$,
$\varepsilon_{\chi_c} = 2$ GeV/fm$^3$ = $\varepsilon_{\Psi^\prime}$ while discarding comover
absorption, {\it i.e.} for $|M_0|^2 = 0$. The figure is taken from Ref.~\protect\cite{Olena.SPS}. }
\label{figure3}
\end{figure}

The results for the `threshold scenario' for $J/\Psi$ as well as
$\Psi'$ are displayed in Fig.~\ref{figure3} in comparison to the
same data for the thresholds $\varepsilon_{J/\Psi} = 16$
GeV/fm$^3$, $\varepsilon_{\chi_c}=2$ GeV/fm$^3$ =
$\varepsilon_{\Psi^\prime}$ while discarding any dissociation with
comovers, i.e. $|M_0|^2$ =0. In this scenario, the $J/\Psi$
suppression is well described for In+In, but the suppression is
slightly too weak for very central Pb+Pb reactions. This result
emerges, since practically all $\chi_c$ and $\Psi^\prime$ dissolve
for $N_{part} >$ 100 in both systems whereas the $J/\Psi$ itself
survives at the energy densities reached in the collision. Since
the nucleon dissociation is a flat function of $N_{part} $ for
central reactions, the total absorption strength is flat, too. The
deviations seen in Fig.~\ref{figure3} might indicate a partial
melting of the $J/\Psi$ for $N_{part} > $ 250, which is not in
line with current lattice QCD calculations claiming at least
$\varepsilon_{J/\Psi} >$ 5 GeV/fm$^3$. In fact, a lower threshold
of 5 GeV/fm$^3$ (instead of 16 GeV/fm$^3$) for the $J/\Psi$ has
practically no effect on the results shown in Fig.~\ref{figure3}.
Furthermore, a threshold energy density of 2 GeV/fm$^3$ for the
$\Psi^\prime$ leads to a dramatic reduction of the $\Psi^\prime$
to $J/\Psi$ ratio, which is in severe conflict with the data
(lower part of Fig.~\ref{figure3}). Also note that due to energy
density fluctuations in reactions with fixed $N_{part} $ (or
$E_T$) there is no step in the suppression of $J/\Psi$ versus
centrality as pointed out before by Gorenstein et al. in
Ref.~\cite{Goren}.

\subsection{Thermal and statistical models at SPS}

It is of interest to compare the results of dynamical (transport) models with the
statistical and thermal fireball models, which assume statistical equilibrium during
the nucleus-nucleus collision.

\begin{figure}
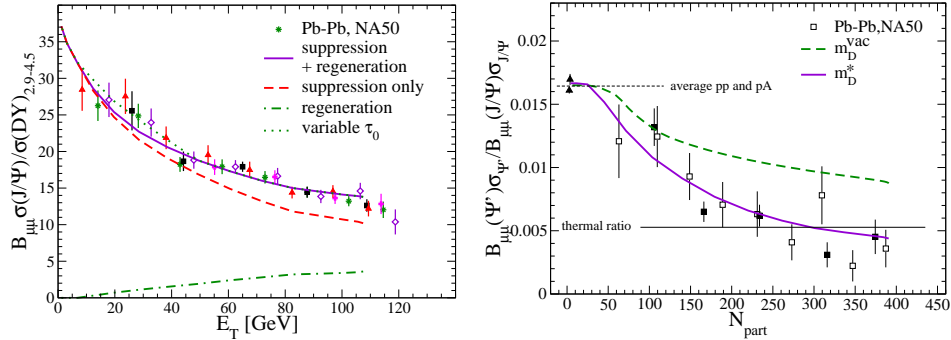

\centerline{\epsfig{file=jpsi-sps.eps,width=0.47\textwidth} \hspace{0.01\textwidth}
\epsfig{file=psip-sps.eps,width=0.49\textwidth}} \caption{NA50
data~\protect\cite{Ramello,na50-psip} on $J/\psi$ (left panel) and $\psi'$ (right panel) production
in Pb(158~AGeV)-Pb collisions at the SPS, compared to solutions of a kinetic rate equation in a
thermal fireball background~\protect\cite{Grandchamp:2003uw} starting from initial yields subject
to primordial nuclear absorption. The figure is taken from Ref.~\protect\cite{CBM.book}.}
\label{fig_na50}
\end{figure}

In Ref.~\cite{Grandchamp:2003uw} the rate equation (\ref{eq:rate})
has been solved for $Pb$(158~AGeV)-$Pb$ collisions using a
fireball evolution fitted to transport calculations. The authors
also include primordial nuclear absorption and suppression in the
hadronic phase. The resulting centrality dependence for $J/\psi$
production (including feed-down form $\chi_c$ and $\Psi'$) is
shown in the left panel of Fig.~\ref{fig_na50}. The only free
parameter in this approach is the strong coupling constant
entering into the quasi-free dissociation cross section in the QGP
which has been fixed to $\alpha_s\simeq0.25$. The NA50
data~\cite{Ramello} are fairly well reproduced, with a small
contribution from regeneration (incorporated also by detailed
balance). The main effect for the direct $J/\psi$'s is their
suppression which is largely restricted to the QGP (after nuclear
absorption as inferred from $p$-A data). On the other hand,
$\Psi'$ suppression is found to be substantially affected by the
hadronic phase, and the NA50 data~\cite{na50-psip} can only be
reproduced in this approach, if in-medium (dropping) $D$-meson
masses are implemented, which accelerate the direct $\Psi'\to
D\bar D$ decays, cf.~right panel of Fig.~\ref{fig_na50}.

\begin{figure}
\centerline{\epsfig{file=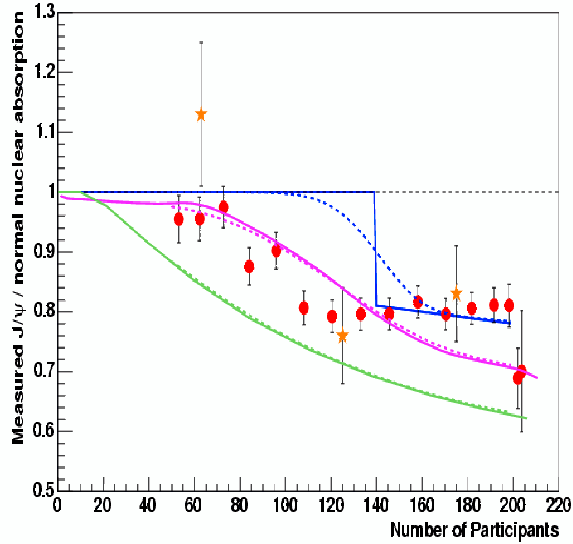,width=0.6\textwidth}}
\caption{NA60 data for $J/\psi$ production in
In(158~AGeV)-In~\protect\cite{Scomparin:2007rt} compared to
theoretical model predictions that are in approximate agreement
with the NA50 $Pb-Pb$ data: percolation model (upper (dashed) blue
line)~\protect\cite{DigalSatz}, kinetic rate equation (middle
magenta line)~\protect\cite{Grandchamp:2004tn} and hadronic
comovers (lower green line)~\protect\cite{Capella:2005cn}. Data
and theory curves are normalized to an ``expected yield" which
includes the effects of primordial nuclear absorption as extracted
from $p$-A data. The figure is taken from
Ref.~\protect\cite{Arnaldi:2006ee}.} \label{fig_na60jpsi}
\end{figure}

A widely debated issue, furthermore, is whether the NA50 data
support the notion of a more or less sharp ``onset behavior" of
$J/\psi$ suppression, possibly related to the formation of a
deconfined medium. The NA60 collaboration has scrutinized this
issue by measuring the $J/\psi$ suppression pattern in a
medium-size system, i.e., In-(158~AGeV)-In. The
data~\cite{Arnaldi:2006ee}, normalized to the $J/\psi$ yield
expected after primordial nuclear absorption, are compared to
theoretical predictions in Fig.~\ref{fig_na60jpsi}: dissociation
of $c\bar c$ states in QGP by Satz, Digal,
Fortunato~\cite{DigalSatz} (blue lines); regeneration of
charmonium in the QGP by Rapp, Grandchamp,
Brown~\cite{Grandchamp:2004tn} (pink lines); comover absorption by
mesons in a Glauber type model (green lines) by Capella and
Ferreiro~\cite{Capella:2005cn}. The latter approaches have been
adjusted to the NA50 Pb+Pb data, but it turns out that none of
them fully describes the In-In measurements. The hadronic comover
scenario~\cite{Capella:2005cn} over-predicts the suppression
throughout, the schematic percolation model~\cite{DigalSatz}
misses the onset significantly, while the kinetic rate equation
approach~\cite{Grandchamp:2004tn} somewhat over-predicts the
suppression for the most central collisions. Overall, the
predictions of the kinetic approach (in a time-dependent thermal
fireball background) do not fare too badly with the data.

\subsection{$J/\Psi$ and $\Psi'$ at RHIC}

\begin{figure}
\centerline{\psfig{file=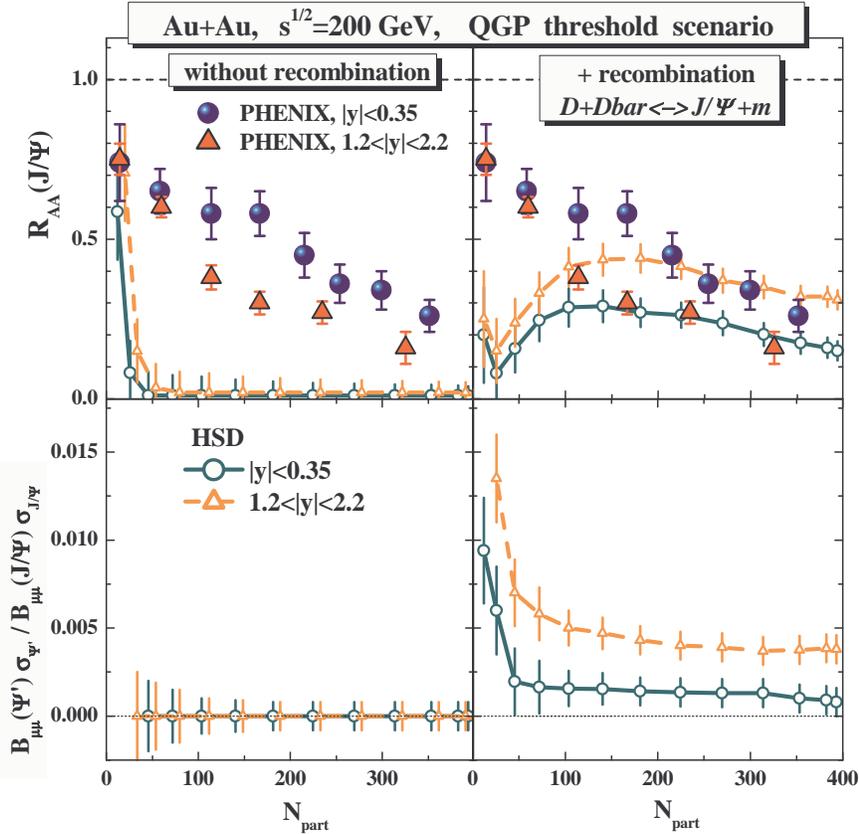,width=0.9\textwidth}}
\caption{ The $J/\Psi$ nuclear modification factor $R_{AA}$ for
Au+Au collisions at $\sqrt{s} = 200$ GeV as a function of the
number of participants $N_{part}$ in comparison to the data from
Ref. ~\protect\cite{PHENIXNov06} for midrapidity (full circles)
and forward rapidity (full triangles). HSD results for the 'QGP
threshold melting' scenarios are displayed in terms of the lower
(green solid) lines for midrapidity $J/\Psi's$ ($|y| \leq 0.35$)
and in terms of the upper (orange dashed) lines for forward
rapidity ($1.2 \leq y \leq 2.2$) without recombination (l.h.s.)
and with a recombination via $D+\bar D$. The error bars on the
theoretical results indicate the statistical uncertainty due to
the finite number of events in the HSD calculations. Predictions
for the ratio $B_{\mu \mu}(\Psi^\prime) \sigma_{\Psi^\prime} /
B_{\mu \mu}(J/\Psi) \sigma_{J/\Psi}$ as a function of the number
of participants $N_{part}$ for Au+Au at $\sqrt{s}$ = 200 GeV are
shown in the lower set of plots. The figure is taken from
Ref.~\protect\cite{Olena.RHIC}.} \label{qgp}
\end{figure}

Up to date a simultaneous description of the seemingly
energy-independent suppression of $J/\Psi$ together with its
narrow rapidity distribution and a strong elliptic flow $v_2$ of
charmed hadrons - as found at RHIC - has presented a challenge to
microscopic theories. The large discrepancies of present studies
are striking in view of the success of the hadron-string transport
theories in describing charmonium data at SPS energies. This has
lead to the conjecture that the sizeable difference between the
measured yields and transport predictions is due to a neglect of
the transition from hadronic to partonic matter, i.e. the
strongly-coupled Quark-Gluon-Plasma (sQGP).

In the RHIC experiments, one defines the nuclear modification
factor $R_{AA}$ as
\begin{equation} R_{AA} = \frac{d N^{J/\Psi}_{AA} / d y }{N_{coll} \cdot d N^{J/\Psi}_{pp} /d y}, \label{raa}
\end{equation}
where  $d N^{J/\Psi}_{AA} / d y$ denotes the final yield of $J/\Psi$ in $A A$ collisions, $d
N^{J/\Psi}_{pp} / d y$  is the yield in elementary $p p$ reactions while  $N_{coll}$ is the number
of initial binary collisions.

In the upper part of Fig.~\ref{qgp} we present a comparison of
$R_{AA} (J/\Psi)$ (from HSD) for $Au+Au$ collisions as a function
of the number of participants $N_{part}$ to the data from
Ref.~\cite{PHENIXNov06}. The results for the `threshold melting'
scenario (without the reformation channels $D+\bar{D} \rightarrow
(J/\Psi, \chi_c, \Psi^\prime)$ + meson) are displayed on the
l.h.s. of Fig. \ref{qgp} in terms of the lower (green) solid line
for mid-rapidity $J/\Psi's$ ($|y| \leq 0.35$) and in terms of the
upper (orange) dashed line at forward rapidity ($1.2 \leq  |y|
\leq 2.2$). The experimental data from PHENIX~\cite{PHENIXNov06}
are given by the full circles at mid-rapidity and by triangles at
forward rapidity. In this simple scenario practically all
charmonia are dissolved for $N_{part} > 50$ due to the high energy
densities reached in the overlap zone of the collision, which is
clearly not compatible with the PHENIX data and indicates that
charmonium reformation channels are important.

\begin{figure}
\centerline{\psfig{file=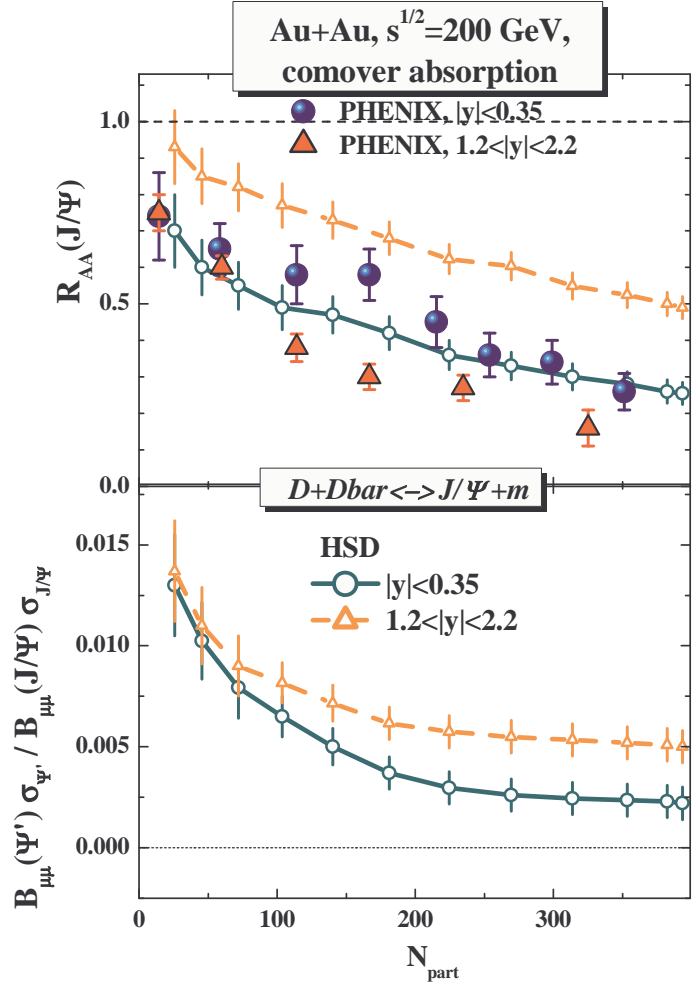,width=0.75\textwidth}}
\caption{Same as Fig.~\ref{qgp} for the `comover absorption
scenario' including the charmonium reformation channels. The
figure is taken from Ref.~\protect\cite{Olena.RHIC}.} \label{com}
\end{figure}

The reformation and dissociation channels ($D+\bar{D}
\leftrightarrow (J/\Psi, \chi_c, \Psi^\prime)$ + meson) are
switched on after a formation time.
The results for this model study are displayed in the upper right
part of Fig.~\ref{qgp} and demonstrate that for $N_{part} > 200$
an approximate equilibrium between the reformation and
dissociation channels is achieved. However, here the calculations
for forward rapidity match the data at mid-rapidity and vice versa
showing that the rapidity dependence is fully wrong. Furthermore,
the $J/\Psi$ suppression at more peripheral reactions is severely
overestimated. Thus one has to conclude that the `threshold
melting + reformation scenario' is incompatible with the PHENIX
data and has to be ruled out at top RHIC energies.

In the lower parts of Fig.~\ref{qgp} the results for the ratio of
the $\Psi^\prime$ and $J/\Psi$ dilepton yields (given by their
cross sections multiplied by the corresponding branching ratios) are shown,
which so far have not been measured. Here the two
recombination models give finite ratios as a function of
centrality but predict a larger $\Psi^\prime$ to $J/\Psi$ ratio at
forward rapidity than at mid-rapidity which is a consequence of
the higher comover density at mid-rapidity. Experimental data on
this ratio should provide further independent information.

The suppression of charmonia by the `comover' dissociation
channels within the model described in Section~\ref{comover} is
presented in Fig.~\ref{com}, where the charmonium reformation
channels by $D+\bar{D}$ annihilation have been incorporated. The
HSD results for $R_{AA}$ in the purely hadronic `comover' scenario
are displayed in the upper part of Fig.~\ref{com} in comparison to
the data from Ref.~\cite{PHENIXNov06} using the same assignment of
the lines as in Fig.~\ref{qgp}. This scenario gives a continuous
decrease of $R_{AA}(J/\Psi)$ with centrality. However, the
rapidity dependence of the comover result is opposite to the one
experimentally observed, as dictated by the higher comover density
at mid-rapidity.

The $\Psi^\prime$ to $J/\Psi$ ratio is displayed in the lower
parts of Fig.~\ref{com} and shows a decreasing ratio with
centrality similar to the results at SPS energies ({\it cf.}
Fig.~\ref{figure1}). As pointed in Ref.~\cite{Olena.SPS}, an
independent measurement of $\Psi^\prime$ will provide further
information on the charm reaction dynamics and final charmonium
formation. For instance, a leveling off of the $\Psi '$ to
$J/\Psi$ ratio with increasing centrality would be a signal for
charm chemical equilibration in the medium \cite{BMS,BMS2,PBM07}.
Additionally, it provides a very clear distinction between the
`threshold melting' scenario and the `comover' approach.

\begin{figure}
\centerline{\psfig{file=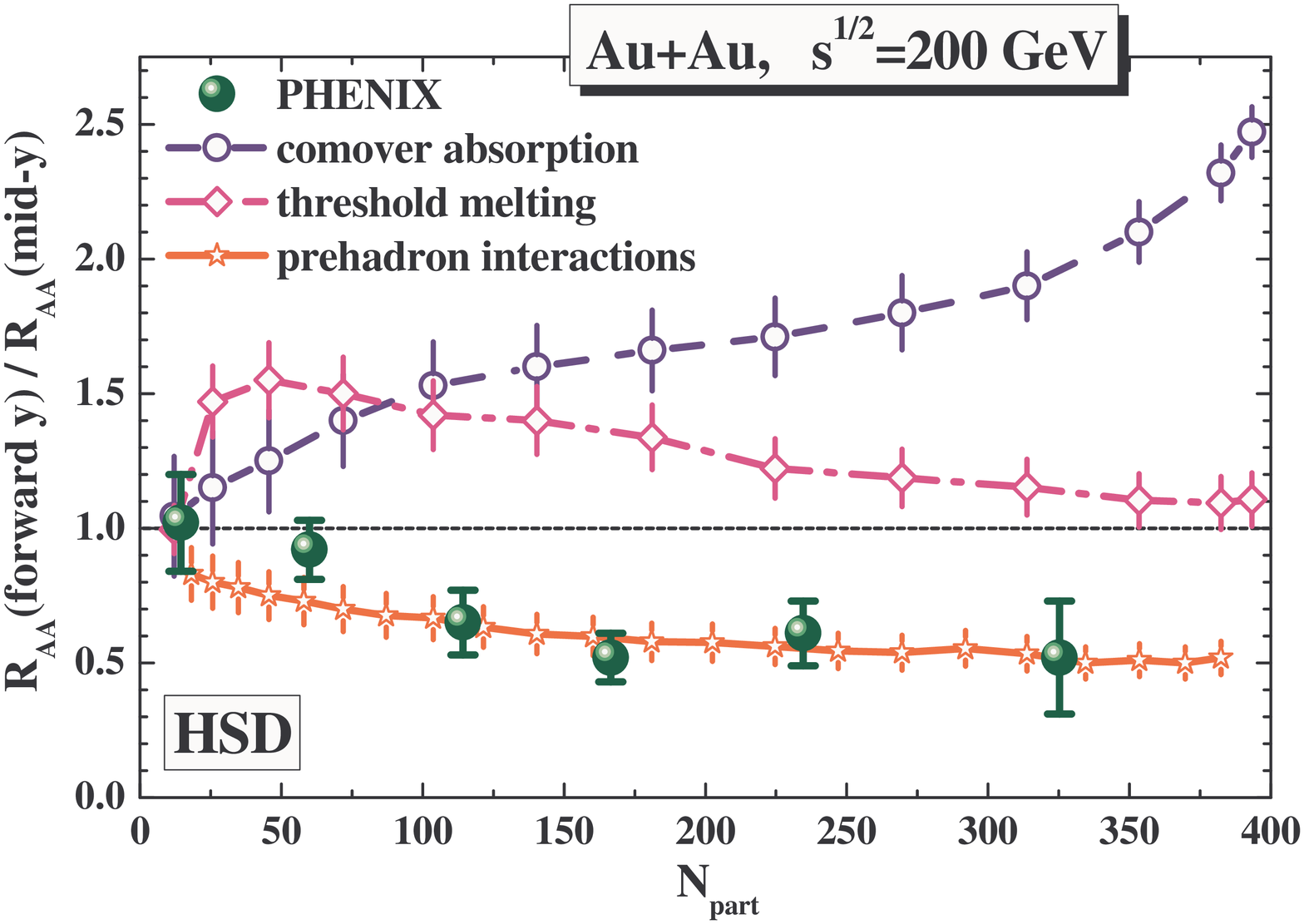,width=0.75\textwidth}} \caption{The ratio of the nuclear
modification factors $R_{AA}$ at mid-rapidity ($|y|<0.35$) and at forward rapidity ($1.2<|y|<2.2$)
{\it vs} centrality in $Au+Au$ collisions at $\sqrt{s}=200$~GeV. The HSD results in the purely
hadronic scenario (`comover absorption') are displayed in terms of the blue dashed line (with open
circles) and in case of the `threshold melting' scenario in terms of the violet dot-dashed  line
(with open squares).  The error bars on the theoretical results
 indicate the statistical uncertainty due to the finite number of
Monte-Carlo events in the calculations. The lower full green dots represent the data of the PHENIX
Collaboration~\protect\cite{PHENIXNov06}. Note that the data have an additional systematic
uncertainty of $\pm 14 \%$. The lower solid (red) line with stars gives the result for the `comover
absorption' scenario when including additional pre-hadronic interactions with charm (see text). The
figure is taken from Ref.~\protect\cite{Olena.RHIC.2}.} \label{Ratio}
\end{figure}

\begin{figure}
\centerline{
\psfig{file=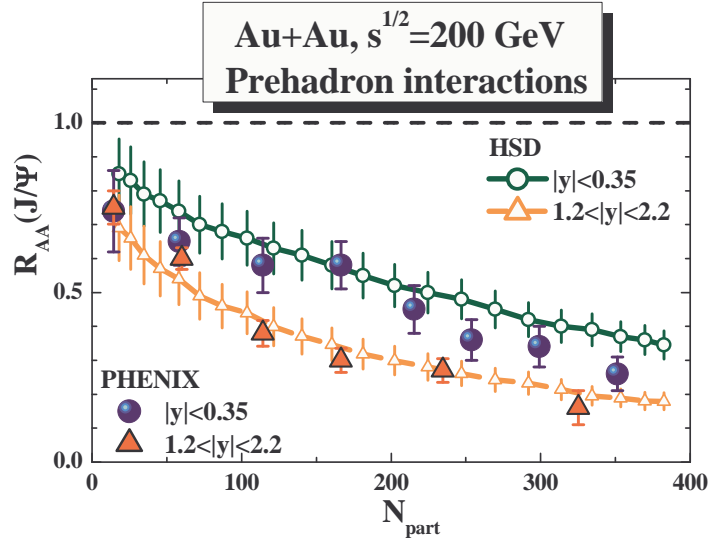,width=0.75\textwidth}}
\caption{The $J/\Psi$ nuclear modification factor $R_{AA}$
(\ref{raa}) for Au+Au collisions at $\sqrt{s} = 200$~AGeV as a
function of the number of participants $N_{part}$ in comparison to
the data from~\protect\cite{PHENIXNov06} for mid-rapidity (full
circles) and forward rapidity (full triangles).
The HSD results for the hadronic `comover'
scenario including additionally pre-hadronic interactions of charm
according to (\ref{sss1}) - (\ref{sss4}) are displayed in terms of
the upper (green solid) line with open circles for mid-rapidity
$J/\Psi's$ ($|y| \leq 0.35$) and in terms of the lower (orange
dashed) line with open triangles for forward rapidity ($1.2 \leq
|y| \leq 2.2$). The figure is taken from
Ref.~\protect\cite{Olena.RHIC.2}. } \label{xs5}
\end{figure}

The non-applicability of the traditional `comover absorption'
model and 'threshold melting' scenario at the top RHIC energy is
most clearly seen in the centrality dependence of the ratio of the
nuclear modification factors $R_{AA}$ at forward rapidity
($1.2<|y|<2.2$) and at mid-rapidity ($|y|<0.35$) as shown in
Fig.~\ref{Ratio}. The HSD results in the purely hadronic scenario
(`comover absorption') are displayed in terms of the blue dashed
line (with open circles) and in case of the `threshold melting'
scenario in terms of the dot-dashed magenta line (with open
squares).  The error bars on the theoretical results indicate the
statistical uncertainty due to the finite number of Monte-Carlo
events in the calculations. The lower full green dots in
Fig.~\ref{Ratio} represent the corresponding data of the PHENIX
Collaboration~\cite{PHENIXNov06} which show a fully different
pattern as a function of centrality (here given in terms of the
number of participants $N_{part}$). The failure of these
`standard' suppression models at RHIC has lead to the conclusion
in Ref.~\cite{Olena.RHIC} that the hadronic `comover absorption
and recombination' model is falsified by the PHENIX data and that
strong interactions in the pre-hadronic (or partonic) phase is
necessary in order to explain the large suppression at forward
rapidities.

\begin{figure}
\centerline{\psfig{file=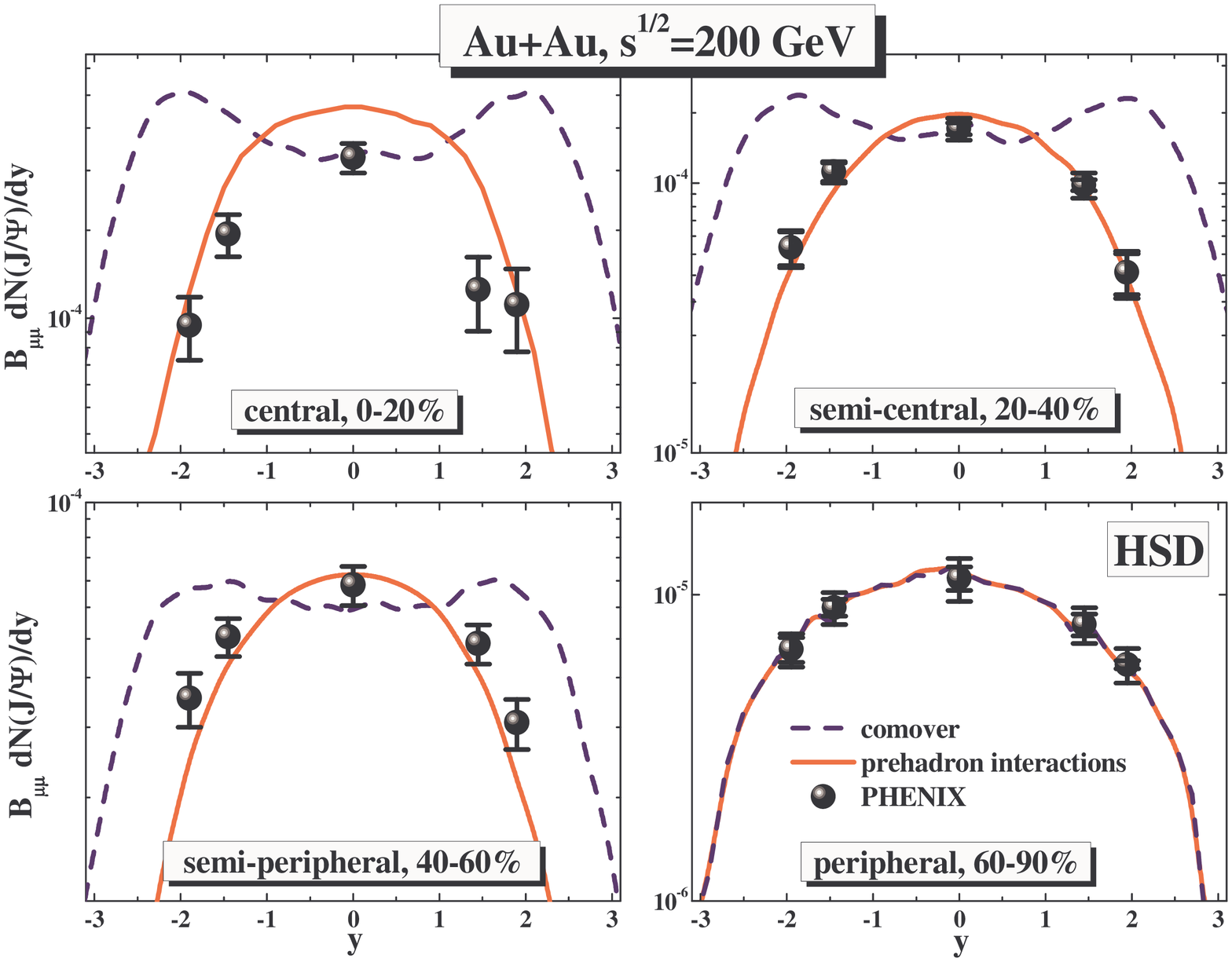,width=\textwidth}} \caption{The rapidity distribution
$dN_{J/\Psi}/dy$ for different centralities  from the standard `comover' model (dashed blue lines)
and the 'comover' model with additional pre-hadronic interactions of charm  according to
(\ref{sss1}) - (\ref{sss4}) (solid red lines). The full dots show the respective data from the
PHENIX Collaboration~\protect\cite{PHENIXNov06}. The calculated lines have been smoothed by a
spline algorithm.  The reactions are Au+Au at $\sqrt{s}=200$~GeV. The figure is taken from
Ref.~\protect\cite{Olena.RHIC.2}.} \label{y1}
\end{figure}

Consequently, in Ref.~\cite{Olena.RHIC.2} additional interactions
of charm with pre-hadrons (as described in Section 6.3) have been
incorporated in the `comover scenario' to have a first glance at
the dominant effects. The $J/\Psi$ suppression pattern in this
case is shown in Fig.~\ref{xs5} in comparison to the same data as
in Fig.~\ref{com} for the hadronic comover model. When including
the pre-hadronic interactions, the suppression pattern for central
and forward rapidities becomes rather similar to the data within
the statistical accuracy of the calculations. Indeed, the ratio of
$R_{AA}$ at forward rapidity to mid-rapidity now follows closely
the experimental trend as seen in Fig.~\ref{Ratio} by the lower
red solid line.

Some further information may be gained from the $J/\Psi$ rapidity
distributions in  Au+Au collisions at RHIC. The latter
distribution is shown in Fig.~\ref{y1} in comparison to the PHENIX
data for central collisions (upper l.h.s.), semi-central (upper
r.h.s.), semi-peripheral (lower l.h.s.)  and peripheral reactions
(lower r.h.s.) for the standard `comover' scenario (dashed blue
lines) and the `comover' model including additionally pre-hadronic
interactions of charm  according to (\ref{sss1}) - (\ref{sss4})
(solid red lines).  Whereas for peripheral reactions these
additional early interactions practically play no role, the additional pre-hadron elastic scattering
lead to a narrowing of the $J/\Psi$ rapidity distribution with the
centrality of the collision (roughly in line with the data). In
the standard `comover' model an opposite trend is seen: here the
interactions of charmonia with formed hadrons produce a dip in the
rapidity distribution at $y\approx0$ which increases with
centrality since the density of formed hadrons increases
accordingly around mid-rapidity. Since the total number of
produced $c{\bar c}$ pairs is the same (for the respective
centrality class) and detailed balance is incorporated in the
reaction rates, we find a surplus of $J/\Psi$ at more forward
rapidities. The net result is a broadening of the $J/\Psi$
rapidity distribution with centrality in the purely hadronic scenario opposite to the trend
observed in experiment.

In summarizing the results at RHIC energies, the hadronic
`comover' dynamics for charmonium dissociation and recreation as
well as the charmonium `melting' scenario do not match the general
dependence of the $J/\Psi$ on rapidity and centrality as seen by
the PHENIX Collaboration.  In fact, a narrowing of the $J/\Psi$
rapidity distribution cannot be achieved by comover interactions
with formed hadrons, since the latter appear too late in the
collision dynamics. Only when including early pre-hadronic
interactions with charm, a dynamical narrowing of the charmonium
rapidity distribution with centrality can be achieved, as
demonstrated  within the pre-hadronic interaction model.
Consequently, the PHENIX data on $J/\Psi$ suppression indicate the
presence and important impact of pre-hadronic or partonic
interactions in the early charm dynamics. This finding is in line
with earlier studies in Refs. \cite{brat03,HPT1,HPT2}
demonstrating the necessity of non-hadronic degrees of freedom in
the early reaction phase of nucleus-nucleus collisions at RHIC
energies for the description of the elliptic flow $v_2$, the
suppression of hadrons at high transverse momentum $p_T$ and
far-side jet suppression.



\subsection{Charmonia excitation functions}

\begin{figure}
\centerline{\psfig{figure=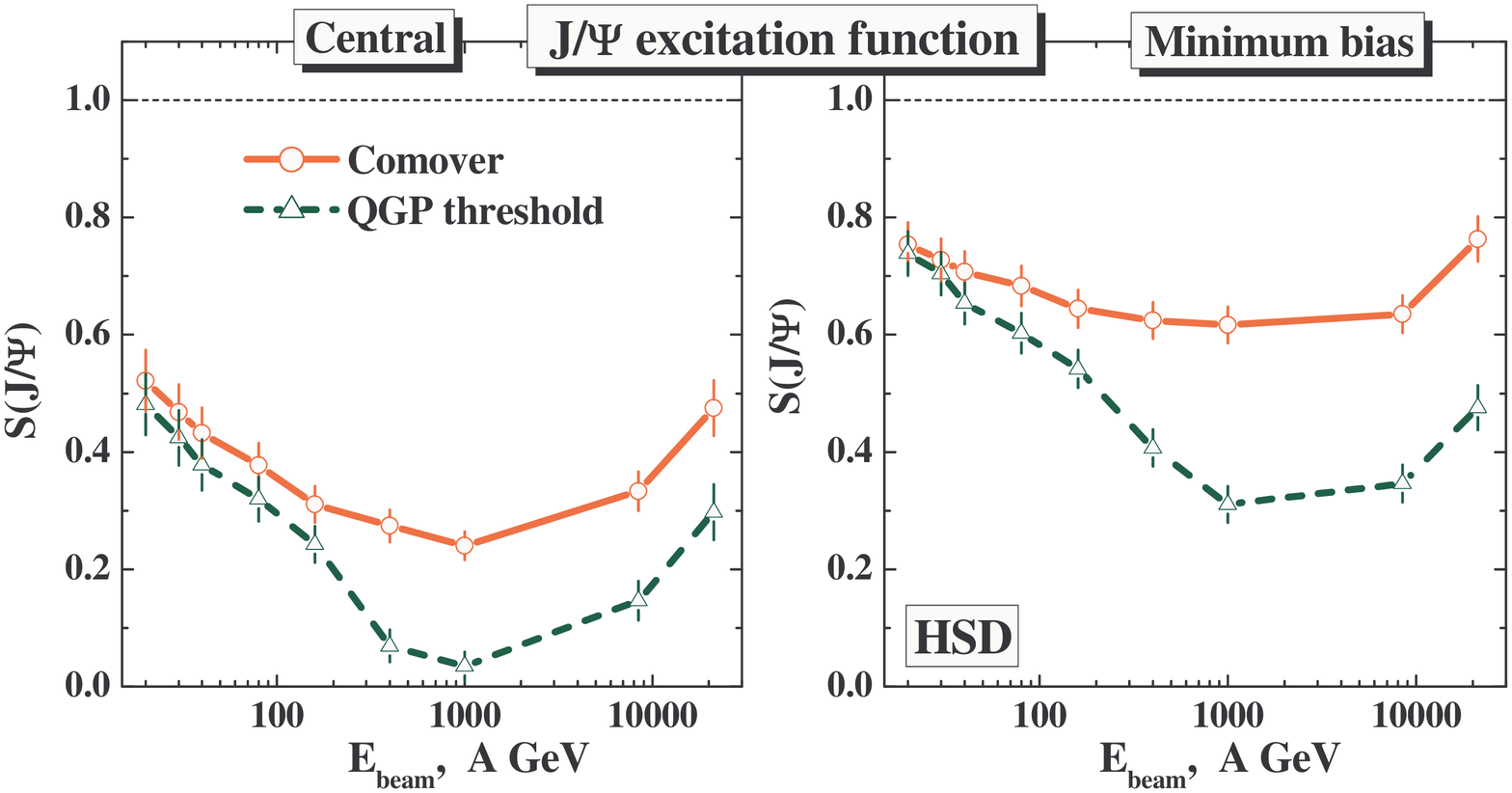,width=\textwidth}}
\centerline{\psfig{figure=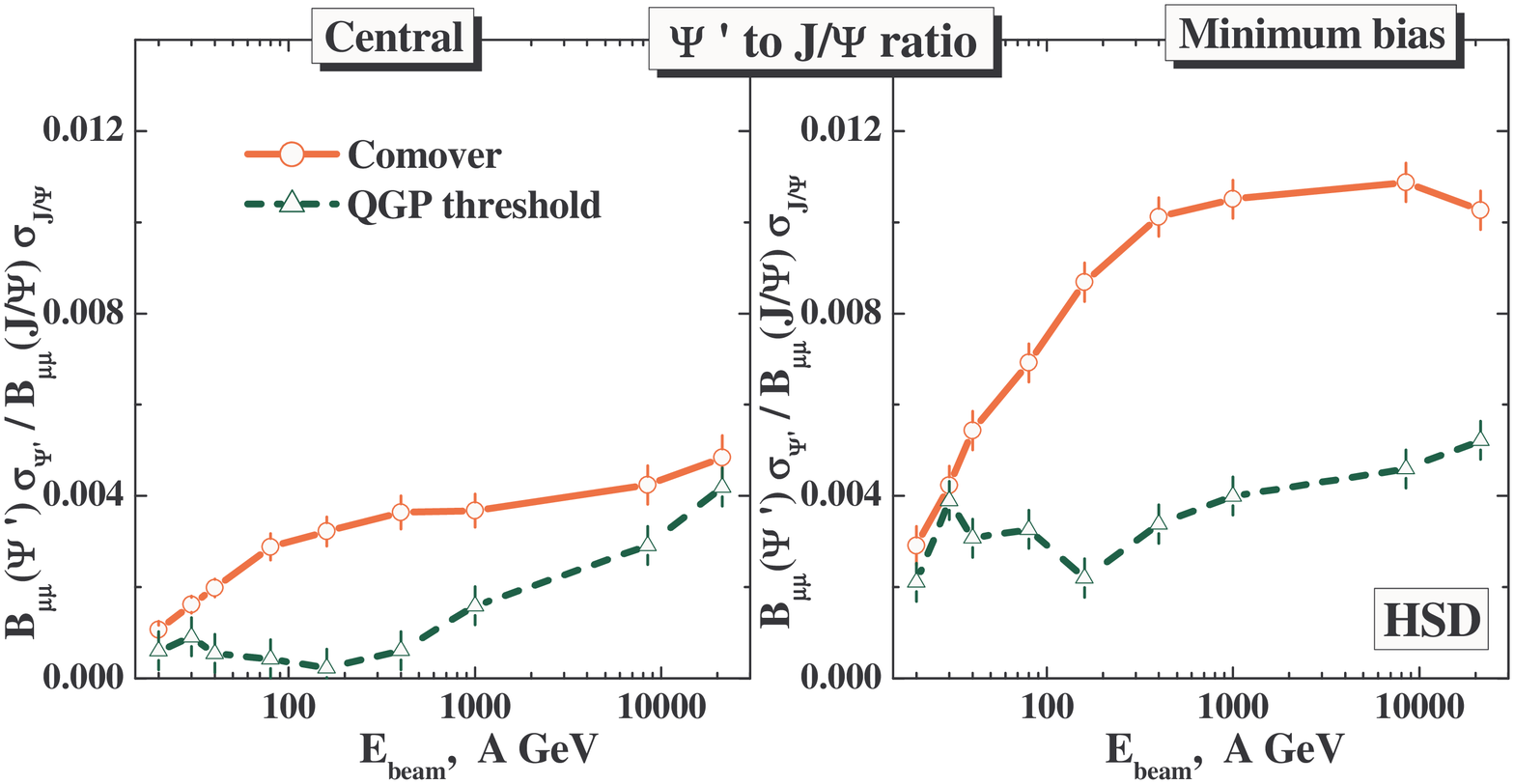,width=\textwidth}} \caption{upper part: The excitation
function for the $J/\Psi$ survival probability in the `QGP threshold melting + hadronic
recombination' scenario (dashed green lines with triangles) and the `comover absorption +
recombination' model (solid red lines with circles) for central (l.h.s.) and minimum bias Au+Au
reactions (r.h.s.) as a function of the beam energy. Lower part: The $\Psi^\prime$ to $J/\Psi$
ratio for the same reactions as in the upper part of the figure in the `QGP threshold melting +
hadronic recombination' scenario (dashed green lines with triangles) and the `comover absorption +
recombination' model (upper solid red lines with circles ). The figure is taken
from Ref.~\protect\cite{Olena.RHIC.2}.} \label{JPsiEx}
\end{figure}

In this Section we present the excitation functions for the
$J/\Psi$ survival probability in Au + Au collisions from FAIR to
top RHIC energies in the different scenarios in order to allow for
a further distinction between the different concepts. The results
of  HSD calculations are presented in the upper part of
Fig.~\ref{JPsiEx} for the `QGP threshold melting + hadronic
recombination' scenario (dashed green lines with open triangles)
and the `comover absorption + recombination' model (solid red
lines with open circles) for central (l.h.s.) and minimum bias
(r.h.s.) Au+Au reactions  as a function of the beam energy.
The $J/\Psi$
survival probability $S_{J/\Psi}$ is defined as
\begin{equation} \label{supp} S_{J/\Psi} = \frac{N^{J/\Psi}_{fin}}{N^{J/\Psi}_{BB}},
\end{equation}
where $ N^{J/\Psi}_{fin}$ and $N^{J/\Psi}_{BB}$ denote the final
number of $J/\Psi$ mesons and the number of $J/\Psi$'s produced
initially by $BB$ reactions, respectively. We find that from FAIR
energies of 20 - 40 A$\cdot$ GeV up to top SPS energies of 158
A$\cdot$GeV there is no significant difference between the two
models for the $J/\Psi$ survival probability in case of central
collisions. The differences here show up mainly in the full RHIC
energy range where the `QGP threshold melting + hadronic
recombination' scenario leads to substantially lower $J/\Psi$
survival probabilities. In case of minimum bias collisions the
`comover absorption + recombination' model (solid lines) gives a
roughly energy independent $J/\Psi$ survival probability, whereas
the `QGP threshold melting + hadronic recombination' scenario
shows lower $J/\Psi$ survival probabilities (lower dashed green
lines) for laboratory energies above $\sim$ 100 A$\cdot$GeV due to
a larger initial melting of $J/\Psi$-mesons at high energy
density.

A clearer distinction between the different concepts is offered by the excitation functions for the
$\Psi^\prime$ to $J/\Psi$ ratio in Au + Au collisions. The calculated results are shown in the
lower part of Fig.~\ref{JPsiEx} for the `QGP threshold melting + hadronic recombination' scenario
(dashed green lines with open triangles) and the `comover absorption + recombination' model (solid
red lines) for central (l.h.s.) and minimum bias reactions (r.h.s.). Here the $\Psi^\prime$ is
already melting away in central Au+Au reactions in the `QGP threshold melting' scenario at
bombarding energies above 40 A$\cdot$GeV, whereas a substantial amount of $\Psi^\prime$ survives in
the `comover absorption + recombination' model. Thus measurements of $\Psi^\prime$ suppression at
the lower SPS or top FAIR energies will clearly distinguish between the different model concepts.

\subsection{FAIR energies}
\label{Fair.S}

\begin{figure}
\centerline{\psfig{file=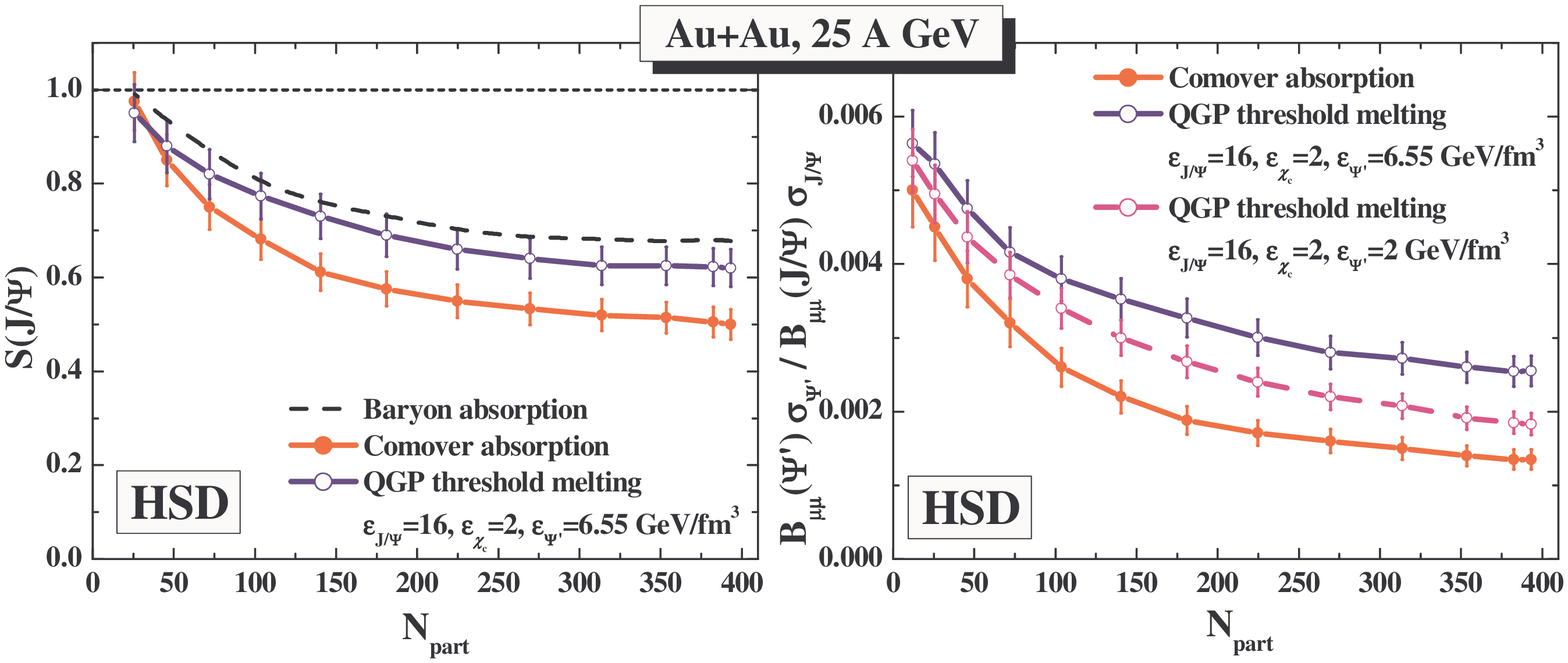,width=0.98\textwidth}} \caption{The survival probability
$S_{J/\Psi}$ (left plot) and ratio $\Psi^\prime$ to $J/\Psi$  (right plot) as a function of the
number of participants $N_{part}$ in Au+Au reactions at 25 A$\cdot$GeV.  The blue lines (with open
dots) reflect the `threshold scenario' for $\varepsilon_{J/\Psi} = 16$ GeV/fm$^3$,
$\varepsilon_{\chi_c} = 2$ GeV/fm$^3$, $\varepsilon_{\Psi^\prime}$ = 6.55 GeV/fm$^3$ while the
violet line (the lower line with open dots on the r.h.s.) stands for the 'threshold scenario' for
$\varepsilon_{J/\Psi} = 16$ GeV/fm$^3$, $\varepsilon_{\chi_c} = 2$ GeV/fm$^3$,
$\varepsilon_{\Psi^\prime}$ = 2 GeV/fm$^3$. The solid red lines (full dots) denote the results for
the comover absorption model with the standard matrix element squared $|M_0|^2$ = 0.18
fm$^2$/GeV$^2$. The dashed line (l.h.s.) represents the HSD calculations including only
dissociation channels with nucleons. The figure is taken from
Ref.~\protect\cite{Olena.SPS}.} \label{figure8}
\end{figure}

The CBM Collaboration at GSI is aiming at charmonium measurements in heavy-ion collisions at the future FAIR
facility~\cite{CBMprop}. This opens up the possibility to explore the charmonium suppression
mechanism at lower bombarding energies of about 25 A$\cdot$GeV in Au+Au collisions.
The corresponding HSD predictions are displayed in
Fig.~\ref{figure8} for the survival probability $S_{J/\Psi}$ (l.h.s.) and the ratio $\Psi^\prime$ to
$J/\Psi$  (r.h.s.) as a function of the number of participants $N_{part}$.
The violet line in Fig.~\ref{figure8} stands for the `threshold scenario' with
$\varepsilon_{\Psi^\prime}$, i.e. $\varepsilon_{J/\Psi} = 16$ GeV/fm$^3$, $\varepsilon_{\chi_c} =
2$ GeV/fm$^3$, $\varepsilon_{\Psi^\prime}$ = 2 GeV/fm$^3$. The solid red lines denote the results
for the comover absorption model with the standard matrix element squared $|M_0|^2$ = 0.18
fm$^2$/GeV$^2$.

We note that in Au+Au reactions at 25 A$\cdot$GeV in the
`threshold scenario' (solid line with open dots in the left plot)
only a very low amount of $\chi_c$ and no $J/\Psi$ are melted at
the energy densities reached in these reactions. On the other hand
the comover density decreases only moderately when stepping down
in energy from 158 A$\cdot$GeV to 25 A$\cdot$GeV such that the
$J/\Psi$ survival probability in the comover absorption model
(lower solid line in the left part) is lower. This is even more pronounced for the $\Psi^\prime$ to $J/\Psi$ ratio versus centrality, which in the `threshold melting scenario' (middle line
in the right part) is clearly above the result achieved in the
comover absorption model (lower line in the right part).
Consequently, the different dissociation scenarios may well be
distinguished in future charmonium measurements at FAIR.

\section{Testing the assumption of statistical hadronization}
\label{chemi}

The assumption of statistical hadronization -- {\it i.e.}  of
$J/\Psi$'s being dominantly produced at hadronization in a purely
statistically fashion according to available phase space and the
number of available $c$ and ${\bar c}$ quarks -- leads to a
scaling of the $\langle J/\Psi\rangle/\langle h\rangle$ ratio with
the system size~\cite{MG**2}, where $\langle h \rangle$ is the
average hadron multiplicity. Since $\langle h \rangle \sim
\langle\pi\rangle$, the ratio $\langle
J/\Psi\rangle/\langle\pi\rangle$ has been calculated in HSD in the
different scenarios for charmonium suppression:
\begin{itemize}
\item `threshold melting' + recombination via $D\bar D\to c\bar c +m$
including the backward reactions $c \bar c + m \to D \bar D $, %
\item
hadronic (`comover') absorption: $D \bar D \to c \bar c  + m$ and the backward reactions $c \bar c
+ m \to D \bar D $;
\item
`prehadron interactions': $D \bar D \to c \bar c  + m$ and the backward reactions $c \bar c + m \to
D \bar D $ as well as early pre-hadronic charm interactions as described in Section 6.3.
\end{itemize}

\begin{figure}
\begin{minipage}[l]{0.50\textwidth}
\psfig{figure=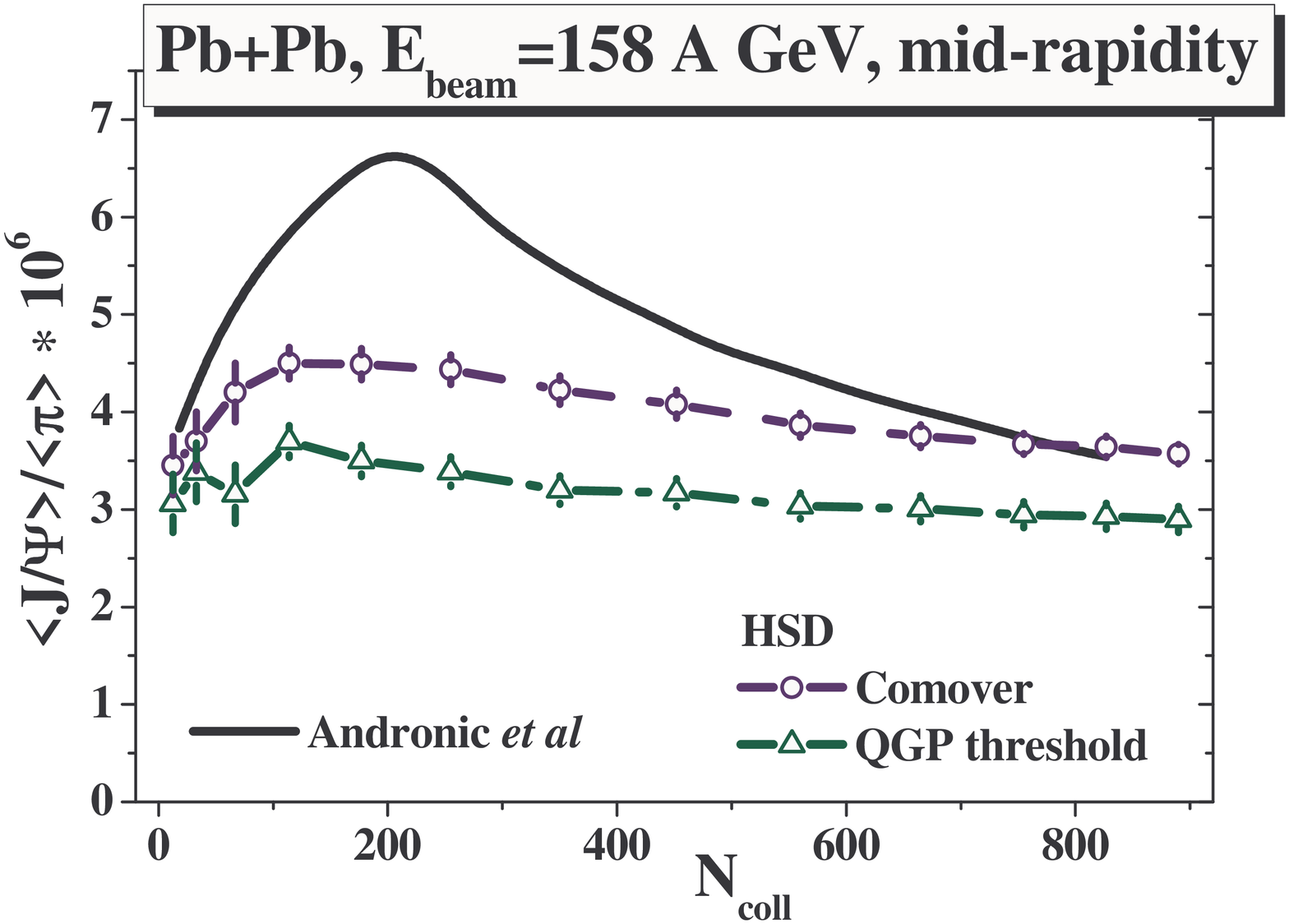,width=\textwidth}
\end{minipage}
\begin{minipage}[l]{0.49\textwidth}
\psfig{figure=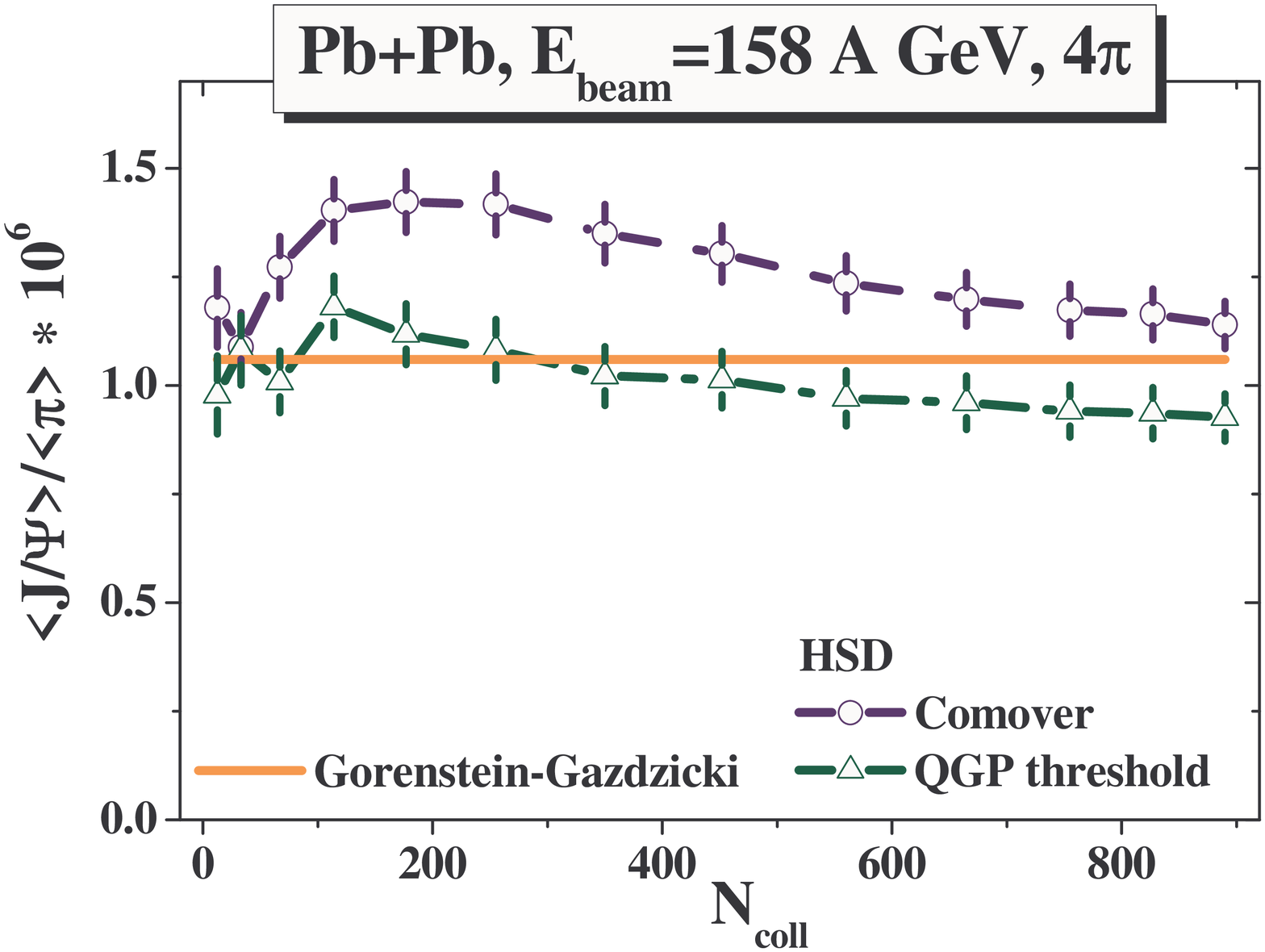,width=\textwidth}
\end{minipage}
\caption{Ratio of the averaged $J/\Psi$ to $\pi$ multiplicity  for $Pb+Pb$ at the SPS beam energy
of 158 A$\cdot$GeV at mid-rapidity (l.h.s.) and in  full $4\pi$ acceptance (r.h.s.) as a function
of the number of binary collisions $N_{coll}$ for the different suppression scenarios implemented
in HSD - the `comover' model (dashed blue line with open circles) and  the `threshold melting'
scenario (green dot-dashed line with open triangles) - in comparison to the statistical model by
Gorenstein and Gazdzicki~\protect\cite{MG**2} (r.h.s.; straight orange line) and the statistical
hadronization model by Andronic~{\it et al.}~\protect\cite{PBM07} (l.h.s.; solid black line). The
figure is taken from Ref.~\protect\cite{Olena.RHIC.2}. } \label{PsiPiSPS}
\end{figure}

\begin{figure}
\begin{minipage}[l]{0.495\textwidth}
\psfig{file=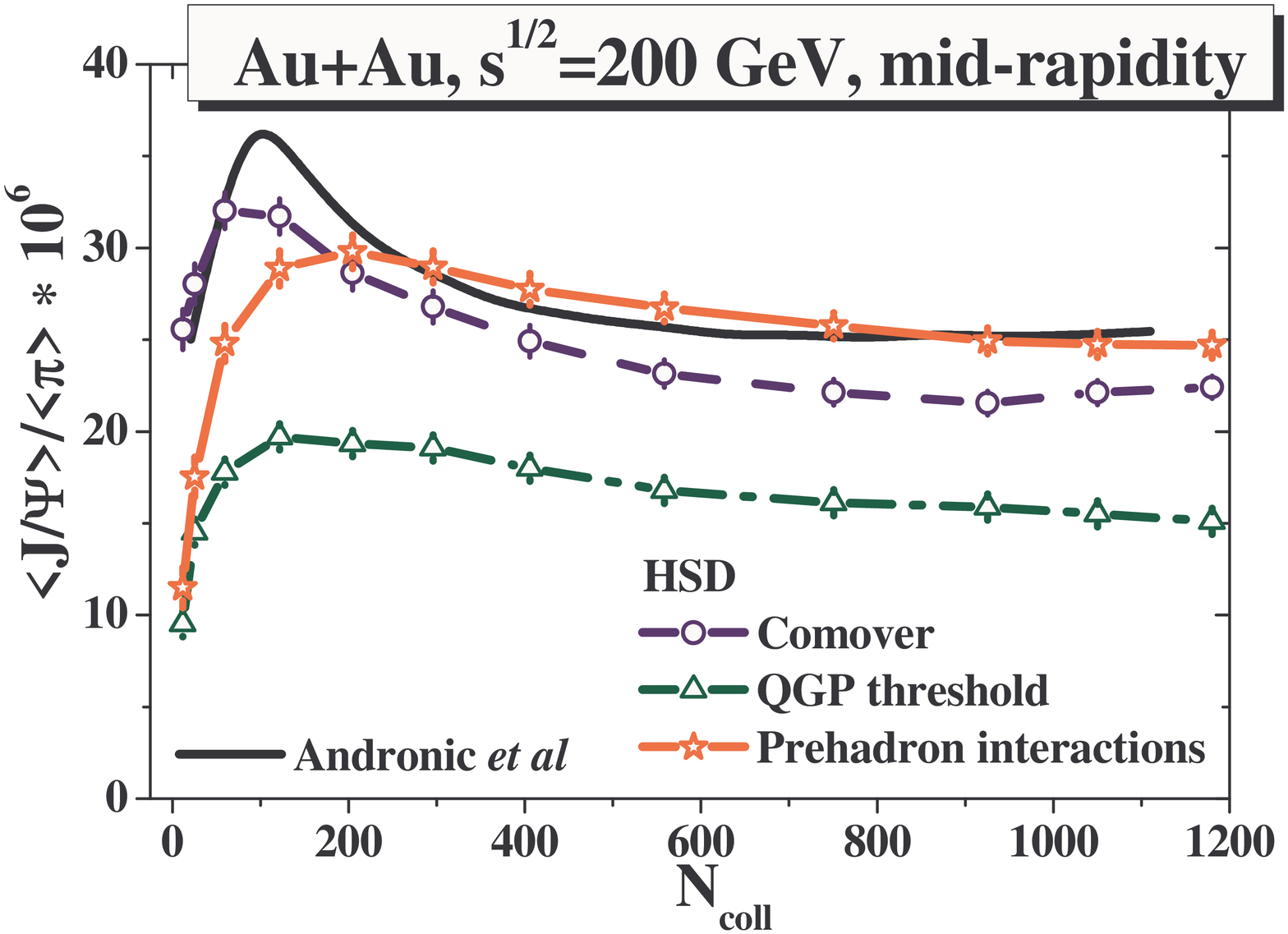,width=\textwidth}
\end{minipage}
\begin{minipage}[l]{0.495\textwidth}
\psfig{file=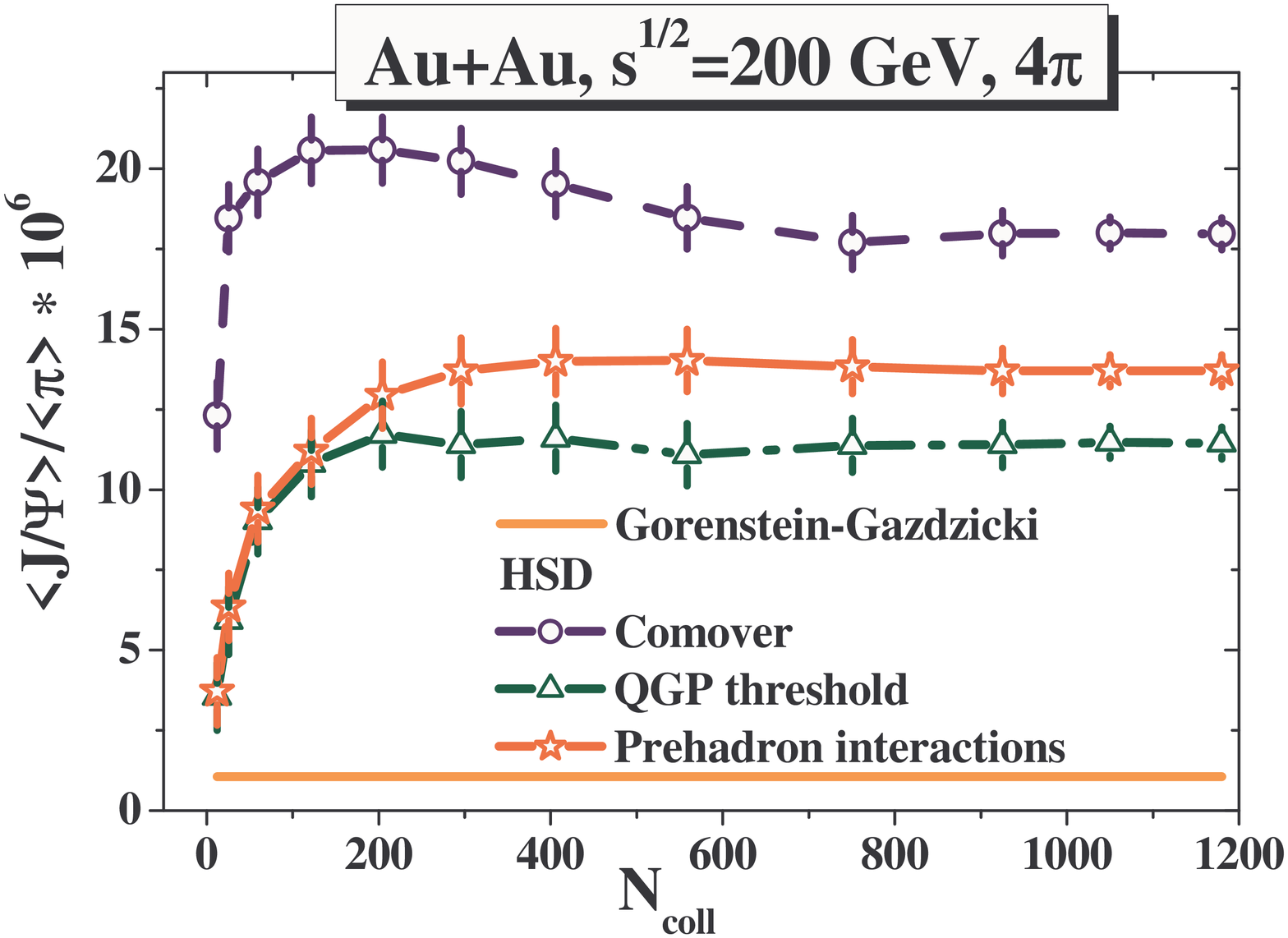,width=\textwidth}
\end{minipage}
\caption{Same as Fig.~\ref{PsiPiSPS} but for $Au+Au$ at the top RHIC energy of $\sqrt{s}$ = 200
GeV. The red solid line shows additionally the result of the `comover' model including the
pre-hadronic charm interactions (see text). The figure is taken from Ref.
~\protect\cite{Olena.RHIC.2}.} \label{PsiPiRHI}
\end{figure}

The results of the calculations are shown in Fig.~\ref{PsiPiSPS}
together with the prediction of the statistical model of
Gorenstein and Gazdzicki~\cite{MG**2} for the full phase space
(straight orange  line; r.h.s.) and the statistical hadronization
model by Andronic~{\it et al.}~\cite{PBM07,AndronicPrivat} for
mid-rapidity (solid black line; l.h.s.) for Pb+Pb at 158
A$\cdot$GeV. The centrality dependence here is given by the number
of initial binary collisions $N_{coll}$. The actual comparison in
Fig.~\ref{PsiPiSPS} indicates that the statistical model by
Andronic {\it et al.}~\cite{PBM07} predicts a sizeably larger
$J/\Psi$ to $\pi$ ratio at midrapidity for peripheral and
semi-peripheral reactions than the microscopic HSD results for the
different scenarios. For central reactions - where an approximate
equilibrium is achieved - all scenarios give roughly the same
ratio. In full $4 \pi$ phase space the HSD results indicate also a
slightly higher $J/\Psi$ to $\pi$ ratio in the `comover' model
relative to the `melting' scenario but both ratios only weakly
depend on centrality - roughly in line with the statistical model
of Gorenstein and Gazdzicki~\cite{MG**2} (orange straight line).
Consequently, only peripheral reactions of heavy nuclei might be
used to disentangle the different scenarios at top SPS energies at
midrapidity (or in full phase space).

The situation is different for Au+Au collisions at the top RHIC
energy as may be extracted from Fig.~\ref{PsiPiRHI} where the
$J/\Psi$ to pion ratio (l.h.s.: at mid-rapidity; r.h.s.: for
$4\pi$ acceptance) is shown as a function of $N_{coll}$. The
standard `comover' model (dashed blue lines) is only presented for
reference but is unrealistic according to the analysis in Section
7. We find that the `comover' model with early pre-hadronic charm
interactions (solid red line with stars, l.h.s.) is very close to
the statistical hadronization model~\cite{PBM07} (solid black
line) at mid-rapidity except for very peripheral collisions. The
`threshold melting' scenario follows the trend in centrality but
is down by about 30\%. Thus at mid-rapidity there is no essential
extra potential in differentiating the scenarios. Considering the
full $4 \pi$ acceptance (r.h.s.) one finds a practically constant
$J/\Psi$ to pion ratio for $N_{coll} > $ 200 from the HSD
calculations as expected from the statistical model, however, the
early model of Gorenstein and Gazdzicki~\cite{MG**2} is down by
about a factor of $\sim$10 (and may be ruled out by present
data). The latter conclusion is in agreement with the independent
analysis in Ref.~\cite{BMS}.

\section{Transverse mass spectra}
\label{mT}

Apart from rapidity dependent particle abundances, also the
dynamics in the transverse direction (to the beam) provides
relevant information.

\subsection{Elementary collisions}

\begin{figure}
 \centerline{\psfig{file=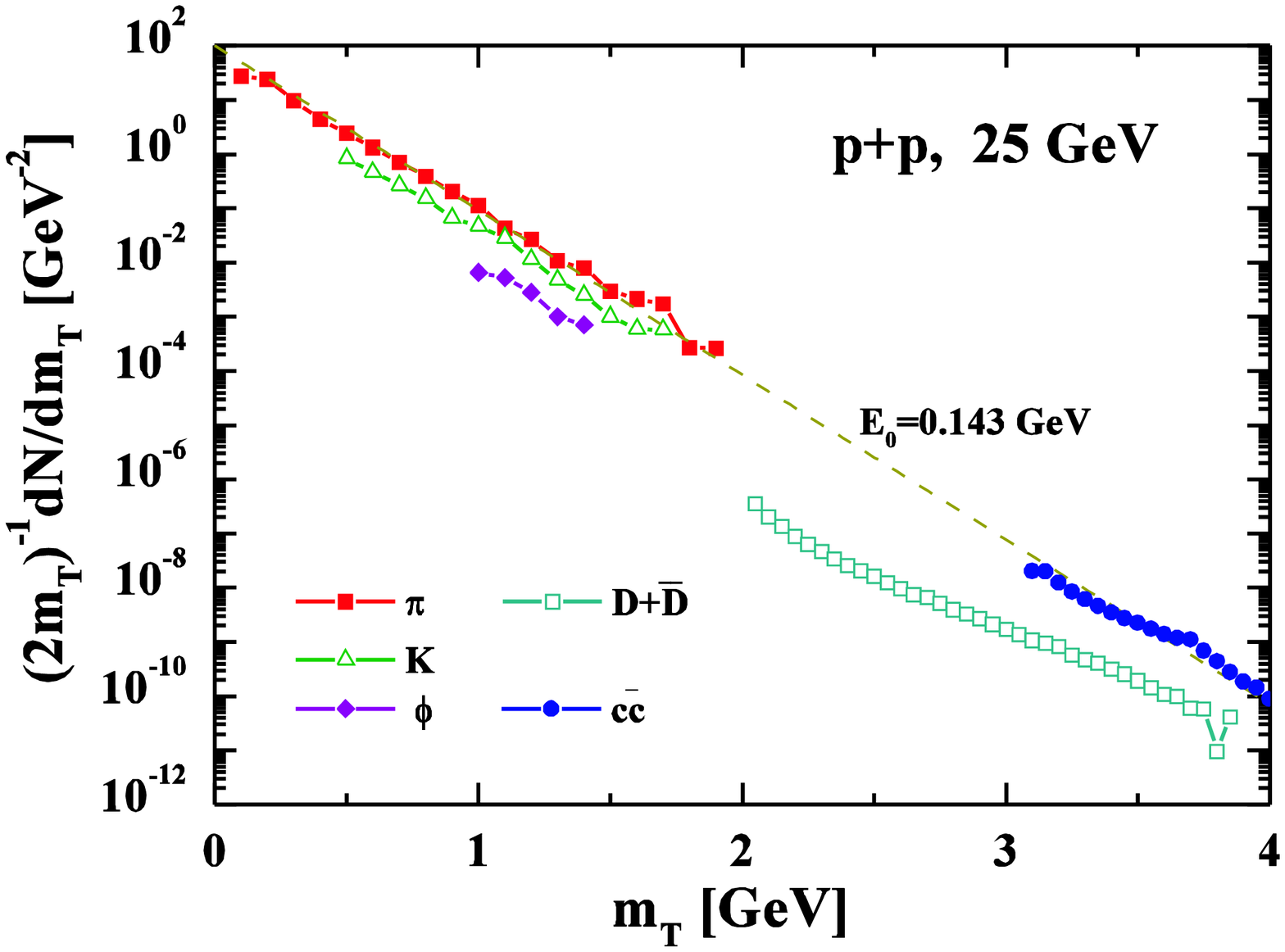,width=0.65\textwidth}}
\centerline{\psfig{file=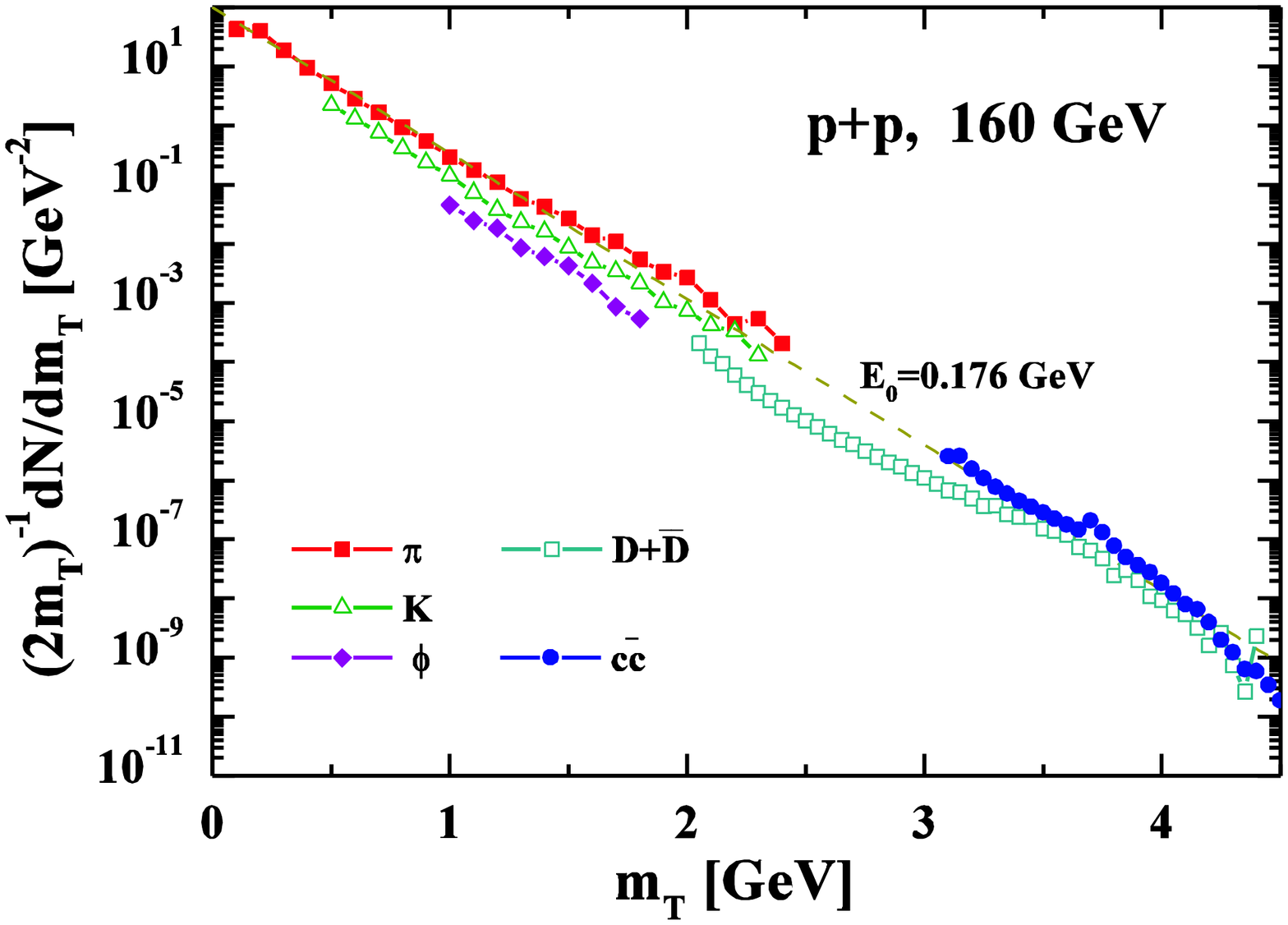,width=0.65\textwidth}}
\centerline{\psfig{file=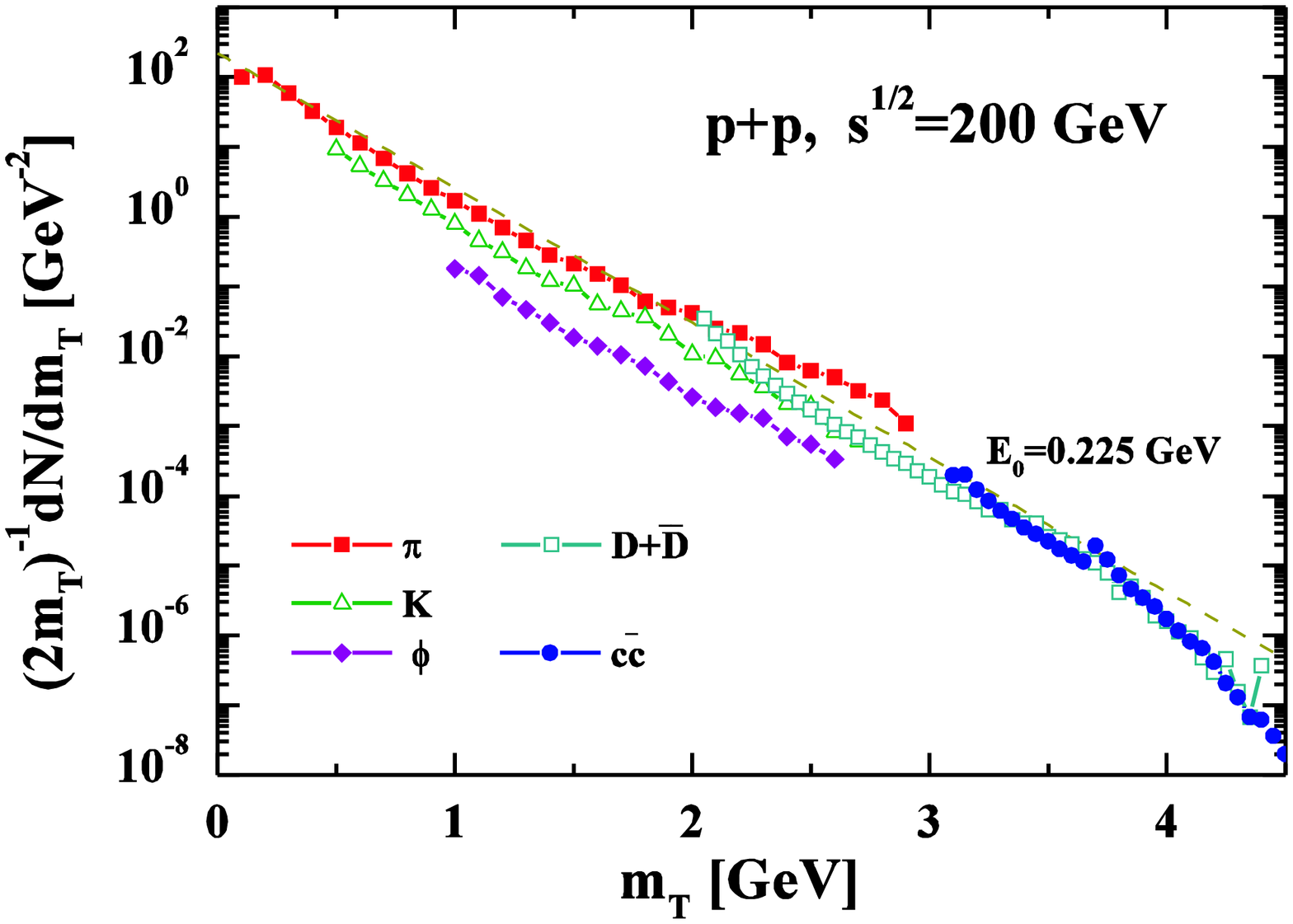,width=0.65\textwidth}} \caption{Upper plot: The transverse
mass spectra from $pp$ collisions at $T_{lab}$ = 25 GeV for pions (full squares), kaons (open
triangles), and $\phi$-mesons (full rhombi) from the string model as implemented in HSD. The
dashed line shows an exponential with slope parameter $E_0$ = 0.143 GeV. Middle plot: The same as
the upper plot, but for $pp$ reactions at $T_{lab}$ = 160 GeV. The dashed line shows an exponential
with slope parameter $E_0$ = 0.176 GeV. Lower plot: The same as the upper plot, but for $pp$
reactions at $\sqrt{s}$ = 200 GeV. The dashed line shows an exponential with slope parameter $E_0$
= 0.225 GeV. The figures are taken from Ref.~\protect\cite{Cass01}.} \label{Sib1}
\end{figure}

We recall again that
HSD successfully reproduces the measured transverse mass $ m_T = \sqrt{p_T^2 + m_X^2}$ spectra of
pions and kaons in $pp$ collisions for bombarding energies of 12~GeV, 24~GeV as well as
for collisions at
$\sqrt{s}=200$~GeV~\cite{Bratkovskaya:2004kv}. In Fig.~\ref{Sib1} we present additionally the
corresponding spectra for $\phi$-mesons, $D + \bar{D}$ mesons and charmonia at SPS, RHIC and FAIR
energies. We display the differential multiplicities $(2 m_T)^{-1} dN_X/dm_T$ in the transverse
mass. The pion spectra describe the sum of $\pi^+, \pi^0, \pi^-$, the kaon spectra the sum of $K^+,
K^0, \bar{K}^0, K^-$, the $D$-meson spectra the sum of all $D, D^*,D_s,D_s^*$ and their
antiparticles while the spectrum denoted by $c\bar{c}$ includes the $J/\Psi$, the $\chi_c$ as well
as the $\Psi^\prime$, where the latter contribution starts at $m_T \approx 3.7$ GeV and becomes
visible as a tiny kink in the $m_T$-spectra. Here the open charm and charmonia results stem from
the parametrizations specified in Section~\ref{elementary} (including the decay $\chi_c \rightarrow
J/\Psi + \gamma$), while the spectra for pions, kaons and $\phi$-mesons are from the LUND string
model as implemented in the HSD transport approach. For orientation, we also show exponential
spectra with slope parameters of 143 MeV, 176 MeV and 225 MeV, respectively, which describe the
$m_T$-spectra of pions rather well. The kaon spectra at all energies are down by a factor of $\sim$
3, the $\phi$ spectra by a factor of 9-10 relative to this line due to strangeness suppression in
$pp$ collisions.  However, it is quite remarkable that the charmonia spectra fit well to this
approximate $m_T$-scaling (within a factor of 2-3) at $\sqrt{s}$ = 7.1, 17.3 and 200 GeV,
respectively. Furthermore, the spectrum of open charm is roughly compatible with $m_T$-scaling at
$\sqrt{s}$ = 17.3 and 200 GeV, while the $D, \bar{D}$ mesons are suppressed relative to the scaling
by a factor $\sim$ 30 close to threshold ($\sqrt{s}$ = 7.1 GeV). Such an `apparent' statistical
production of mesons in elementary reactions has been advocated before by
Becattini~\cite{Becattini}. We have to stress, however, that all these observations on the charm
sector are based on our extrapolations (Section~\ref{elementary}) and have to be checked
experimentally.

\subsection{SPS and RHIC energies}

\begin{figure}
\psfig{file=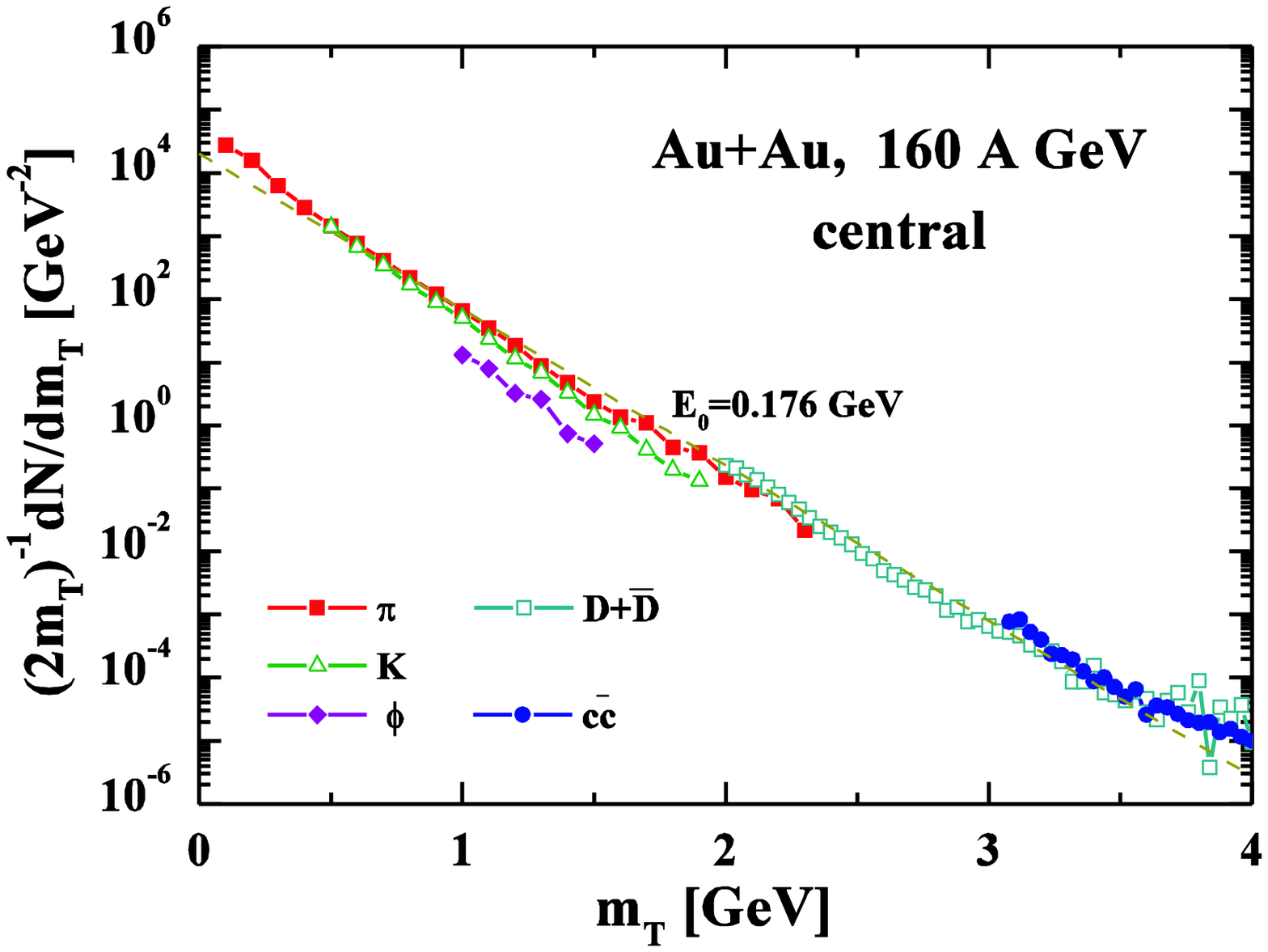,width=0.49\textwidth} \psfig{file=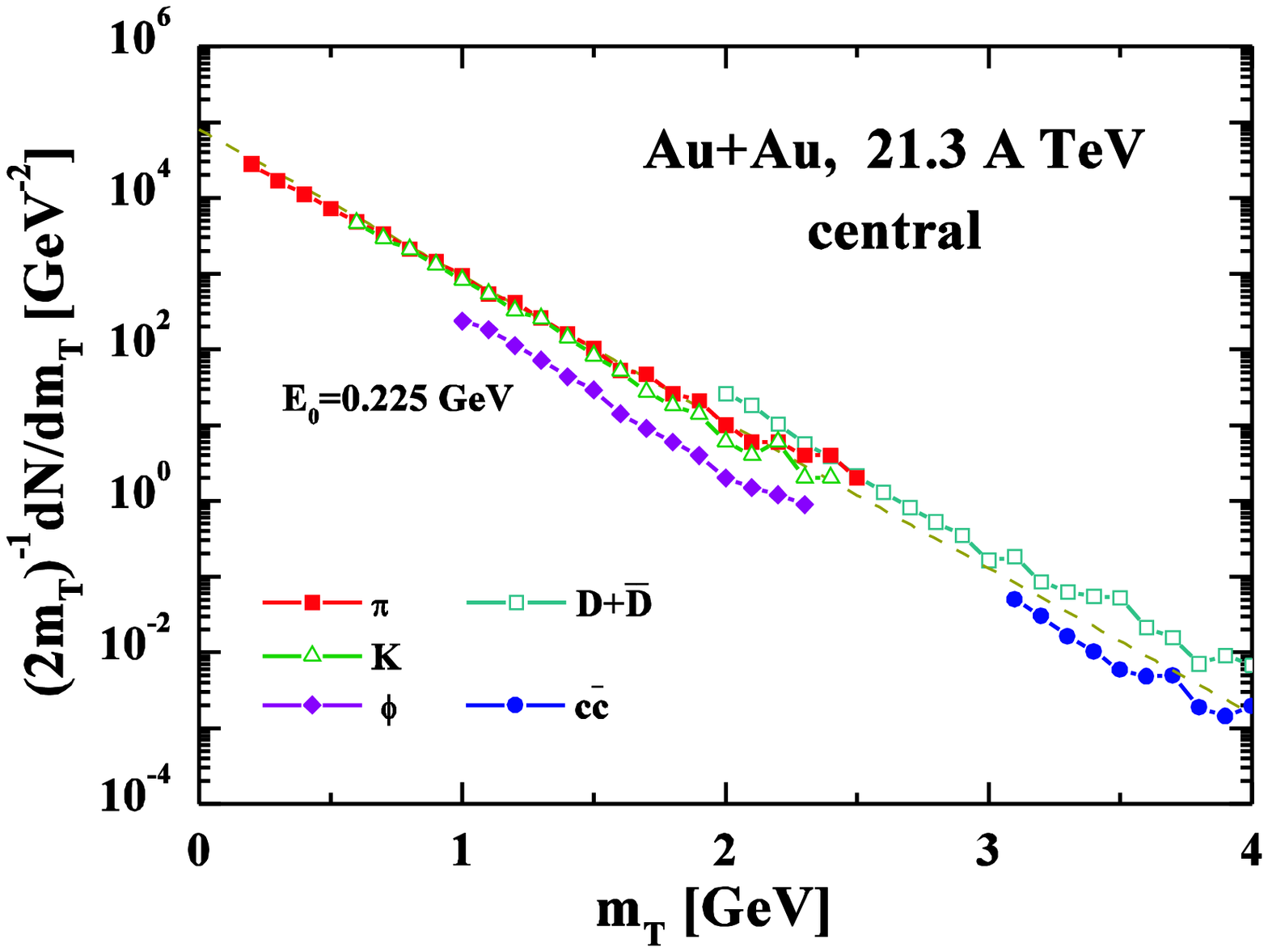,width=0.49\textwidth}
\caption{Left panel: The transverse mass spectra of pions (full squares), kaons (open triangles),
$\phi$-mesons (full rhombi), $D+\bar{D}$ mesons (open squares) and $J/\Psi, \Psi^\prime$ mesons
(full dots) in the HSD approach for a central $Au+Au$ collision at 160 A$\cdot$GeV. The thin dashed
line shows an exponential with slope parameter $E_0$ = 0.176 GeV. Right panel: Same as left panel,
but at top RHIC energy of 21.3 A$\cdot$TeV. The thin dashed line shows an exponential with slope
parameter $E_0$ = 0.225 GeV. Both figures are taken from Ref.~\protect\cite{Cass01}.} \label{Sib2}
\end{figure}

As mentioned before (cf. Ref.~\cite{Bratkovskaya:2004kv}) the
hadron spectra  at high $m_T$ are underestimated both in HSD and
UrQMD compared to the data at energies from SPS to RHIC.
Nevertheless, it is illustrative to have a global view on the
transverse mass spectra of hadrons with different flavor in case
of central Au+Au collisions. In this respect Fig.~\ref{Sib2}
displays the results of the hadronic transport at SPS and RHIC
energies which illustrates an approximate meson $m_T$-scaling for
all flavors (cf . Ref.~\cite{Cass01}). The $m_T$-scaling for
pions, kaons, $D$-mesons and $J/\Psi$  in central collisions of Au
+ Au at top  SPS and RHIC energies is essentially due to an
approximate $m_T$-scaling in $pp$ collisions at $\sqrt{s}$ = 17.3
GeV and substantial $D, \bar{D}$ and $J/\Psi$ final state
interactions in the nucleus-nucleus case.

We recall that the measured transverse mass spectra of hadrons
(heavier than pions) at  AGS, SPS and RHIC energies show a
 `hardening' in central Au+Au collisions relative to $pp$
interactions ({\it cf.} Ref.~\cite{NA49_T}) for low transverse mass or
momentum.  This hardening is commonly attributed to  collective
flow, which is absent in the respective $pp$ or $pA$ collisions.
Consequently it is important to get precise data on open charm and
charmonium transverse momentum ($p_T$) spectra, since their slope might
give information on the pressure generated in a possible
early partonic phase \cite{Xu:2002hq}. This argument is expected
to hold especially for $J/\Psi$ mesons, since their elastic rescattering cross
section with hadrons should be small in the
hadronic expansion phase~\cite{vanHecke:1999jh,Heinz95} (see also
Section~\ref{quenching}).

\begin{figure}
\centerline{\psfig{file=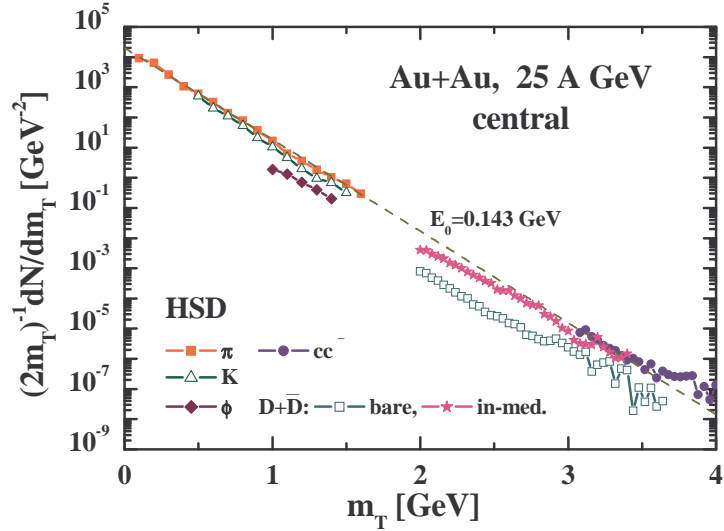,width=0.75\textwidth}}
\caption{The transverse mass spectra of pions (full squares),
kaons (open triangles), $\phi$-mesons (full rhombi),$D+\bar{D}$
mesons (open squares) and $J/\Psi, \Psi^\prime$ mesons (full dots)
in the HSD approach for a central Au+Au collision at 25
A$\cdot$GeV without including self energies for the mesons. The
crosses stand for the $D$-meson $m_T$ spectra when including an
attractive mass shift of $- 50 \rho/\rho_0$ MeV. The thin dashed
line shows an exponential with slope parameter $E_0$ = 0.143 GeV.
Note that final state elastic scattering of kaons and
$\phi$-mesons with pions has been discarded in the calculations.
The figure is taken from Ref.~\protect\cite{CBM.book}.}
\label{bild16}
\end{figure}
\subsection{FAIR energies}

In contrast to the observation at SPS energy, the approximate
$m_T$-scaling for pions, kaons, $D$-mesons and $J/\Psi$ no longer
holds  for central collisions of Au+Au at  25 A$\cdot$GeV as
demonstrated in Fig.~\ref{bild16}. Here the HSD calculations show
a suppression of $D$-mesons by a factor of $\sim$ 10 relative to
the global $m_T$-scaling - characterized by a slope of 143~MeV -
if no $D$-meson self energies are accounted for.

On the other hand, attractive mass shifts of $D, \bar{D}$ mesons
of -50~MeV at $\rho_0$  might be expected due to hadronic
interaction models when extending $SU(3)_{flavor}$ to
$SU(4)_{flavor}$ symmetry~\cite{Mishra}.  At the densities of
5-8 $\rho_0$ chiral symmetry should be restored, i.e. the large
$<q\bar{q}>$ condensate of the nonperturbative vacuum should have
disappeared. Accordingly, the production of $c\bar{c}$ pairs in
the `new' perturbative vacuum might be enhanced since only the
invariant mass of a $c\bar{c}$ has to be produced  e.g. by
gluon-gluon fusion. Accordingly, the threshold for $D\bar{D}$
production - which is $\sim$ 3.739 GeV in vacuum
 - might be reduced by $\approx2\cdot0.35$ GeV = 0.7 GeV in the
 chirally restored phase to about 3~GeV, only.
Such a reduction of the $c\bar{c}$ production threshold leads to
an enhancement of open charm mesons by about a factor of 7
(magenta crosses in Fig.~\ref{bild16}) such that an approximate
$m_T$-scaling for all mesons is regained. Thus, a global $m_T$
scaling of all mesons may be regarded as a strong medium effect on
the charmed hadrons and as a signature for a chirally restored
phase -- if observed by CBM.

\section{High $p_T$ quenching of open charm and charmonia}
\label{quenching}

\begin{figure}
\centerline{\psfig{file=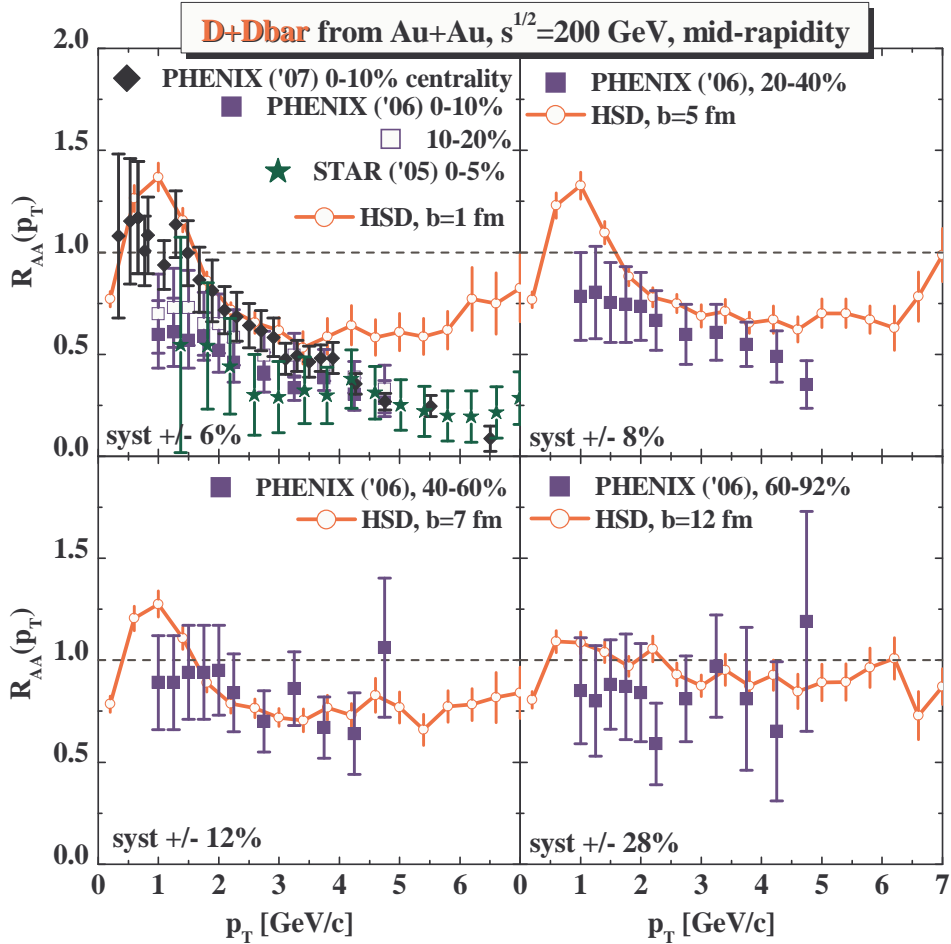,width=\textwidth}}
\caption{HSD predictions for the ratio of the final to the initial
(i.e. at the production point) transverse momentum spectra of $D+\bar
D$-mesons (solid lines with open dots, color: blue) from Au + Au
collisions at $\sqrt{s}=200$ GeV for $b=1, 5, 7$ and 12 fm at
mid-rapidity as calculated in Ref.~\protect\cite{brat05}. The
PHENIX data from Ref.~\protect\cite{LuizdaSilva:2006vw} (denoted
'06) and from Ref.~\protect\cite{PHENIXv2D} (denoted '07) on
$R_{AA}$ of non-photonic electrons as well as the STAR
data from Ref.~\protect\cite{Bielcik:2005wu} have been added later.} \label{D-Raa-PT}
\end{figure}

A significant suppression of high transverse momentum hadrons in
Au+Au collisions compared to $pp$ is observed at RHIC energies of
$\sqrt{s} =$ 200 GeV~\cite{survey,PHENIX_AA,STAR_AA,BRAHMS_dAu}
and
 is attributed to the energy loss of highly
energetic particles in a hot colored medium
(QGP)~\cite{Wang,Baier}.  In fact, the recent observation by the
PHENIX~\cite{PHENIX_dAu}, STAR~\cite{STAR_dAu} and
BRAHMS~\cite{BRAHMS_dAu} collaborations that a similar suppression
is not observed in d+Au interactions at mid-rapidities at the same
energy supports this idea.

Vigorous theoretical efforts are under way to understand parton
energy loss in terms of perturbative QCD (pQCD). Various groups
have described the suppression of light hadrons in terms of
radiative energy loss by gluon bremsstrahlung. According to such
calculations, charm and beauty quarks should be absorbed
significantly less than light quarks and gluons due to their
higher mass. However, data from the PHENIX and STAR experiments,
which compare the production in nucleus-nucleus to proton-proton
collisions of high-$p_T$ "non-photonic" electrons (which originate
mainly from heavy-flavor decays) seem to indicate that heavy
quarks loose energy in a comparable fashion as light quarks.

\begin{figure}
\centerline{\psfig{file=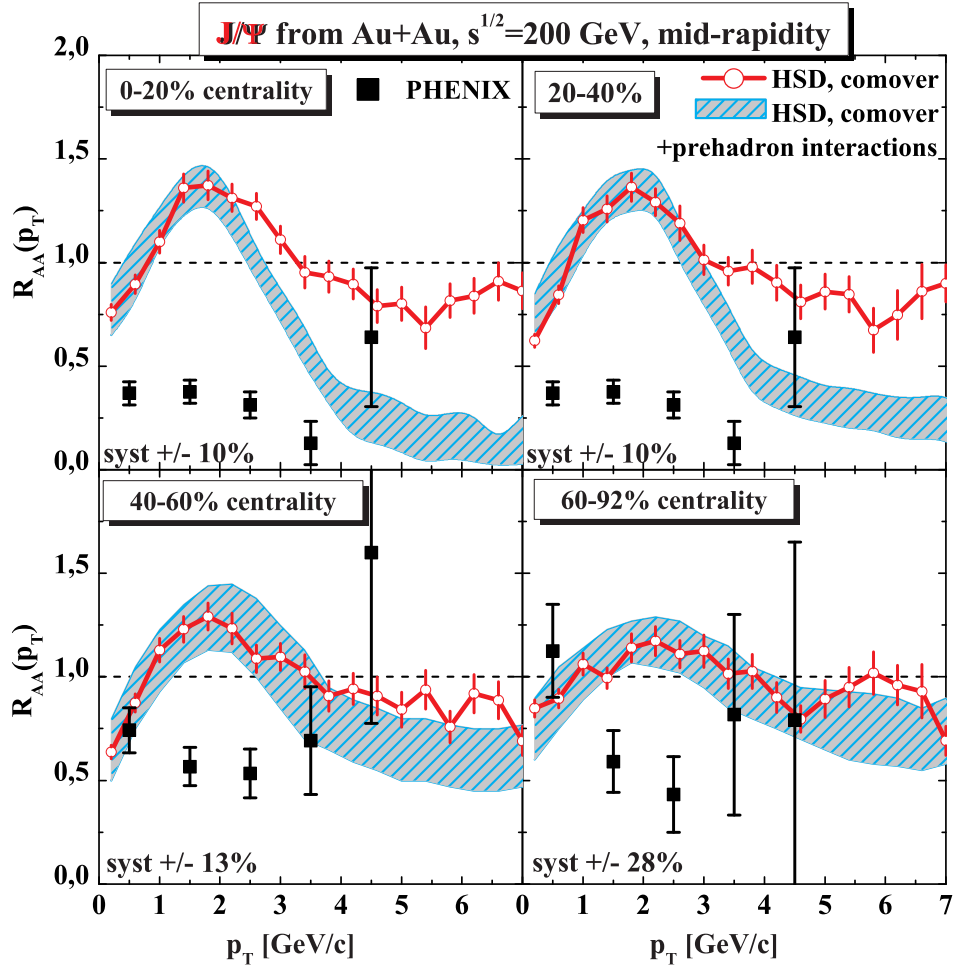,width=\textwidth}}
\caption{HSD predictions for the ratio of the final to the initial
$p_T$ spectra of
$J/\Psi$-mesons (solid lines with open dots, color: red) from Au +
Au collisions at $\sqrt{s}=200$~GeV for $b=1, 5, 7$ and 12 fm at
mid-rapidity as calculated in Ref.~\protect\cite{brat05}. The
preliminary PHENIX data for $R_{AA}(J/\Psi)$ from
Ref.~\protect\cite{PHENIXNov06} have been added later. The grey
dashed bands represent the results from the extended version of the HSD
comover approach, in which prehadron interactions are taken into
account as described in Section~\protect\ref{prehadron}.}
\label{JP-Raa-PT}
\end{figure}

In order to quantify the effect of hadronic final state
interactions, we show in Fig.~\ref{D-Raa-PT} the HSD predictions
from Ref.~\cite{brat05} for the ratio of the final to the initial
transverse $p_T$ spectra of $D+\bar D$-mesons (solid lines with
open dots, color: blue) from Au + Au collisions at $\sqrt{s}=200$
GeV calculated for impact parameter $b=1, 5, 7$ and 12 fm at
midrapidity. HSD predicts an enhancement of $D, \bar{D}$ mesons at
low momenta with a maximum at $p_T \approx $ 1 GeV/c and a
relative suppression for $p_T > $ 2 GeV/c. These effects increase
with the centrality of the Au+Au collision. We note that the
maxima in the ratios disappear when switching off the rescattering
with mesons in the transport approach. Thus a collective
acceleration of the $D+\bar{D}$ mesons occurs also via elastic
scattering with mesons. As seen in Fig.~\ref{D-Raa-PT} the
suppression seen by PHENIX may well be explained by hadronic
comover interactions up to transverse momenta about 4 GeV/c. Only
for higher $p_T$ a clear signal for parton energy loss - either
gluon bremsstrahlung or parton elastic scattering - may be
extracted in comparison to the data!

The suppression pattern $R_{AA}(J/\Psi)$ from  HSD (in the comover
scenario) is quite analogous to that of $D$-mesons for different
centrality showing a slight maximum for transverse momenta of
$\sim$ 2 GeV/c and a steady decrease for higher $p_T$.
In Fig.~\ref{JP-Raa-PT} the predictions (from Ref.~\cite{brat05}) for the
ratio of the final to the initial
transverse $p_T$ spectra of $J/\Psi$-mesons (solid red lines with
open dots) as well as new calculations in the extended version of
the comover approach -- as described in Section~\ref{prehadron} (grey
dashed bands) -- from Au + Au collisions at $\sqrt{s}=200$~GeV for
four different centrality regions at mid-rapidity are displayed.
The preliminary PHENIX data for $R_{AA}(J/\Psi)$ from
Ref.~\cite{PHENIXNov06} - added later - show a substantially
different pattern especially for non-peripheral interactions. The
strong suppression for low $p_T$ $J/\Psi$ mesons seen experimentally suggests that not
primordial $J/\Psi$'s are accelerated during the dynamical
evolution but that at least a part of initially formed $J/\Psi$'s
are dissolved and created later e.g. by $c\bar{c}$ coalescence. We
stress that the reformation of charmonia in the hadronic phase (by
$D + \bar{D}$ etc.) carries the flow from the $D$-mesons and thus
does not lead to suppression at small $p_T$ as seen
experimentally. This observation supports the idea that part of
the charmonia are produced in the hadronization process!

\section{Collective flow}
\label{flow}

A further possible way to disentangle hadronic from partonic
dynamics is the elliptic flow $v_2(y,p_T)$ as a function of the
rapidity $y$ and transverse momentum $p_T$.  The flow $v_2(y,p_T)$
is driven by different pressure gradients in case of non-vanishing
spatial anisotropy
\begin{equation}
\epsilon_2 = < \frac{y^2 - x^2}{y^2 + x^2}>. \end{equation} Since
$\epsilon_2$ decreases fast during the expansion of a noncentral
nucleus-nucleus reaction, the magnitude of $v_2$ gives information
about the interaction strength or interaction rate of the early
medium.

The phenomenon of collective flow can generally be characterized in terms of anisotropies of the
azimuthal emission pattern, expressed in terms of a Fourier series
\begin{equation}
\frac{dN}{d\phi}(\phi) \propto  1+ 2v_1\cos(\phi) +2 v_2\cos(2\phi) + \dots \label{dndphi}
\end{equation}
which allows a transparent interpretation of the coefficients
$v_1$ and $v_2$. The dipole term $v_1$ arises from a collective
sideward deflection of the particles in the reaction plane and
characterizes the transverse flow in the reaction plane (the
``bounce-off''). The second harmonic describes the emission
pattern perpendicular to the reaction plane. For negative $v_2$
one has a preferential out-of-plane emission, called {\it
squeeze-out}. Pions at SIS energies exhibit a clear out-of-plane
preference~\cite{brill93,venem93} which is due to shadowing by
spectator nucleons~\cite{Larionov}.

Presently, the most employed flow observables are the in-plane and
elliptic flows~\cite{Voloshin}:
\begin{eqnarray}
v_1 = \left< \frac{p_x}{p_T} \right>, \ \ \ v_2 = \left< \frac{p_x^2 - p_y^2}{p_x^2 + p_y^2}
\right> \, . \label{v1def}
\end{eqnarray}
Here, $p_x$ denotes the momentum in $x$-direction, i.e. the
transverse momentum within the reaction plane and $p_y$ the
transverse momentum out of the reaction plane. The total
transverse momentum is given as $p_T = \sqrt{p_x^2 + p_y^2}$; the
$z$-axis is in the beam direction. The bounce-off, the squeeze-out
and the antiflow~\cite{Stocker79,Hofmann74} (third flow
component~\cite{Csernai99}) have been suggested as differential
barometers for the properties of compressed, dense matter from SIS
to RHIC. In particular, it has been
shown~\cite{Hofmann76,Stocker79} that the disappearance or
``collapse'' of flow might be a direct result of a first order
phase transition.

\subsection{SPS energies}

\begin{figure}
\centerline{\psfig{file=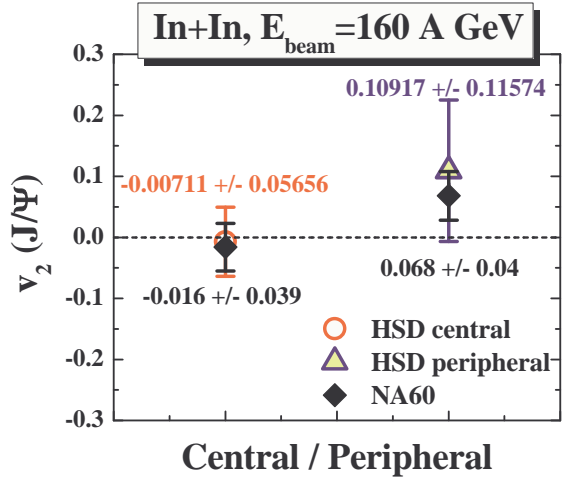,width=0.6\textwidth}} \caption{Elliptic flow $v_2$ of $J/\Psi$'s
produced in central and peripheral  $In+In$ collisions at 158~A$\cdot$GeV beam energy in the
hadronic `comover' mode of HSD (open circle and open triangle) compared to the NA60
data~\protect\cite{NA602007} represented by black diamonds. The figure is taken from
Ref.~\protect\cite{Olena.RHIC.2}.} \label{NA60v2}
\end{figure}

In Fig.~\ref{NA60v2} we compare the HSD result for $v_2 (J/\Psi)$
at SPS in the purely hadronic `comover' scenario in comparison to
the data for $v_2 (J/\Psi)$ of the NA60 collaboration for In+In
collisions~\cite{NA602007}. In central collisions the elliptic
flow is practically zero both in the calculation as well as in the
experiment, whereas in peripheral reactions a nonzero flow
emerges. The agreement (within error bars) between the theory and
the data indicates that, in line with the reproduction of the
$J/\Psi$ suppression data~\cite{Olena.SPS} (see
Section~\ref{dataSPS}), the low amount of $v_2$ does not point
towards additional strong partonic interactions at SPS energies.
Consequently, the present measurements of $J/\Psi$ elliptic flow
(at SPS energies) do not provide further constraints on the model
assumptions.

\subsection{RHIC energies}

\begin{figure}
\centerline{\psfig{file=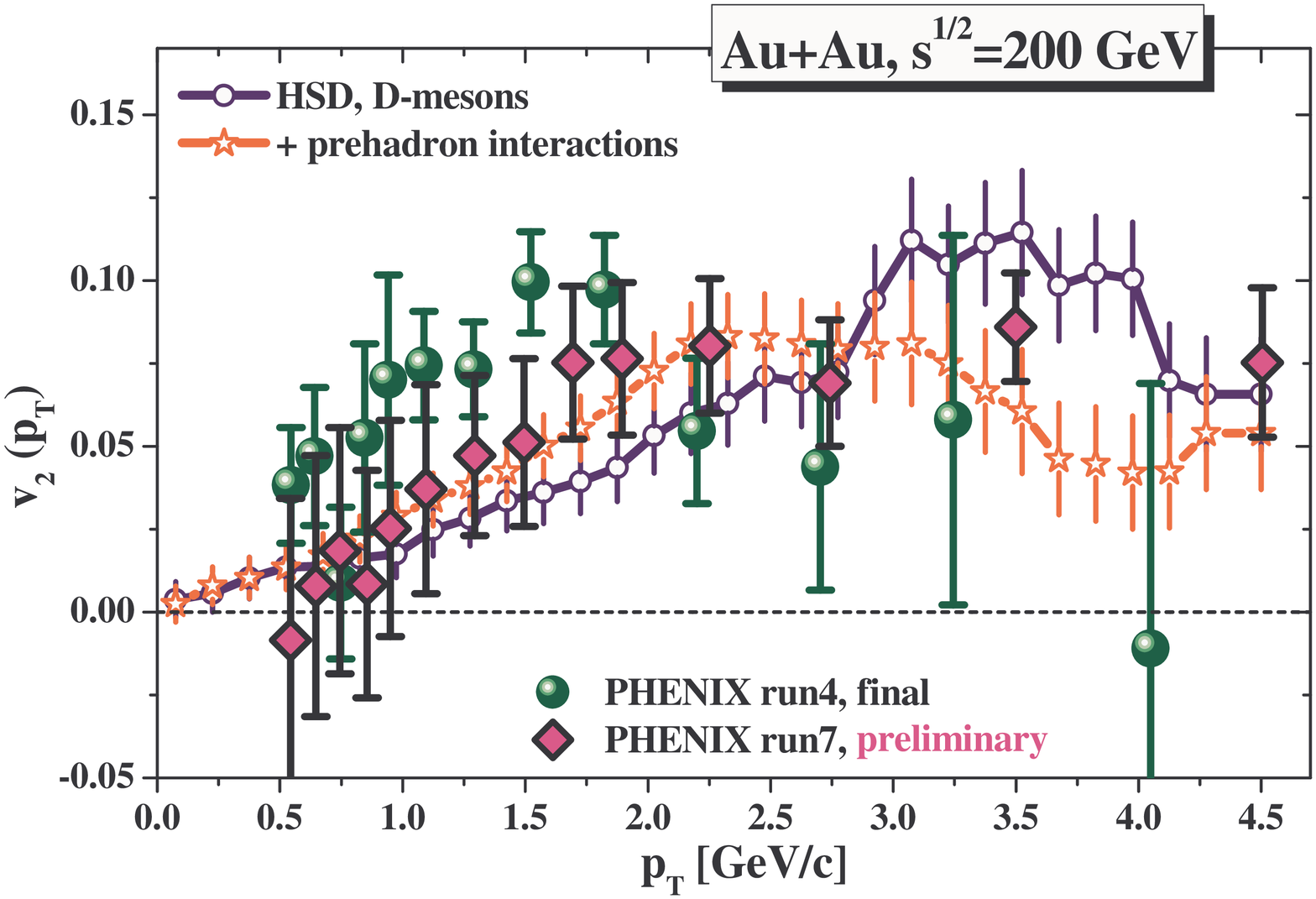,width=0.75\textwidth}} \caption{Elliptic flow of $D$-mesons
produced in $Au+Au$ collisions at $\sqrt{s}=200$~GeV as a function of $p_T$ from HSD (solid blue
line with open circles) in comparison to the PHENIX data~\protect\cite{PHENIXv2D} on $v_2$ of
non-photonic electrons. The red line with open stars shows the HSD result for the $v_2$ of
$D$-mesons when including additionally pre-hadronic charm interactions as described in
Section~\protect\ref{prehadron}. The figure is taken from Ref.~\protect\cite{Olena.RHIC.2}.}
\label{Dv2}
\end{figure}

\begin{figure}
\centerline{\psfig{file=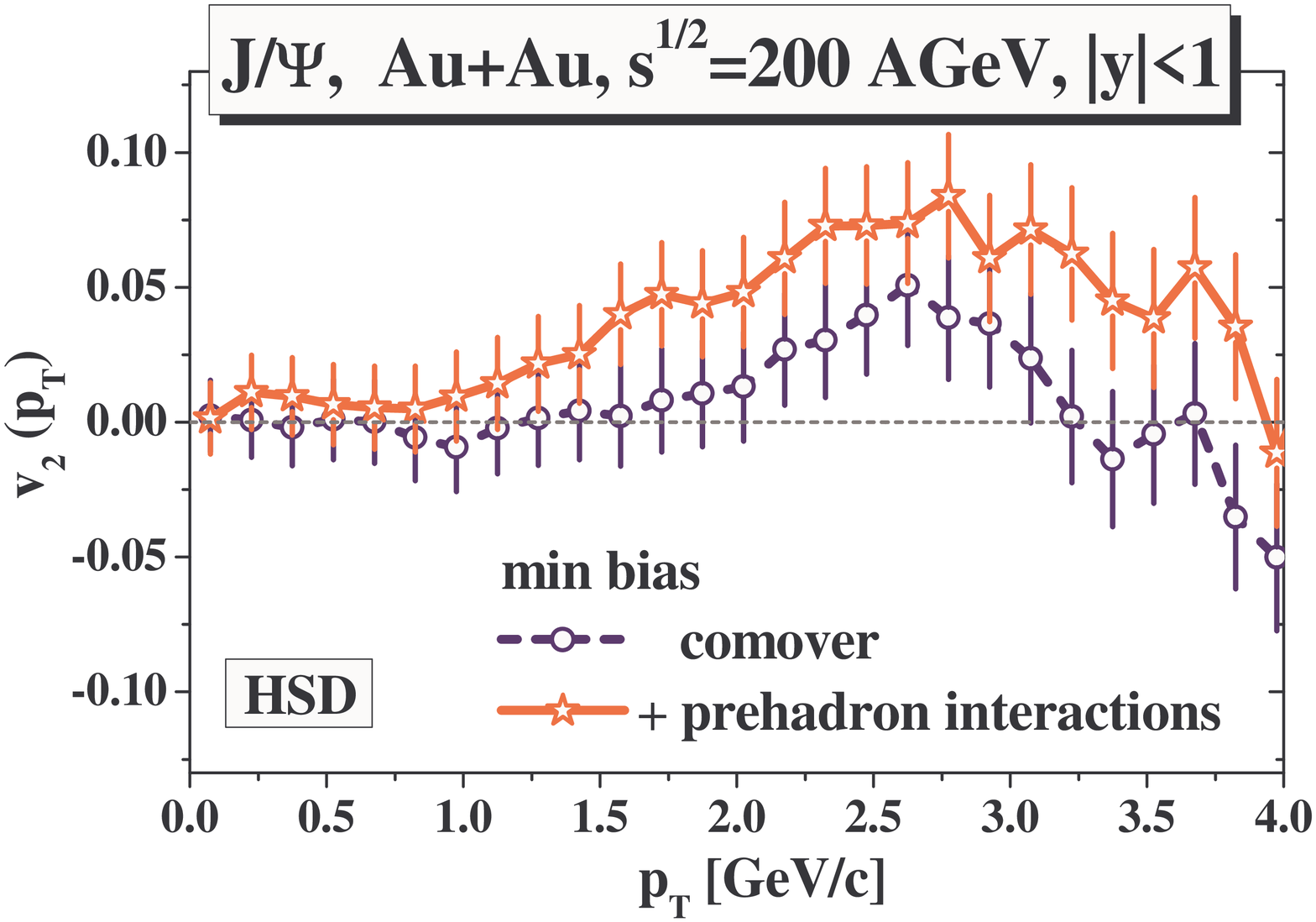,width=0.75\textwidth}}
\caption{Elliptic flow of $J/\Psi$-mesons produced in $Au+Au$
collisions at $\sqrt{s}=200$~GeV as a function of $p_T$ from HSD.
The red line with open stars shows the HSD result for the $v_2$ of
$D$-mesons when including additionally pre-hadronic charm
interactions as described in Section~\protect\ref{prehadron}. }
\label{JPsiv2}
\end{figure}

\begin{figure}
\centerline{\psfig{file=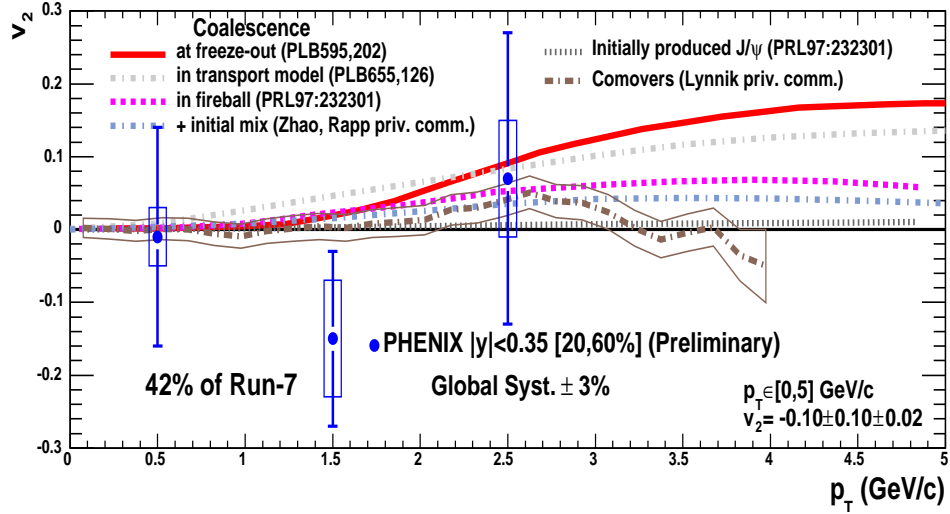,width=\textwidth,height=7cm}} \caption{Elliptic flow of
$J/\Psi$-mesons produced in $Au+Au$ collisions at $\sqrt{s}=200$~GeV as a function of $p_T$ from
HSD and other models. Preliminary PHENIX data~\protect\cite{GdC08,Silvestre:2008tw,Atomssa:2008dn}
are shown by symbols. The figure is taken from Ref.~\protect\cite{Silvestre:2008tw}.}
\label{otherJPsiv2}
\end{figure}

The situation, however, is different for the collective flow of
$D$-mesons at top RHIC energies. Though collective flow can be
described very elegantly in hydrodynamics ({\it cf.}
Refs.~\cite{Kolb,Teaney,Shuryak1,Nara}) by a proper choice of
initial conditions, one has to be very careful, since most
hydrodynamical calculations -- describing flow -- fail to
reproduce the hadron spectra with the same initial conditions (and
vice versa). Hydrodynamic flow and shock formation has been
proposed early~\cite{Hofmann74} as the key mechanism for the
creation of hot and dense matter during relativistic heavy-ion
collisions. However, the full three-dimensional hydrodynamical
flow problem is much more complicated than the one-dimensional
Landau model~\cite{Landau} used in many of the present
hydrodynamical calculations. The 3-dimensional compression and
expansion dynamics yields complex triple differential
cross-sections, which provide quite accurate spectroscopic handles
on the equation of state. In this respect, it is important to
consider also microscopic multi-component (pre-) hadron transport
theory (see Section~\ref{elementary}) as control models for
viscous hydro and as background models to extract interesting
non-hadronic effects from data.

In Fig.~\ref{Dv2} we show the elliptic flow of $D$-mesons produced
in $Au+Au$ collisions at $\sqrt{s}=200$~GeV as a function of the
transverse momentum $p_T$ in HSD (solid blue line with open
circles) compared to the PHENIX data~\cite{PHENIXv2D} on $v_2$ of
non-photonic electrons. Here the elliptic flow of $D$-mesons is
clearly underestimated in the default (purely hadronic) HSD model
({\it cf.} Ref.~\cite{brat05}). Only when including pre-hadronic
charm interactions - as described in Section 6.3 - the elliptic
flow moderately increases (red line with open stars), but still
stays below the PHENIX data. We thus conclude that the modeling of
charm interactions by pre-hadronic interactions - as described in
Section 6.3 - accounts for part of the non-hadronic generation of
the $v_2$, but does not provide enough interaction strength in the
early phase of the collision. Quite remarkably, this finding is
again fully in line with the underestimation of high $p_T$ hadron
suppression~\cite{HPT1} as well as far-side jet
suppression~\cite{HPT2} in the pre-hadronic interaction model.
Independently, also the charm collective flow points towards
strong partonic interactions in the early reaction phase beyond
the pre-hadronic scattering incorporated so far.

Since a large fraction of $J/\Psi$'s in central Au+Au collisions
at RHIC are created by $D- {\bar D}$ recombination, the elliptic
flow of $J/\Psi$'s  obtained from HSD in the comover (purely
hadronic) case is comparatively small, too (cf.
Fig.~\ref{JPsiv2}). The preliminary PHENIX
data~\cite{GdC08,Silvestre:2008tw,Atomssa:2008dn} are additionally
shown by blue symbols in Fig.~\ref{otherJPsiv2}, where the HSD
prediction is represented by a brown (dash-dot) line and a band.
The accuracy of the preliminary data so far does not allow for a
differentiation between the different model predictions presented
in Fig.~\ref{otherJPsiv2}.

In principle, partonic cascade
simulations~\cite{Geiger,Zhang,DMolnar1,DMolnar2,Bass,AMPT,Carsten}
are promising tools in approaching the mechanism of the early
generation of elliptic flow. However, unexpectedly high parton
cross sections of $\sim$ 5--10 mb have to be assumed in parton
cascades~\cite{DD4} in order to reproduce the elliptic flow seen
experimentally. These cross sections (per constituent quark) are
roughly the same in the partonic and hadronic phase (or even
higher in the partonic phase). In the new version of the parton
cascade model of Xu and Greiner~\cite{carsten08} lower binary
cross sections are employed but additional $2 \leftrightarrow 3$
gluonic reactions are incorporated that are very sensitive to the
screening masses (parameters) used. Furthermore, a strong elliptic
flow of particles may also be generated by repulsive mean-fields
as known from the nuclear physics context \cite{Larionov}. In all
the parton cascades addressed above no mean-fields or parton
selfenergies are incorporated. In fact, a recent study within the
Parton-Hadron-String-Dynamics (PHSD) approach suggests that a
large fraction of the final elliptic flow might stem from strong
repulsive partonic mean-fields \cite{PHSD.transport}.

\begin{figure}
\centerline{\psfig{file=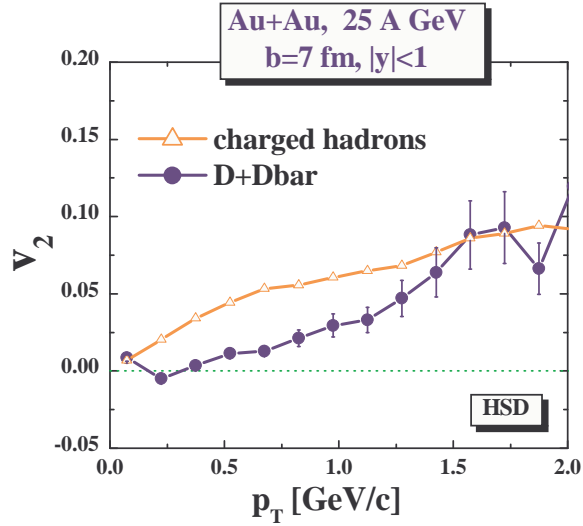,width=0.6\textwidth}} \caption{
The HSD predictions for the elliptic flow $v_2$ of $D+\bar
D$-mesons (solid lines with full dots) and charged hadrons (solid
lines with open triangles) from Au + Au collisions at
25~A$\cdot$GeV at $b=7$ fm versus $p_T$ for $|y| < 1.$ The figure
is taken from Ref.~\protect\cite{CBM.book}.} \label{Fig_v2DD}
\end{figure}

\subsection{FAIR energies}

The expected results for the $D$-meson elliptic flow $v_2$ at
mid-rapidity for Au+Au reactions at 25 A$\cdot$GeV from HSD is
compared in Fig.~\ref{Fig_v2DD} with the $v_2$ of charged hadrons.
As seen from \ref{Fig_v2DD} the $D,\bar{D}$ elliptic flow is
smaller than the $v_2$ of the lighter hadrons following
essentially the mass ordering. Such a low elliptic flow is due to
the moderate interaction rates of $D,\bar{D}$ mesons in the
hadronic expansion phase. A measurement of a sizeably larger
elliptic flow of $D$-mesons at FAIR would indicate the presence of
partonic degrees of freedom already at bombarding energies of 25 -
35 A$\cdot$GeV.

\section{Conclusions}
\label{conclusions}

This review has been devoted to an update of the information on
charm dynamics gained from p+A and A+A reactions at SPS and RHIC
energies in comparison to microscopic transport studies by
incorporating different scenarios for the anomalous suppression of
charmonia seen experimentally in central heavy-ion reactions. We
have found that present data on charmonium suppression for Pb+Pb
and In+In reactions at top SPS energies (158 A$\cdot$GeV) compare
well with transport calculations in the hadronic comover model
involving only a single parameter for the average matrix element
squared $|M_0|^2$ that fixes the strength of the charmonium cross
sections with comoving hadrons. This holds for the $J/\Psi$
suppression as well as the $\Psi^\prime$ to $J/\Psi$ ratio versus
collision centrality. The bare `QGP threshold scenario' gives
satisfying results for the $J/\Psi$ suppression for both systems
(In+In and Pb+Pb) at 158 A$\cdot$GeV but fails in the
$\Psi^\prime$ to $J/\Psi$ ratio since too many $\Psi^\prime$
already melt away for a critical energy density of 2 GeV/fm$^3$ at
158 A$\cdot$GeV. These findings suggest that the charmonium
dynamics in heavy-ion reactions is dominantly driven by hadronic
interactions in the SPS energy regime. Since energy densities
above 1 GeV/fm$^3$ are reached in central nucleus-nucleus
collisions at 158 A$\cdot$GeV, our observation indicates that
hadronic correlators (with quantum numbers of the familiar
hadrons) still persist above the critical energy density for the
formation of a QGP. This finding is independently supported by
lattice QCD calculations that show a survival of strongly bound
hadronic states up to temperatures of 2 $T_c$, which corresponds
to an energy density of about 30 GeV/fm$^3$. In fact, such high
energies densities (or temperatures) are not reached even in
central Au+Au (or Pb+Pb) collisions at the SPS.

The situation is found to be essentially different at top RHIC
energies of $\sqrt{s}$ = 200 GeV (or 21.3 A$\cdot$TeV). The study
of the formation and suppression dynamics of $J/\Psi$, $\chi_c$
and $\Psi^\prime$ mesons within the HSD transport approach for
Au+Au reactions at the top RHIC energy has demonstrated that both
scenarios i.e. `charmonium melting' in a QGP state as well as the
hadronic `comover absorption and recreation model' fail severely
at RHIC energies. This is found in the $J/\Psi$ suppression
pattern versus the centrality of the collision, in the $J/\Psi$
rapidity distribution as well as in the differential elliptic flow
of $J/\Psi$ and the charmonium nuclear modification factor
$R_{AA}$ as a function of transverse momentum $p_T$ (in comparison
to recent data from the PHENIX Collaboration \cite{PHENIXNov06}).
Especially the latter observable  indicates that at least part of
the final $J/\Psi$'s are created by coalescence of $c\bar{c}$
pairs in the hadronization phase. Only when including pre-hadronic
degrees in the early charm reaction dynamics, the general
suppression pattern of charmonia may be reasonably described;
though, the elliptic flow $v_2$ is still (slightly)
underestimated.
On the other hand, $R_{AA}(p_T)$ for $J/\Psi$ mesons cannot be described in the comover
approach, even when incorporating the early prehadron interactions.
 This analysis demonstrates that the dynamics of
$c, \bar{c}$ quarks in heavy-ion reactions at RHIC energies are
dominated by partonic interactions in the strong QGP (sQGP) and
can neither be modeled by `hadronic' interactions nor described
appropriately by color screening alone.

The different scenarios of `charmonium melting' and `hadronic
comover dissociation' may clearly be distinguished at FAIR
energies (of about 25 - 35 A$\cdot$GeV), where the centrality
dependence of the $J/\Psi$ survival probability and the
$\Psi^\prime$ to $J/\Psi$ ratio are significantly lower in the
`comover absorption'  model than in the `charmonium melting'
scenario. This result comes about since the average comover
density decreases only moderately with lower bombarding energy,
whereas the region in space-time with energy densities above
critical values of {\it e.g.} 2 GeV/fm$^3$ decreases rapidly and
ceases to exist below about 20 A$\cdot$GeV even in central
collisions of Au+Au.

\section{Open questions and perspectives}
\label{open}

Although substantial progress has been achieved on
the field of charmonium physics in the last years - both
theoretically and experimentally - the description of charmonium
and open charm dynamics in relativistic nucleus-nucleus collisions
remains a challenging task. So far we have just obtained a very
rough picture of the collective flow and attenuation pattern of
charmed and hidden-charm mesons and a profound microscopic
understanding is still lacking.

On the theoretical side the basic open problem - and future
challenge - is to incorporate explicit partonic degrees of freedom
in the description of relativistic nucleus-nucleus collisions and
their transition to hadronic states in a microscopic transport
approach. The available string/hadron transport models only
schematically incorporate some parton dynamics (by modeling
interactions of pre-hadrons) but definitely fail in describing the
partonic phase in accordance with an equation of state from
lattice QCD. On the other side the present partonic cascade
simulations (propagating massless partons) fail in describing the
reaction dynamics when employing cross sections from perturbative
QCD. One might argue that this is due to the neglect of partonic
selfenergies which even in case of small interaction strength
should be large due to the high parton densities reached in
ultra-relativistic nucleus-nucleus collisions. In fact, the
studies of Peshier in a dynamical quasiparticle approach
\cite{Andre04,Andre05} indicate that the effective degrees of
freedom in a partonic phase should be quite massive and have a
width in the order of the pole mass already slightly above $T_c$.
Such short lived degrees of freedom have to be propagated in
off-shell transport approaches which also allow for a description
of hadronization in terms of `local' rate
equations~\cite{PHSD.transport}.

On the experimental side, further differential spectra of
charmonia and open charm mesons will become available from RHIC-II
and the LHC  in the next years.  While studies at RHIC most likely
probe the `partonic fluid' produced in the heavy-ion reactions,
experiments at LHC - especially for high $p_T$ degrees of freedom
- might provide further surprising results with respect to the
properties of the partonic system at extreme energy densities.
Also the future charm measurements at FAIR are expected to provide
new insight into the charm dynamics especially at high baryon
density. Accordingly, the field of open charm and charmonium
physics is far from being closed and will remain exciting for a
long time.

\section*{Acknowledgements}

The authors gratefully acknowledge stimulating discussions with
A.~Andronic, P.~Braun-Munzinger,
H.~van~Hees,
J.~Stachel,
R.~Rapp,
P.~Senger,
L.~Tolos, A.~Ramos,
V.~D.~Toneev,
M.~I.~Gorenstein,
R.~Granier de Cassagnac, E.~Scomparin.
We are especially indebted to H.~St\"ocker for his continuous support, valuable suggestions and lively discussions.


\end{document}